\documentclass[usegraphicx,useAMS,usenatbib]{mn2e}
\usepackage{graphicx}
\usepackage{ulem}
\usepackage{amssymb}
\def\phox{PHOX}
\def\fig{Fig.\,}
\def\eq{Eq.\,}
\def\sec{Section~}

\def\mfive{M_{500}}
\def\rfive{R_{500}}

\def\xspec{{\tt XSPEC}}

\def\apec{{\tt APEC}}
\def\bapec{{\tt BAPEC}}

\def\vapec{{\tt VAPEC}}

\def\vmekal{{\tt VMEKAL}}
\def\wabs{{\tt WABS}}

\def\zsun{\rm{~Z_{\odot}}}
\def\msun{\rm{~M_{\odot}}}
\def\cm{\rm{~cm}}
\def\ergs{\rm{~erg/s}}
\def\kms{\rm{~km/s}}
\def\mpc{\rm{~Mpc}}
\def\kpc{\rm{~kpc}}
\def\kev{\rm{~keV}}

\def\om{\Omega_0}
\def\omb{\Omega_b}

\title[Observing simulated clusters with \phox{}]{Investigating the velocity structure and X--ray observable
  properties of simulated galaxy clusters with \phox{}}
\author[V. Biffi et al.]{V. Biffi$^{1,2}$\thanks{E--mail:
biffi@mpa-garching.mpg.de}, K. Dolag$^{3,1}$, H. B\"ohringer$^2$\\
$^{1}$Max--Planck--Institut f\"ur Astrophysik, Karl--Schwarzschild--Strasse 1, 85748 Garching bei M\"unchen, Germany\\
$^{2}$Max--Planck--Institut f\"ur extraterrestrische Physik, Giessenbachstrasse 1, 85748 Garching bei M\"unchen, Germany\\
$^{3}$University Observatory Munich, Scheinerstrasse 1, 81679 M\"unchen, Germany}

\begin{document}

\date{Accepted ... Received ... ; ...}
\pagerange{\pageref{firstpage}--\pageref{lastpage}} \pubyear{...}
\maketitle
\label{firstpage}
%--------------------------------------------------
%------------------ ABSTRACT ----------------------
\begin{abstract}
Non--thermal motions in the intra--cluster medium (ICM) are believed
to play a non--negligible role in the pressure 
support to the total gravitating mass of galaxy clusters.
Future X--ray missions, such as ASTRO--H and ATHENA, will eventually allow us to
directly detect the signature of these motions from
high--resolution spectra of the ICM.
In this paper, we present a study on a set of clusters extracted from
a cosmological hydrodynamical simulation, devoted to explore the role
of non--thermal velocity amplitude in characterising the cluster state
and the relation between observed X--ray properties.
In order to reach this goal, we apply the X--ray virtual telescope
\phox{} to generate synthetic observations of the simulated clusters
with both Chandra and ATHENA, the latter used as an example for the performance of
very high--resolution X--ray telescopes.
From Chandra spectra we extract global properties, e.g. luminosity and
temperature, and we accurately estimate the gas velocity
dispersion along the line of sight achievable from the broadening of emission lines from heavy
ions (e.g. Fe) resolved in ATHENA spectra.
Given the good agreement found between simulations (true, intrinsic
solution) and mock observations (detectable amplitude of 
non--thermal velocities), we further extend the analysis to the
relation between non--thermal velocity dispersion of the gas and the
$L_X-T$ scaling law for the simulated clusters. Interestingly, we find
a clear dependence of slope and scatter on the selection criterion for
the clusters, based on the level of significance of
 non--thermal motions. Namely, the
scatter in the relation is significantly reduced by the exclusion of
the clusters, for which we estimate the highest turbulent velocities.
Such velocity diagnostics appears therefore as a promising independent
way to identify disturbed clusters, in addition to the commonly used
morphological inspection.
\end{abstract}
%--------------------------------------------------
%--------------------------------------------------
\section{INTRODUCTION}\label{sec:intro}
During the mass assembly in galaxy clusters, interactions between
sub--haloes and merging events can generate substantial streaming motions and
turbulence in the hot gas filling the cluster potential well.
Additionally, a number of processes taking place in galaxy clusters,
especially in the
inner regions, are most likely responsible for the
transfer of energy from large modes into smaller modes, causing
rotation, streaming and, mainly, turbulent motions to establish in the
intra--cluster medium (ICM).
Among these physical processes, mergers and sloshing of dark matter
cuspy cores are believed to cause large scale motions, and, in addition,
the AGN activity and its interaction with the
surrounding gas can cause turbulence in the central region. 

Numerically, the non--thermal components of ICM motions have been
investigated by means of hydrodynamical simulations of galaxy clusters, which uniquely
provide full 3D information on the gas velocity field.
The establishment of ICM bulk, streaming and rotational motions
during the growth and assembly of simulated galaxy clusters is
believed to contribute, especially in the central part of the system,
to the cluster pressure support and therefore to the virial estimate
of the total
mass up to significant fractions \cite[e.g.][]{pawl2005,fang2009,lau2009,lau2010,biffi2011}.
Moreover, several studies on SPH and AMR simulations of cluster--like
haloes have addressed the turbulent velocity field in
clusters \cite[e.g., see recent work by][]{paul2011,vazza2011,valdarnini2011},
estimating the pressure support due to chaotic motions to be of order of
$\sim 20-30\%$ of the total pressure
\cite[e.g.][]{norman1999,dolag2005,iapichino2008,iapichino2011,vazza2009,vazza2011}.
Tighter constraints on the gas velocity field are necessary in order to obtain precise measurements of the total
gravitating mass, which is the most important, intrinsic 
quantity to determine. 
In particular, the account for non--thermal motions is
essential for mass estimates based on X--ray global
properties (e.g. gas density and temperature), via the assumption of
hydrostatic equilibrium \cite[e.g.][]{rasia2006,rasia2012,piffaretti2008}.

The presence of non--thermal motions in the gas velocity field within
galaxy clusters is also suggested by several observational evidences,
coming from radio observations of
polarized synchrotron emission in cluster radio galaxies \cite[e.g.][]{cassano2005,bonafede2010}, measurements
of the resonant scattering effects \cite[][]{churazov2010,zhuravleva2010} and study of the
fluctuations in pressure \cite[][]{schuecker2004} and surface brightness \cite[][]{churazov2011} maps, obtained with
X--ray telescopes.
However, mainly indirect indications of ICM turbulence have been
possible so far.
Only with the XMM--Newton RGS spectrometer weak upper limits have been
set on the turbulent velocities in a set of galaxy clusters, as recently
discussed in \cite{sanders2011}.

Future X--ray missions, like ASTRO--H, will allow us to achieve
direct estimations of the ICM non--thermal velocities with great
accuracy, thanks to the high spectral resolution expected to be reached.
In fact, X--ray high--precision spectroscopy potentially offers one of the most promising ways
to directly measure such gas motions, detectable from the detailed
study of the shape and centroid of resolved spectral emission lines.

Theoretically, the non--thermal component of the gas velocity, along
the line of sight (l.o.s.), can be very well constrained by studying the shape
of heavy--ion emission lines in the X--ray spectra, for which the broadening
can be significantly more sensitive to non--thermal velocities of the gas
rather than to thermal motion \cite[e.g.][]{rebusco2008}.
The expectations for such line diagnostics are related in particular
to the most prominent emission line in X--ray spectra, 
namely the $\sim 6.7 \kev$ line from helium--like iron. 
In fact, the large atomic mass of the FeXXV ion significantly reduces
the thermal line broadening and the line width turns out to be
definitely more sensitive to turbulent gas motions \cite[][]{inogamov2003,sunyaev2003}.

Here we discuss the prospect of using high resolution spectra to
detect the amplitude of non--thermal gas motions with the aim of
characterising more precisely galaxy clusters and observed relations
between their X--ray properties, like the $L_X-T$ scaling relation.
 
To this scope, we employ a set of numerically simulated clusters,
obtained with the TreePM/SPH parallel code P--GADGET3 including a
large variety of physical processes to describe the baryonic component, 
and perform X--ray synthetic observations of the haloes with \phox{}
\cite[][]{biffi2012}.
The paper is structured as follows: first, we will describe the
simulated dataset of galaxy clusters (\sec\ref{sec:sims}) and the generation of mock X--ray
spectra with the Chandra telescope and the X--ray spectrometer originally planned for the
ATHENA satellite, as prototype
for X--ray spectroscopy at very high energy resolution.
In \sec\ref{sec:data}, we will describe the analysis performed
to obtain global properties, such as luminosity and temperature, from
the Chandra spectra and
to estimate the gas non--thermal velocity dispersion, from the
velocity broadening of the iron lines in the high--resolution ATHENA
spectra.
The analysis on the gas velocity dispersion, calculated directly from
the simulation, is presented in \sec\ref{sec:results_sim}.
The comparison against the detectable velocities obtained from the
synthetic high--resolution spectra is then discussed in
\sec\ref{sec:results_mock}.
Using the Chandra mock observations, we explore the $L_X-T$ scaling
relation for the simulated clusters in \sec\ref{sec:lt}.
In \sec\ref{sec:discussion} we discuss
the effects on the best--fit relation, 
due to the non--thermal fraction of the ICM motions,
and the relation between velocity diagnostics and cluster internal state.
Our conclusions are finally summarized in \sec\ref{sec:conclusion}.

% -------------------------------------------------
\section{The sample of simulated clusters}\label{sec:sims}
The sample of simulated cluster--like haloes analysed here has been
extracted from a cosmological, hydrodynamical simulation performed
with the TreePM/SPH, parallel code P--GADGET3.
In this extended version of GADGET2 \cite[][]{springel2001,springel2005}, a vast
range of baryonic physics, at a high level of detail, is included,
such as cooling, star formation and supernova--driven winds \cite[][]{springel2003},
chemical enrichment from stellar population, AGB stars and SNe \cite[][]{tornatore2004,tornatore2007},
low--viscosity SPH \cite[][]{dolag2005}, 
black--hole growth and feedback from AGN \cite[][]{springel2005b,fabjan2010}.
The simulated box has a side of $352h^{-1}\mpc$, in comoving units,
resolved with $2\times 594^3$ particles, which results in a mass
resolution of $m_{DM} = 1.3\times10^{10}$ and $m_{gas} = 5.2\times10^{8},$ for dark matter (DM) and gas
particles, respectively.
For the simulation, and throughout the following, the cosmology
assumed refers to the 7--year WMAP estimates \cite[][]{komatsu2011},
i.e. $\om = 0.268$, $\Omega_{\Lambda}=0.728$, $\omb = 0.044$, $\sigma_8
= 0.776$ and $h = 0.704$.

The sample of clusters consists of $43$ objects, selected among the
most massive haloes in the simulated box, for a snapshot of the
simulation at $z=0.213$. The selection criterion adopted requires the cluster
mass encompassed by $\rfive$\footnote{Note that $\rfive$ is defined
  here as the radius enclosing the region of the cluster whose mean density
  is $500$ times the mean density of the Universe. This
  encompasses therefore a larger region with respect to the usual
  definition of $\rfive$, where the overdensity is instead defined
  with respect to the {\it critical} density of the Universe.}, $\mfive,$ to be $> 3\times
10^{14}h^{-1}\msun$, at the redshift considered.
\section{X--ray synthetic observations}\label{sec:data}
The mock X--ray observations have been performed for the selected
haloes of the sample by means of the virtual X--ray simulator \phox{}.
For a detailed description of the method implemented in \phox{}, we
refer to \cite{biffi2012}.
\subsection{Generation of the virtual photon lists}
The snapshot of the hydrodynamical simulation considered,
referring to redshift $z=0.213,$ has been first processed with Unit 1 of
\phox{} as a whole, generating the virtual photon cube associated to
the X--ray emitting gas component in the simulated cosmic volume.

For all the gas particles, model spectra have been calculated and
sampled with packages of photons, each of them in the restframe of the
corresponding emitting particle. 
The model adopted to calculate each theoretical spectrum was an
absorbed, single--temperature
\vapec{}\footnote{\scriptsize{See http://heasarc.gsfc.nasa.gov/xanadu/xspec/manual/XSmodelApec.html.}}
model \cite[][]{apec2001}, implemented in
\xspec{}\footnote{\scriptsize{See
    http://heasarc.gsfc.nasa.gov/xanadu/xspec/.}}
\cite[][]{xspec1996}. 
Here, we make use of the parameter within \xspec{} (v.12) to include the
thermal broadening of emission lines in the model spectra. 
Temperature, density and chemical abundances (see
Appendix~\ref{app:metallicity}, for details on the implementation of
metal composition in \phox{})
are obtained directly from the hydro--simulation output.
Additionally, we fixed the redshift at the value of
the simulated data cube ($z=0.213$) and the equivalent hydrogen column density 
for the \wabs{} absorption model \cite[][]{wabs1983} to a fiducial value of $10^{20} \cm^{-2}.$
At the end of this first stage, the photon database associated to the simulation
output contains roughly $10^7$ photon packages ($\sim10^9$ photons, in total), for fiducial values of
collecting area and exposure time of $1000\cm{}^2$ and $1{\rm Ms}$, respectively.

With \phox{} Unit 2 we then assume to observe the photons from a line
of sight (l.o.s.) aligned with the $z$--axis of the simulation cube and correct
photon energies for the Doppler Shift due to the emitting--particle
motions along this l.o.s..
During this geometrical stage of the process, we also select
cylindrical sub--regions along the l.o.s., corresponding to the
$43$ selected cluster--like haloes.
For each cluster halo, the selected sub--region is centered on the centre of
mass and encloses the region within
$\rfive$, in the $xy$ plan, throughout the box depth (along the $z$--axis). 
For the time frame considered here and the cosmology used, 
the luminosity distance between the observer, positioned along the
positive $z$--axis, and the observed region is $1047.6\mpc$, in
physical units.
Typically, we obtain $1-2\times 10^6$ photons per halo, for $100{\rm
  ks}$ exposure.

The photon lists produced in this general way, are
convolved by \phox{} with real instrumental responses of Chandra and
also for a high--energy resolution response, originally planned for 
the X--ray spectrometer of 
ATHENA.
The synthetic spectra are fitted by means of the X--ray package
\xspec{} v.12.6.0 \cite[][]{xspec1996}.
\subsection{Chandra synthetic spectra}
To create Chandra synthetic spectra, we use the ARF and RMF of the
Chandra ACIS--I3 detector aimpoint. 
The FoV of Chandra, which is $17'\times17'$, corresponds to a physical
scale of $3.52\mpc$ per side, for the given cosmology and redshift.
This encloses typically the region within $1-1.2\rfive{}$ for most of the clusters in
the sample,
except for the $7$ most--massive haloes, for which the Chandra FoV
captures a region slightly smaller than the one out to $\rfive{}$.
In such cases, however, one can in principle compose a mosaic with
multiple Chandra pointings in order to cover the clusters up to $\rfive{}$.
Therefore, we convolve the photon lists corresponding to the whole
$\rfive{}$ region with the response of
Chandra and analyse the corresponding spectra, for all the clusters in
the sample.

\subsubsection{Temperature and bolometric luminosity}
We fit Chandra spectra
either with an absorbed, single--temperature
or with a two--temperature \apec{} model, depending on the
goodness of the single--temperature fit.
The spectra are re--grouped requiring a
minimum of $10$ counts per energy bin. Whenever the spectral fit was
still poor, spectra were instead re--grouped with a minimum of $20$ counts.
The absorption is fixed to the fiducial value adopted to generate
photons. Given the characteristics of the Chandra response, we assumed
the redshift to be fixed at the initial value of the simulation and
only temperature, metallicity and normalization were free in the fit.

From the spectral best fit in the $0.5-10\kev$ range we directly obtain
the temperature of the cluster\footnote{In the case of haloes fitted
  with a two--component model, we assume here the hotter temperature
  to be representative of the dominant gas component.}.

The total X--ray luminosity, $L_X$, is extrapolated from the best--fit spectral model out
to the entire energy range defined by the ACIS--I3 detector response,
to obtain an approximate bolometric luminosity.
\subsection{High--resolution synthetic spectra}
In the present work, we consider the ideal instrumental response of
the spectrometer designed for the ATHENA satellite, which has, however, not been chosen as the next
ESA L--class mission in the recent mission selection process. 
Nevertheless, we use this as proptotype and goal for next--generation
high--resolution X--ray spectrometers (similar to ASTRO--H),
fundamental to explore the details of hot plasma emission.

Therefore, the mock spectra are simulated from the ideal photon lists of the $43$
haloes in the sample by using the latest RSP response matrix
planned for the ATHENA X--ray Microcalorimeter Spectrometer (XMS).
For the given cosmological parameters and redshift, the small FoV of
such instrument ($2.4'\times 2.4'$) encloses a region of $497.52\kpc$
per side.
We center the field of view on the centre of the cluster, probing therefore the
very central part.
\subsubsection{Velocity broadening}\label{sec:mock_sig}
\begin{figure}
          \centering
          \includegraphics[width=0.45\textwidth]{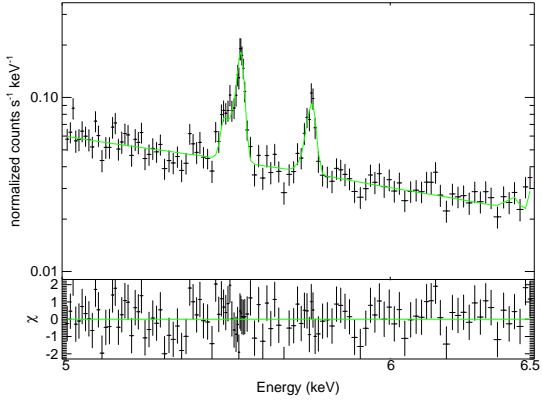}
          \caption{Zoom onto the $5-6.5\kev$ energy
            band, containing 
            the He-- and H--like iron emission lines (at the rest--frame energy $\sim 6.7\kev$ and
            $\sim 6.96\kev$),
            from the synthetic ATHENA XMS spectrum of one of the
            sample cluster. In the figure we show the best--fit
            \bapec{} model to the data (green curve).}
          \label{fig:fe_spec}
	\end{figure}

Given the high energy resolution provided by the XMS spectrometer,
spectra emission lines are very well resolved.
In fact, we restrict the spectral analysis to the observed $5-6.5\kev$ energy band,
containing the iron K$\alpha$ complex around $6.7\kev{}$ (rest--frame
energy), and fit with an absorbed \bapec{}\footnote{\scriptsize{See
    http://heasarc.nasa.gov/xanadu/xspec/manual/XSmodelBapec.html.}} 
model (see \fig\ref{fig:fe_spec} for an example with one of the
presented haloes), i.e. a velocity-- and
thermally--broadened emission spectrum for collisionally--ionised diffuse gas.
The model assumes the distribution of the gas non--thermal velocity
along the l.o.s. to be Gaussian and the velocity broadening is
quantified by the standard deviation, $\sigma,$ of this distribution.
%--------------------------------------------------
\section{Results}

\subsection{Velocity statistics}\label{sec:results_sim}
Our first goal is to investigate the properties of the ICM velocity field
in the simulated data directly, in order to establish the level of reliability of
the mock data results, and therefore the validity of the expectations
for high--resolution spectroscopy.
Gas particles in the simulated haloes have been selected to reside
within $\rfive{}$ in the $xy$ plane, excluding the low--temperature
($T_{gas}<3\times10^4~{\rm K}$)
and star forming (gas particles with $>10\%$ of cold fraction) phase of
the gas.
This selection is done to resemble in the most faithful way possible the X--ray
emitting gas residing in the region observed with the X--ray virtual
telescope \phox{}.
In the following, we will always refer to
regions in the $xy$ plane, considering the cylinder--like volume  
along the line of sight, aligned with the $z$--axis of the simulated box.

By definition, we calculate the weighted value of the standard
deviation of the gas velocities along the l.o.s., $\sigma_w$ as
\begin{equation}\label{eq:sigma}
\sigma_w = \frac{\Sigma_i w_i (v_{i} - \overline{v}_{w})^2}{\Sigma_i w_i},
\end{equation}
where $w$ represents the quantity used to weight the velocity
(e.g. the mass or the emission measure, EM, of the gas particle),
$v_i$ is the l.o.s. component of the velocity for each particle and
$\overline{v}_{w}$ is the weighted mean value of the l.o.s. velocity.

\begin{figure}
          \centering
          \includegraphics[scale=0.46]{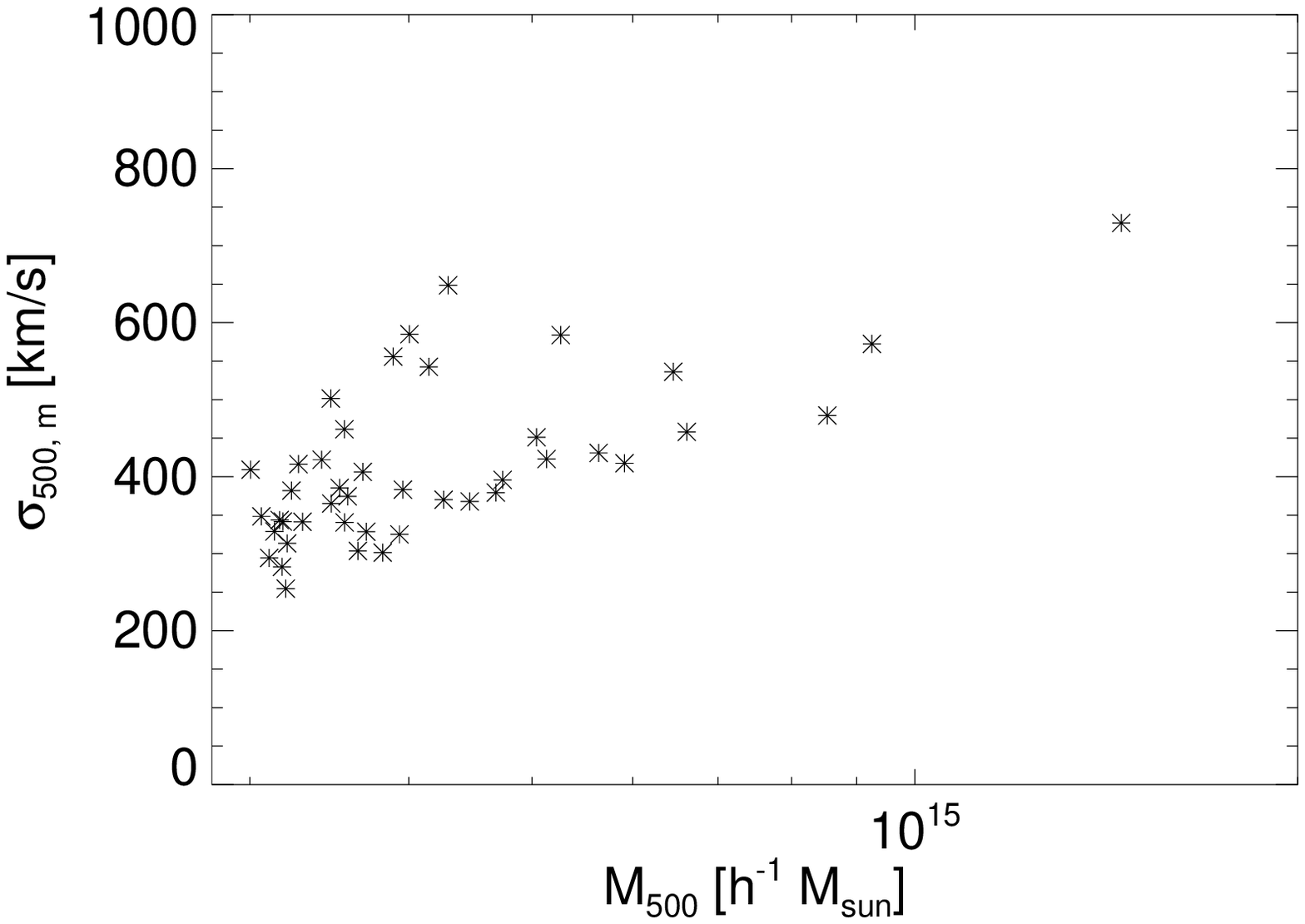}
          \caption{Theoretical value of the mass--weighted velocity
            dispersion, $\sigma_m$, calculated within $\rfive{}$ (in
            the plane perpendicular to the l.o.s. direction), reported
            as function of the halo mass $\mfive$, in $h^{-1}\msun$.}
          \label{fig:sig_mass}
%-------------------------------------------
          \includegraphics[scale=0.55]{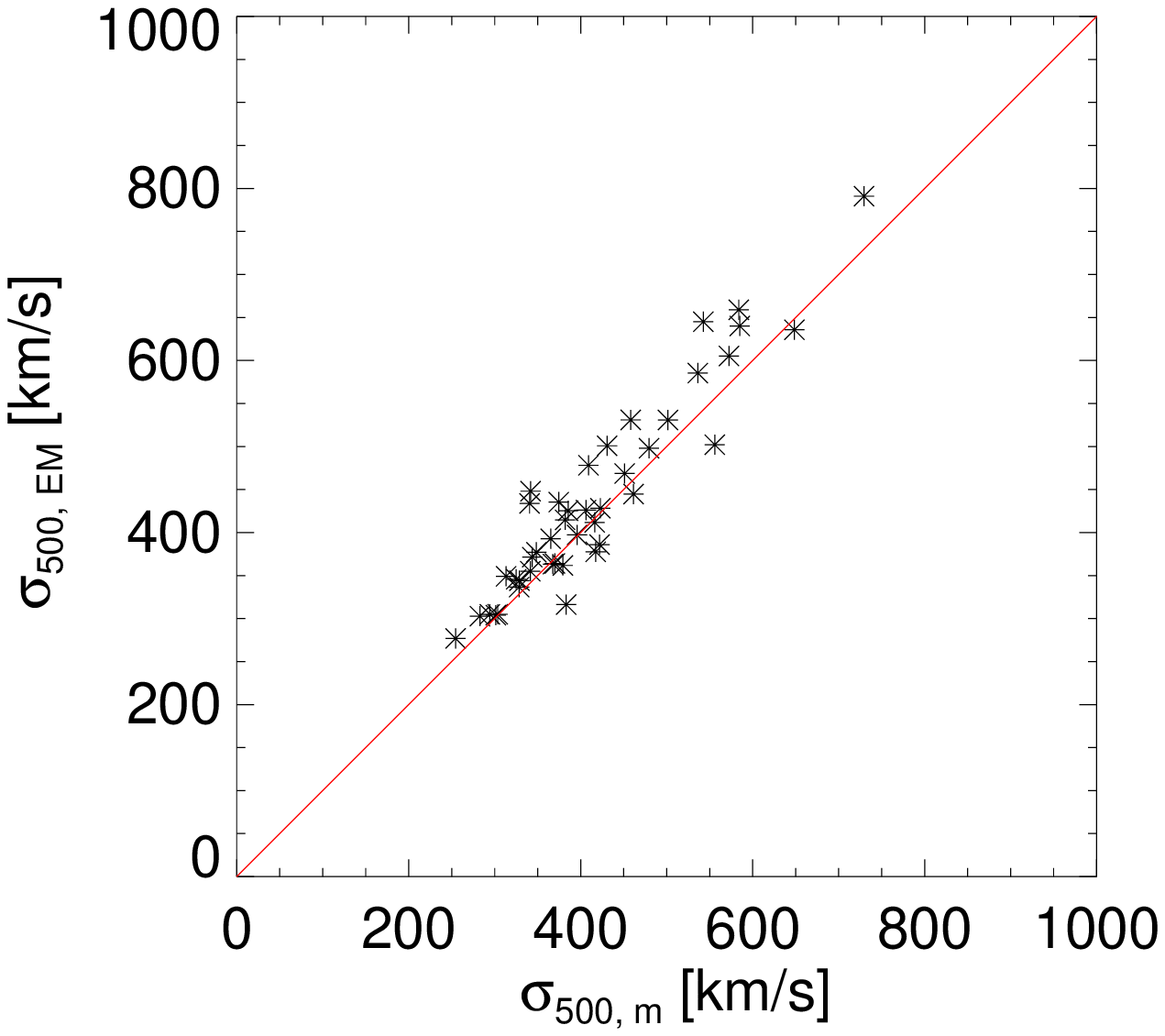}
          \caption{Theoretical value of the EM--weighted velocity 
            dispersion, $\sigma_{500,EM}$ versus the mass--weighted
            value, $\sigma_{500,m}$, in $\kms$. Both
            values are calculated for the region within $\rfive{}$, in
            the plane perpendicular to the
            l.o.s. direction. Overplotted in red is the curve
            referring to the one--to--one relation.}
          \label{fig:sigm_sigem}
	\end{figure}

From \fig\ref{fig:sig_mass}, the relation between the value of
$\sigma_{500,m}$ (i.e. weighted by the particle mass) and $\mfive{}$ shows that, in general, more significant motions of
the gas are expected in more massive systems.
Ideally, the mass--weighted values should
trace in the most faithful way the intrinsic
quantities characterising the cluster, since closely related to the potential well of the halo. 
However, X--ray observations of the ICM are more
sensitive to the emission properties of the gas, like the
emission measure (EM)\footnote{We recall here that ${\rm EM} = \int n_e n_H dV$.}, and  
rather provide estimates for an EM--weighted--like velocity
dispersion.
We therefore investigate the relation between mass--weighted and
EM--weighted velocity dispersion, $\sigma_{500,m}$ and
$\sigma_{500,EM}$ respectively.

The comparison is shown in \fig\ref{fig:sigm_sigem}, wherein
both values of $\sigma_{500,w}$ are calculated for the gas residing within
$\rfive$, in the plane perpendicular to the l.o.s..
The relation found between the two definitions of
$\sigma_{500,w}$ is not coincident with the one--to--one relation
(overplotted in red in the figure), as the EM--weighted value is
likely to be affected by the thermo--dynamical status of the
gas, that is by processes such as turbulence, merging and substructures.
Despite this, the two values are fairly well correlated. 

For the purpose of our following analysis and the comparison against
synthetic X--ray data, however, we decide to use
the EM--weighted velocity dispersion, $\sigma_{500,EM}$, which is more
directly related to the X--ray emission of the 
gas, because of the proportionality between the normalization of the X--ray
spectrum and the gas EM itself. 

\begin{figure}
          \centering
          \includegraphics[scale=0.55]{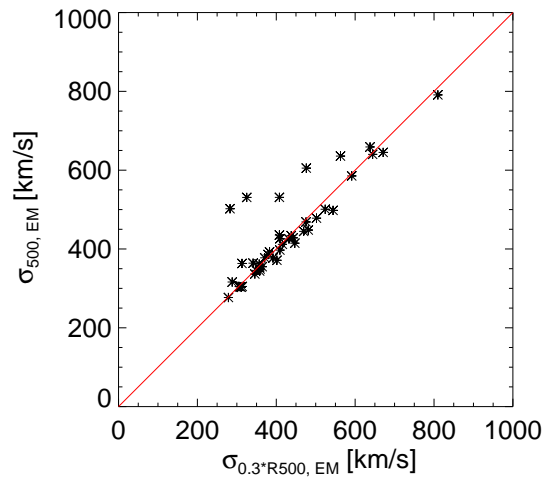}
          \includegraphics[scale=0.55]{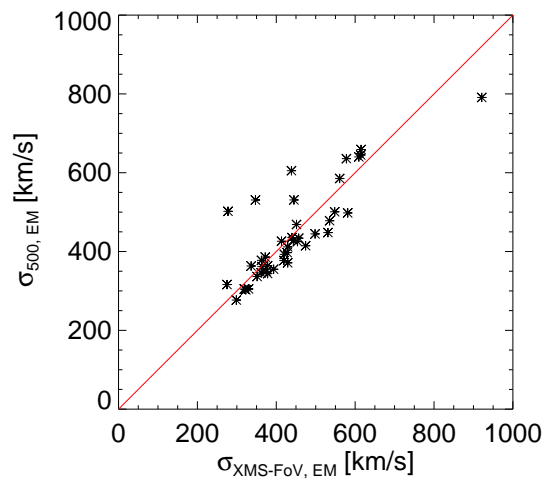}
          \caption{Relation between the EM--weighted velocity
             dispersion within $\rfive{}$, $\sigma_{500,EM}$, and the
             analogous values calculated for: (top panel) the region within
             $0.3\rfive{}$ and (bottom panel) the
             region covered by the FoV of ATHENA,
             (i.e. $\sim0.15\rfive{}$). Overplotted in red is the curve
             referring to the one--to--one relation.}
          \label{fig:sig2}
	\end{figure}

In order to probe the {\it global}, dynamical structure of the ICM we would need to
observationally measure the gas velocity dispersion within the whole
$\rfive{}$ region.
Therefore, we explore the relation between the estimated
value of the velocity broadening along the line of sight in different
regions of the cluster, shown in \fig\ref{fig:sig2}. 
It is evident from the figure that the value
calculated for the gas within $\rfive$ correlates linearly with the
value computed in smaller, internal regions, namely for
$r<0.3\rfive{}$ (upper panel) and for the region covered by the FoV of
ATHENA, $\sim0.15\rfive{}$ (lower panel).
With respect to the one--to--one correlation (red line in \fig\ref{fig:sig2}), however, outliers are
present in this sample, showing that prominent
substructures in the velocity field of the gas must be present in the
observed regions around the clusters. 
Here, as in \fig\ref{fig:sig_mass}, deviations can also be related to
subhaloes residing 
in the projected $\rfive{}$, i.e. in the region
observed along the line of sight, but not necessarily comprised within
the three--dimensional $\rfive{}$.

Therefore, the level of complexity in the spatial structure of the ICM
velocity field can be singled out by the comparison between the
velocity dispersion calculated in the $\rfive{}$ region and the values
corresponding to smaller, inner regions. \\

\noindent Nevertheless, the relations discussed ensure that, for the
purposes of this work, we can safely:\\
(i) assume the EM--weighted velocity dispersion instead of the
mass--weighted value to trace the intrinsic velocity structure;\\
(ii)
focus on the expected value for the whole $\rfive{}$ region of
the cluster. 
This traces fairly well the internal motions also within smaller
  regions, as those covered by instruments with a smaller FoV,
with the exception of those cases where self--bound substructures
in the outskirts, or more likely along the line of sight, are clearly
present and might be important to consider.

%------------ SEC:VELOCITY_STATISTICS
%------------------------

%------------
%------------

%--------------------------------------------------------------------
\subsection{Comparison against synthetic data}\label{sec:results_mock}
The ICM velocity dispersion calculated directly from the simulation
can here be used to compare against the mock ATHENA data, which
is the prototype for the high--resolution spectroscopy (X--ray
microcalorimeter spectrometer, XMS) required to
measure the gas velocity dispersion along the line of sight from resolved emission lines.

This will allow us to (i) test whether predictions from simulations are
indeed comparable to observational data and, in turn, (ii) constrain
the reliability of high--resolution X--ray spectroscopy to derive
direct measurements of the ICM velocity field.

\begin{figure}
          \centering
          \includegraphics[width=0.45\textwidth]{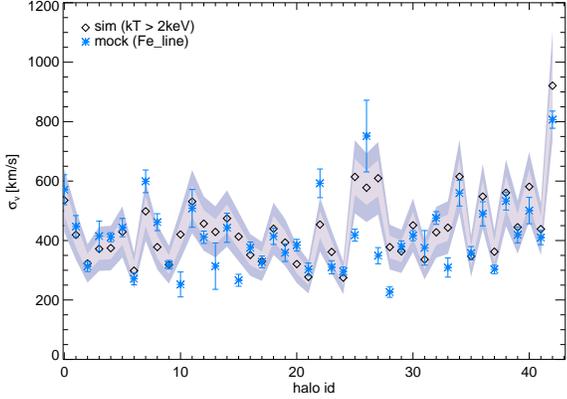}
          \caption{Comparison between the theoretical expectation of
            the velocity dispersion $\sigma_v$, calculated directly
            from the simulation (black diamonds and shaded areas), and the value obtained from the
            spectral fitting of the synthetic XMS spectra obtained
            with \phox{} (blue asterisks with error bars). The id numbers of the $43$ haloes in the
            sample ($x$--axis) are ordered according to the increasing
          halo mass, $\mfive{}.$}
          \label{fig:sigma_xms}
	\end{figure}

\fig\ref{fig:sigma_xms} shows the comparison between expectations
provided by the simulated data (black diamonds) and results from
analysis of the synthetic ATHENA spectra (blue asterisks).

The expected velocity dispersion of the gas particles
residing in the region corresponding to the ATHENA FoV is calculated
according to \eq\ref{eq:sigma}, weighted by the EM, while the values
derived from the X--ray spectra are obtained as described in
\sec\ref{sec:mock_sig}, with error bars corresponding to the $1\sigma$ errors to the best--fit values.
As shown in the Figure, we find very good agreement between simulation
(intrinsic, ``true'' solution) and synthetic spectral data
(observational detections), namely for
$\sim74\%$ of the haloes the spectral analysis of the iron lines
provides a measure of the gas velocity dispersion, along the
l.o.s., within $20\%$ from the expected value (purple, shaded area).
We find, in particular, that $\sim50\%$ of the halos show agreement at a level
better than $\sim10\%$ (internal, pink shaded area in
\fig\ref{fig:sigma_xms}).
We remark here that the reference number, or halo {\it id} in the
Figure, is associated to the haloes of the sample in an increasing
order for increasing $\mfive{}$. 

\begin{figure}
	\centering
	\includegraphics[width=0.45\textwidth]{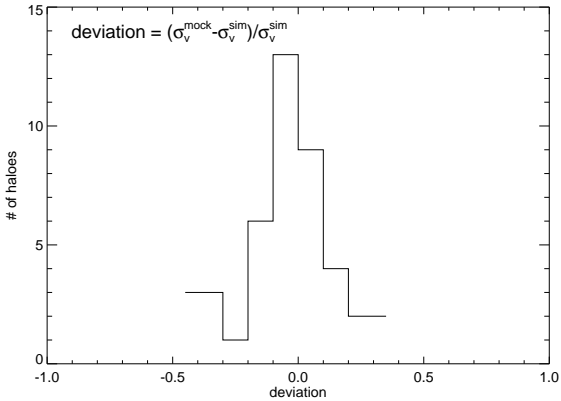}
	\caption{Deviation of the best-fit $\sigma_v$ from the
          expected value for all the 43 haloes in the
          sample.}
        \label{fig:dev}
	\end{figure}

The deviation between expected and measured velocity dispersion,
referring to the XMS FoV as in \fig\ref{fig:sigma_xms}, has been quantified as
\begin{equation}
  \delta = \frac{\sigma_v^{mock} - \sigma_v^{sim}}{\sigma_v^{sim}},
\end{equation}
and its distribution for the sample is reported in
\fig\ref{fig:dev}. 
Clearly, the distribution of $\delta$ is peaked
around the zero value, reflecting the very good agreement between the
two measurements.

However, we also find outliers in the sample that show deviations up
to $\sim43\%$.
%
%-------------------------------------------------------------------------------------------
\subsubsection{Extreme cases in the sample} 
Given the distribution of the deviations (\fig\ref{fig:dev}), we focus
onto two subsets of haloes in the sample for 
which the deviation between simulation and mock data is very minor and
most prominent, respectively.

\begin{figure*}
          \centering
          {\Large 5 MOST--DEVIATING HALOES \hspace{5cm} 5 LEAST--DEVIATING HALOES}\\
          \vspace{0.5cm}
          % halos deviating from expectation
          \includegraphics[width=0.22\textwidth]{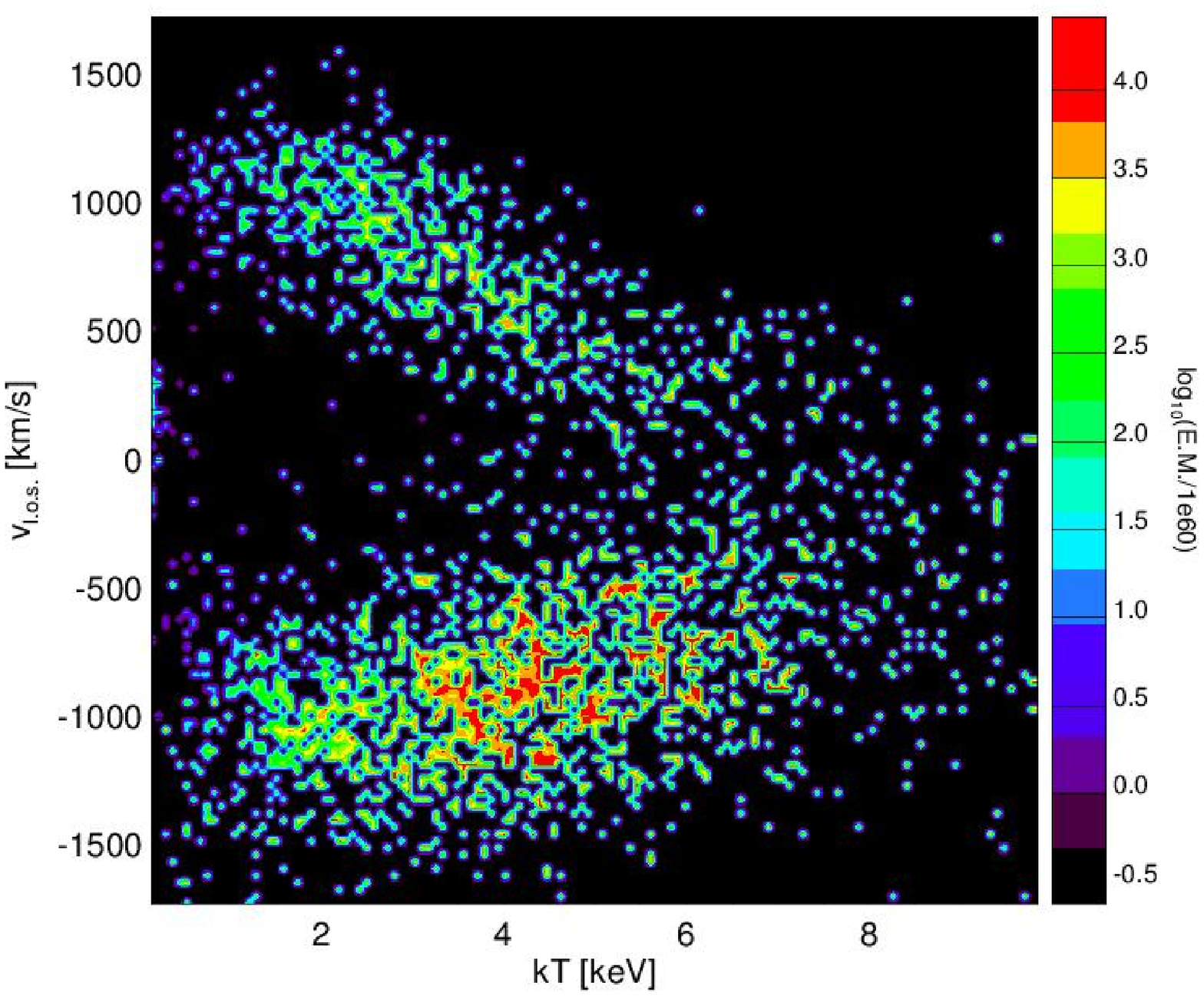}
          \includegraphics[width=0.24\textwidth]{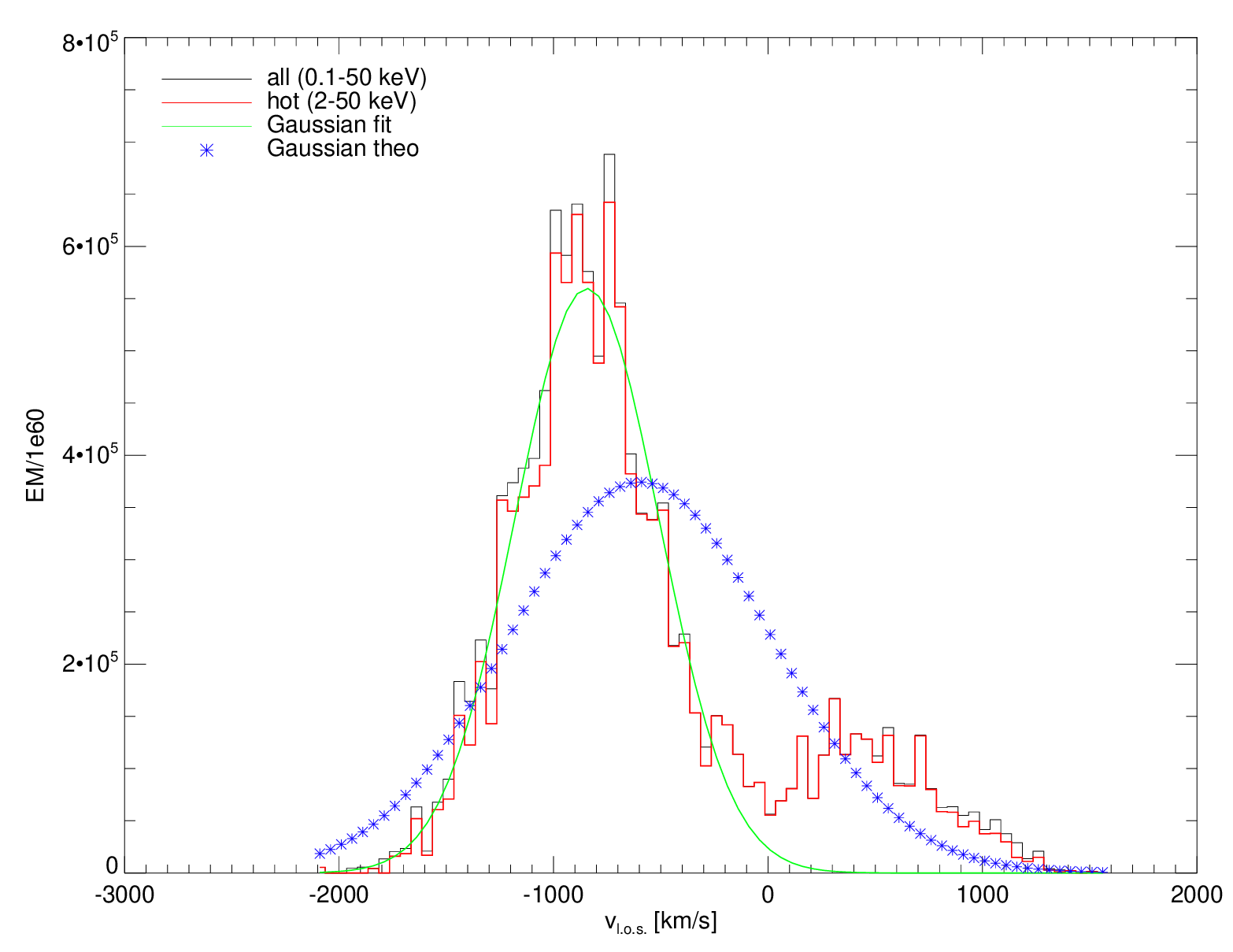}
          \includegraphics[width=0.24\textwidth]{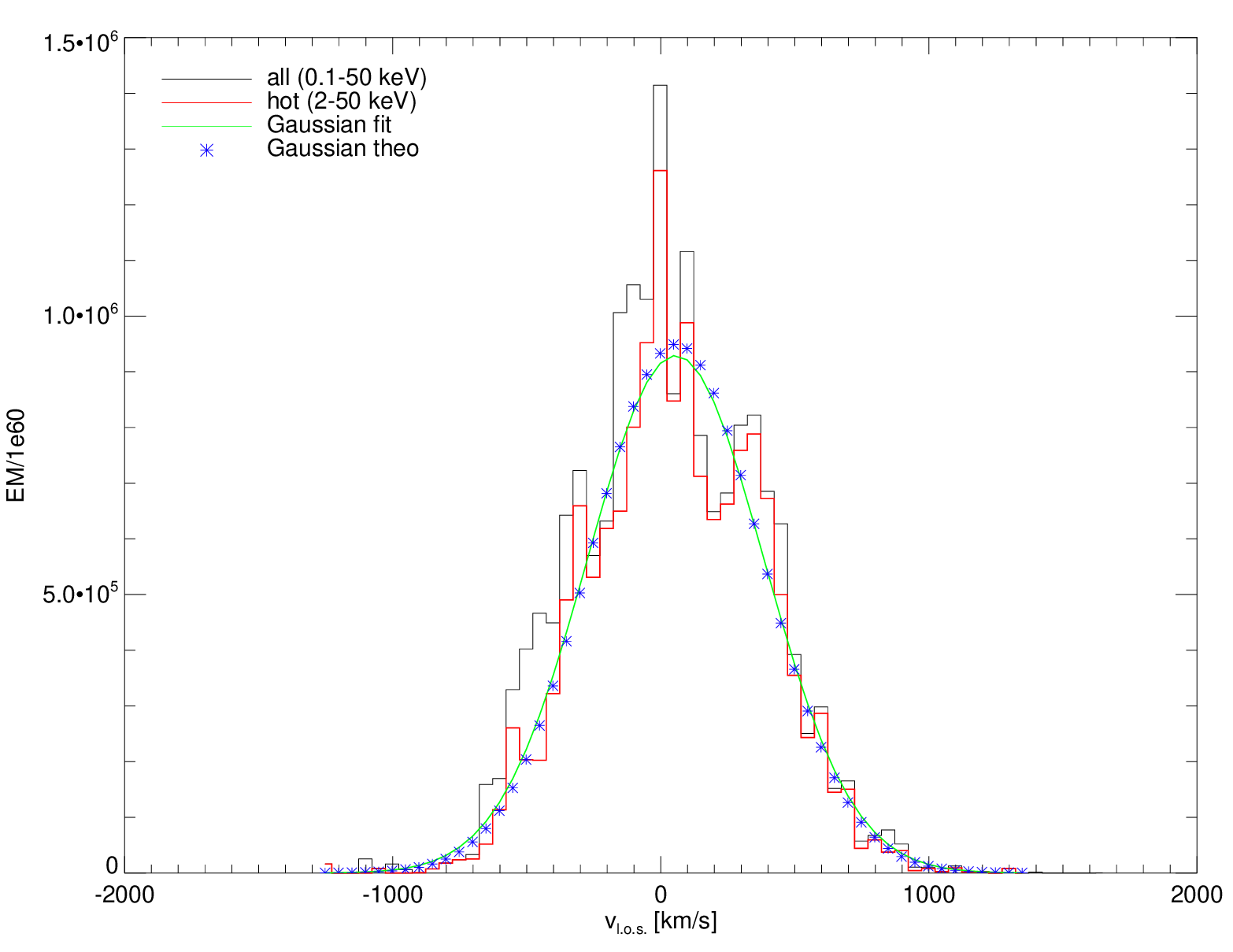}
          \includegraphics[width=0.22\textwidth]{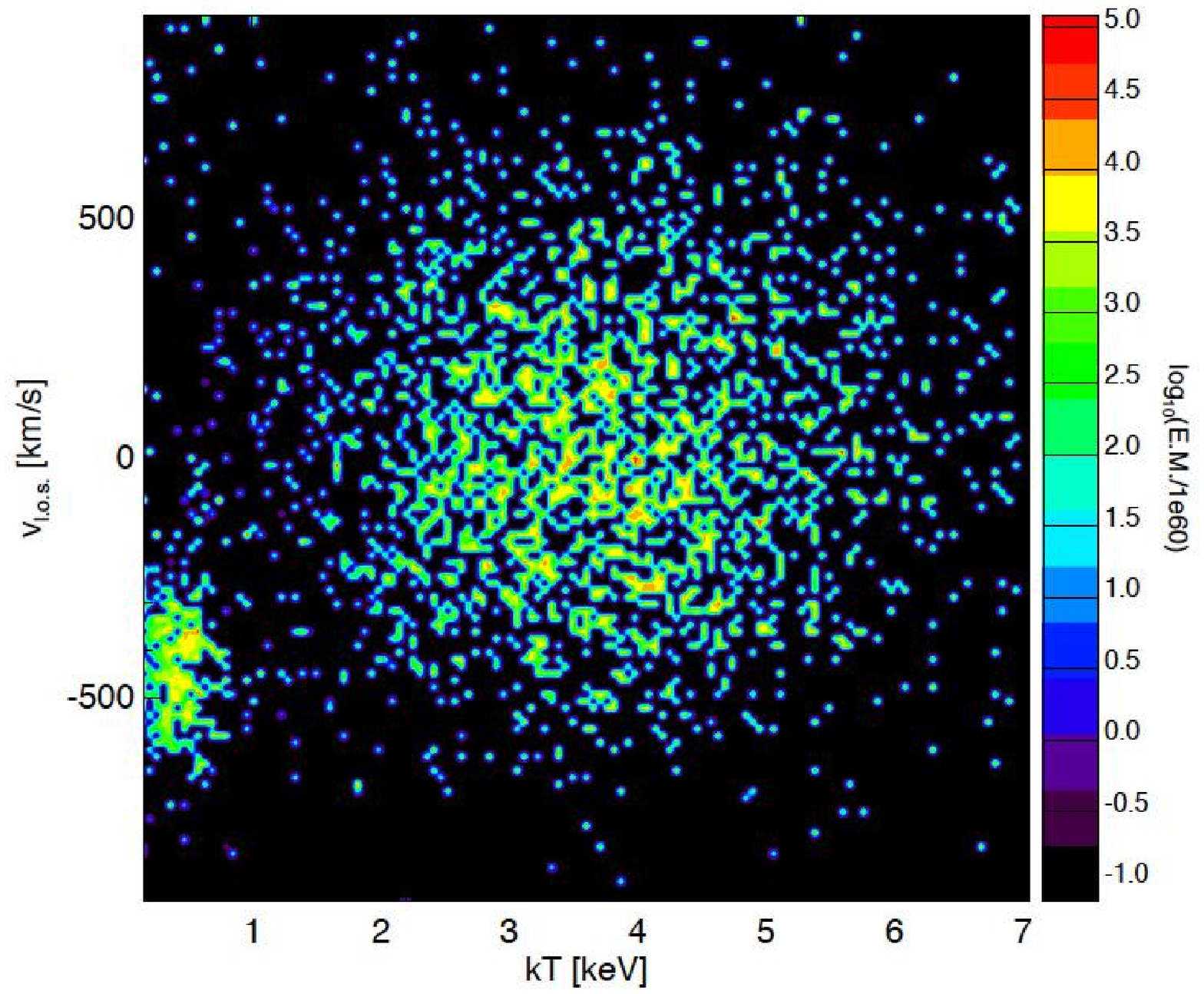}\\

          \includegraphics[width=0.22\textwidth]{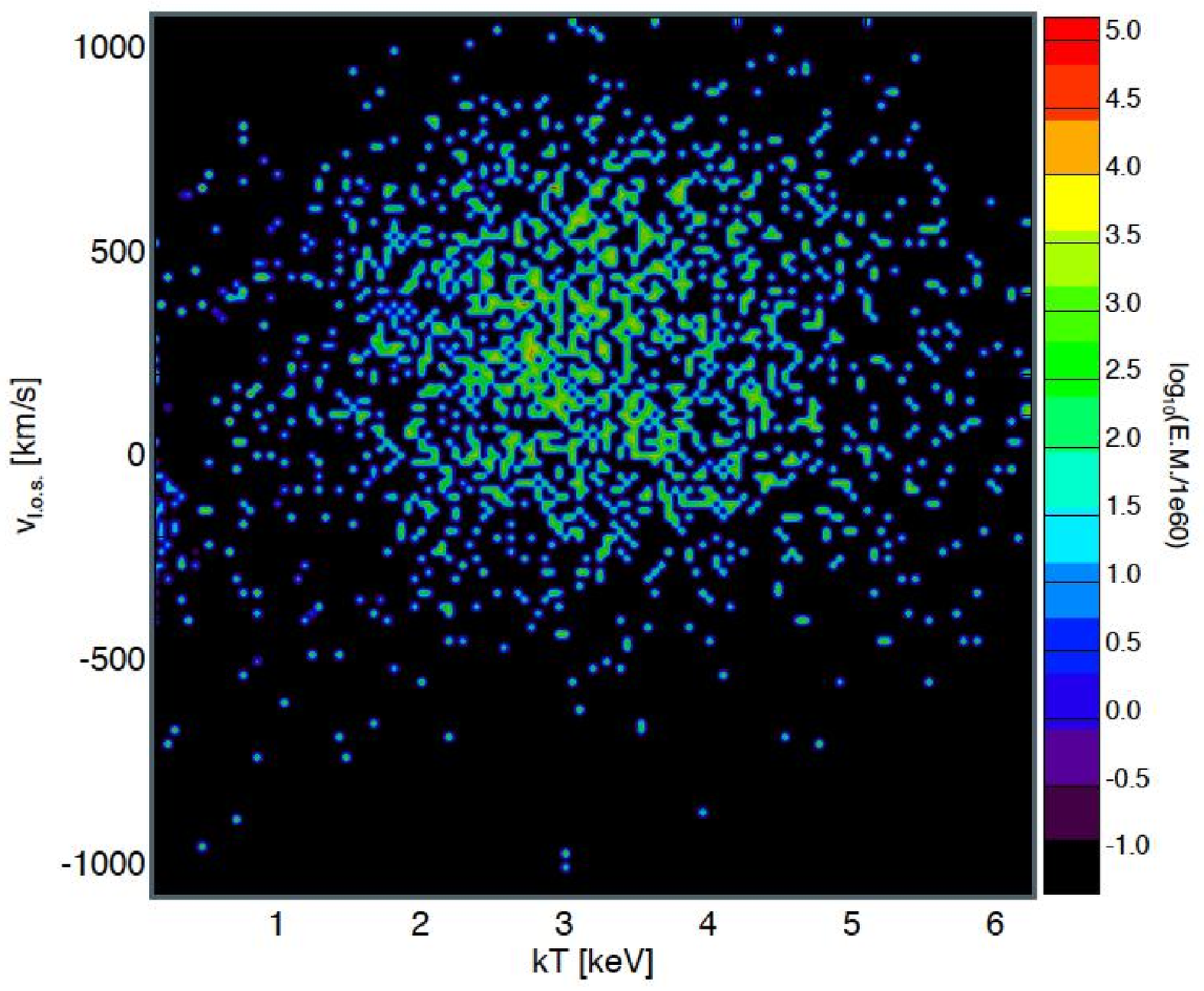}
          \includegraphics[width=0.24\textwidth]{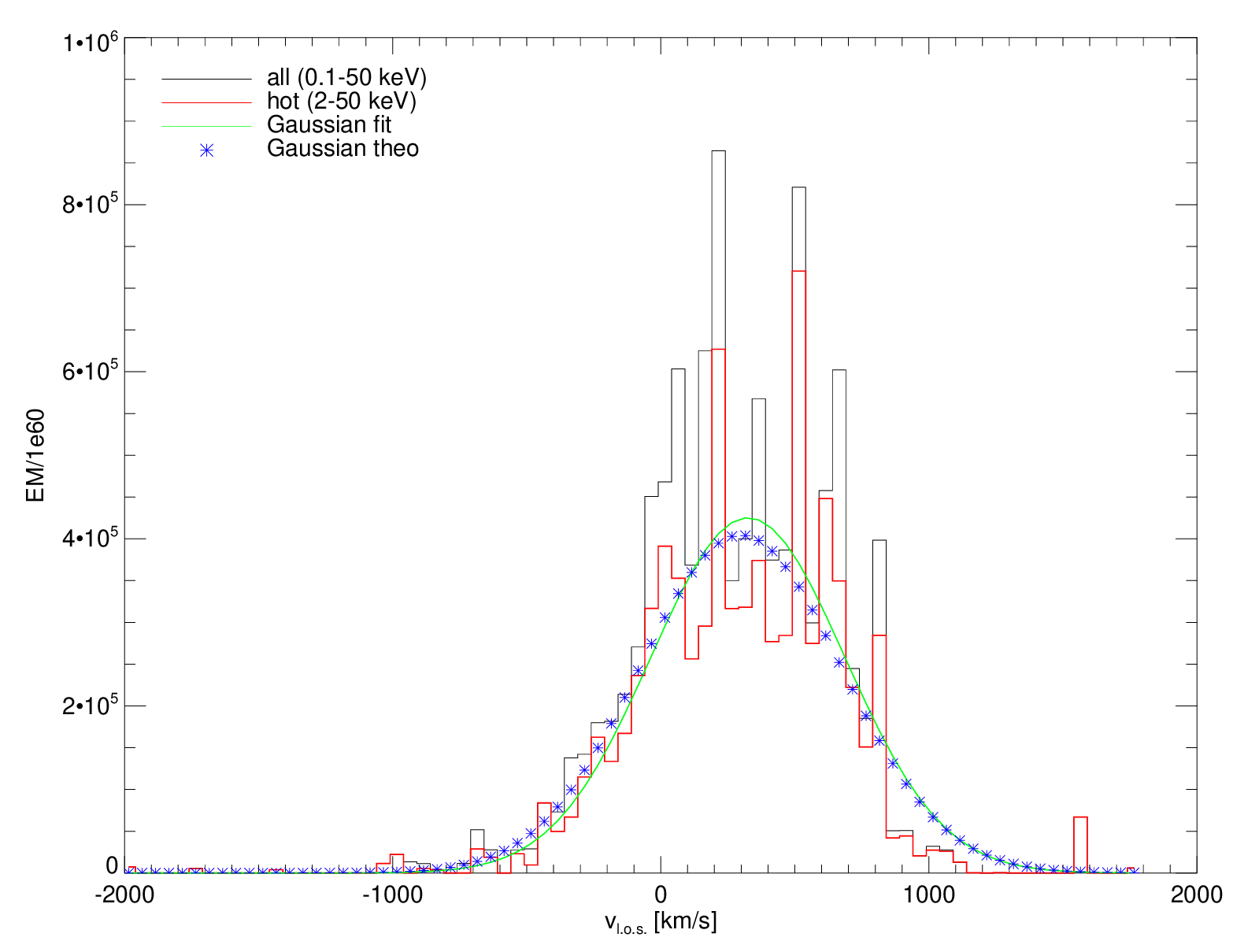}
          \includegraphics[width=0.24\textwidth]{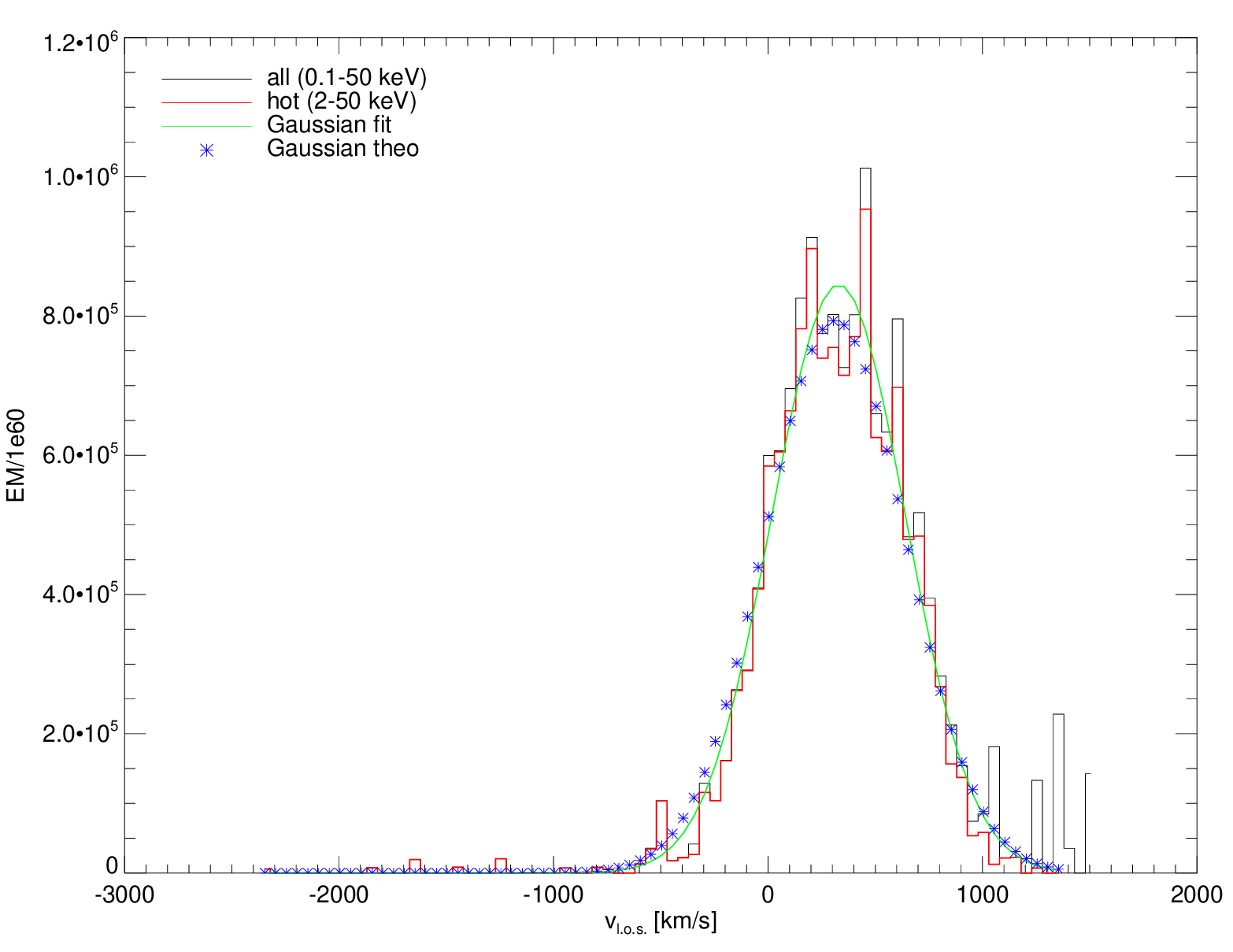}
          \includegraphics[width=0.22\textwidth]{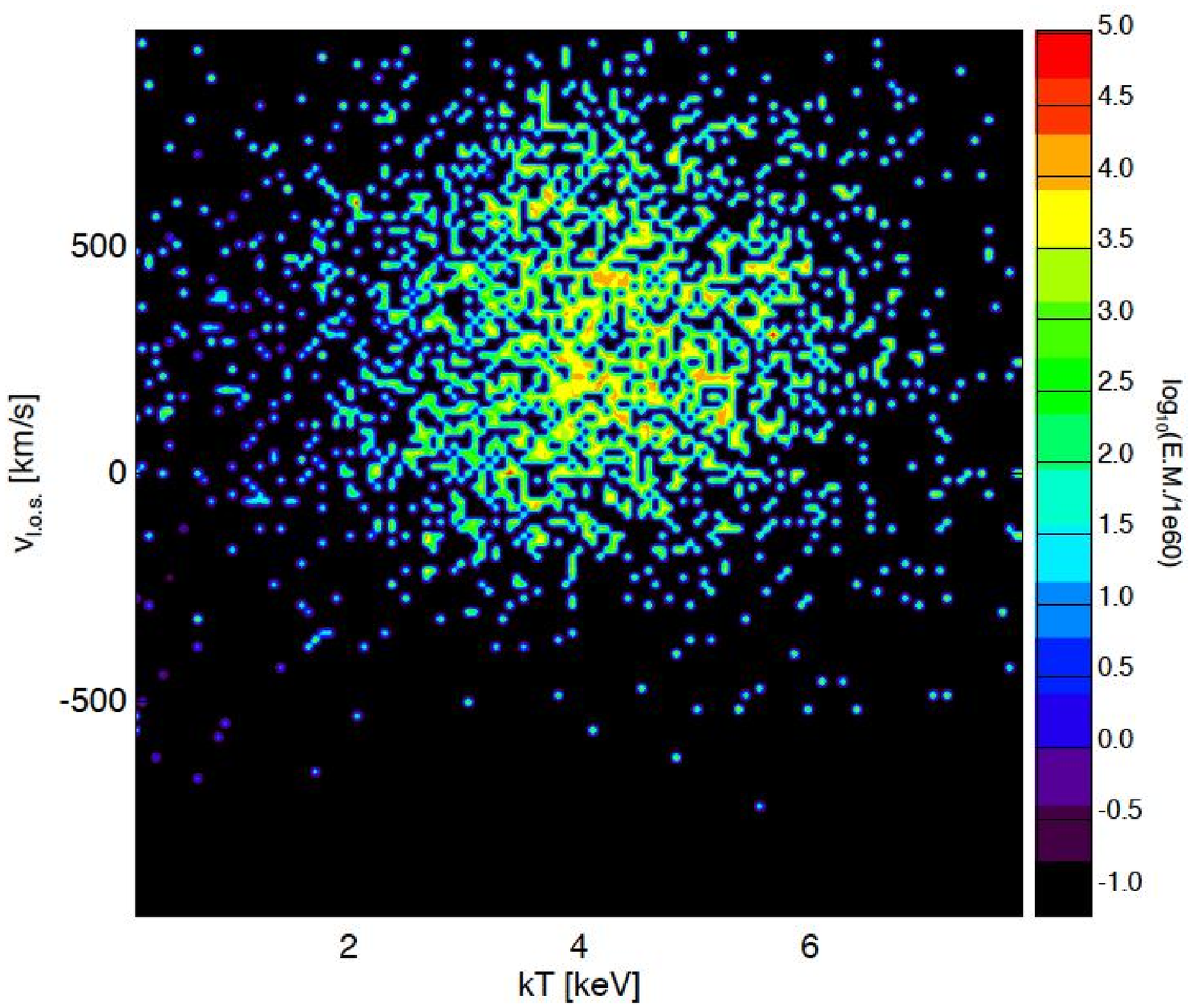}\\

          \includegraphics[width=0.22\textwidth]{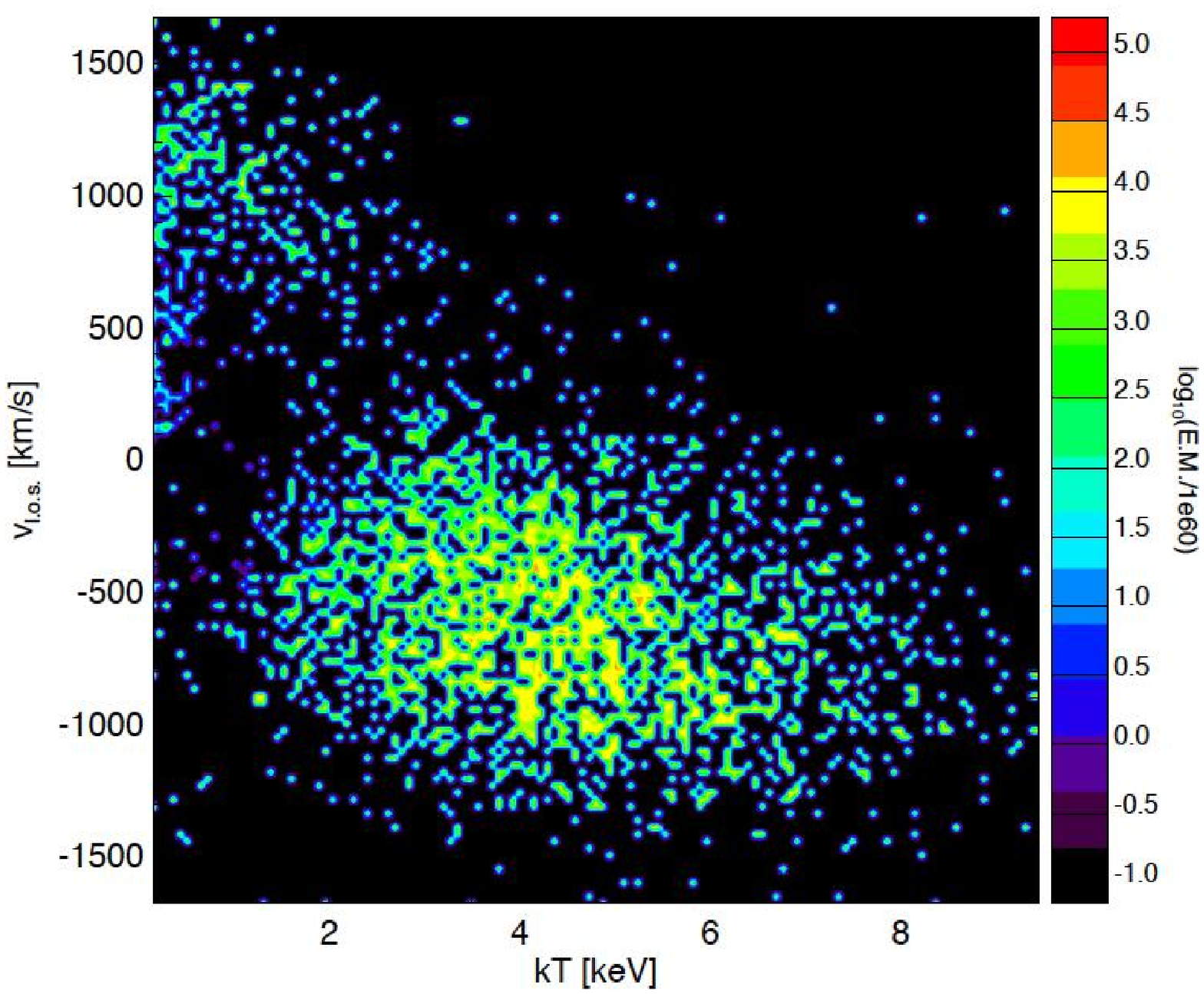}
          \includegraphics[width=0.24\textwidth]{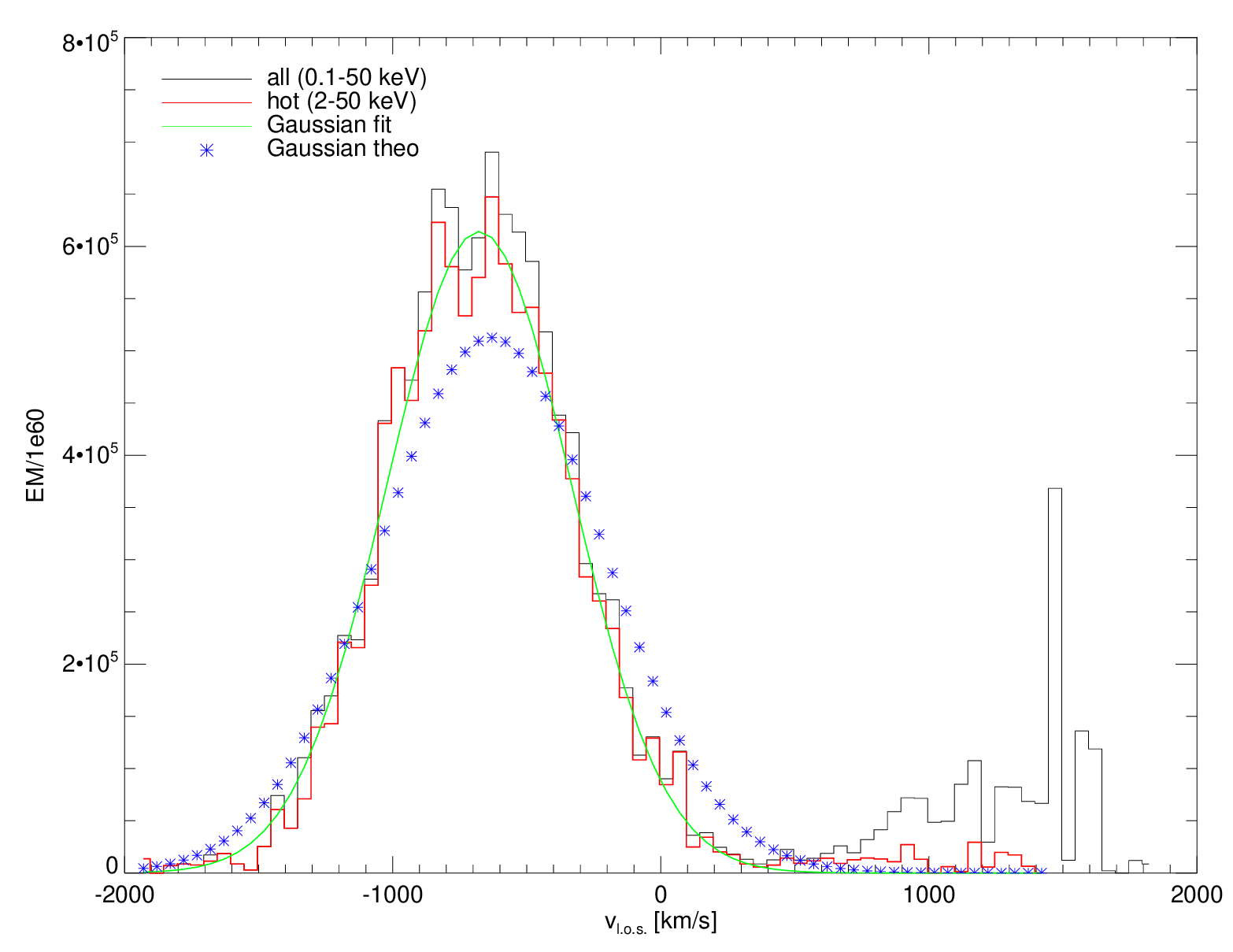}
          \includegraphics[width=0.24\textwidth]{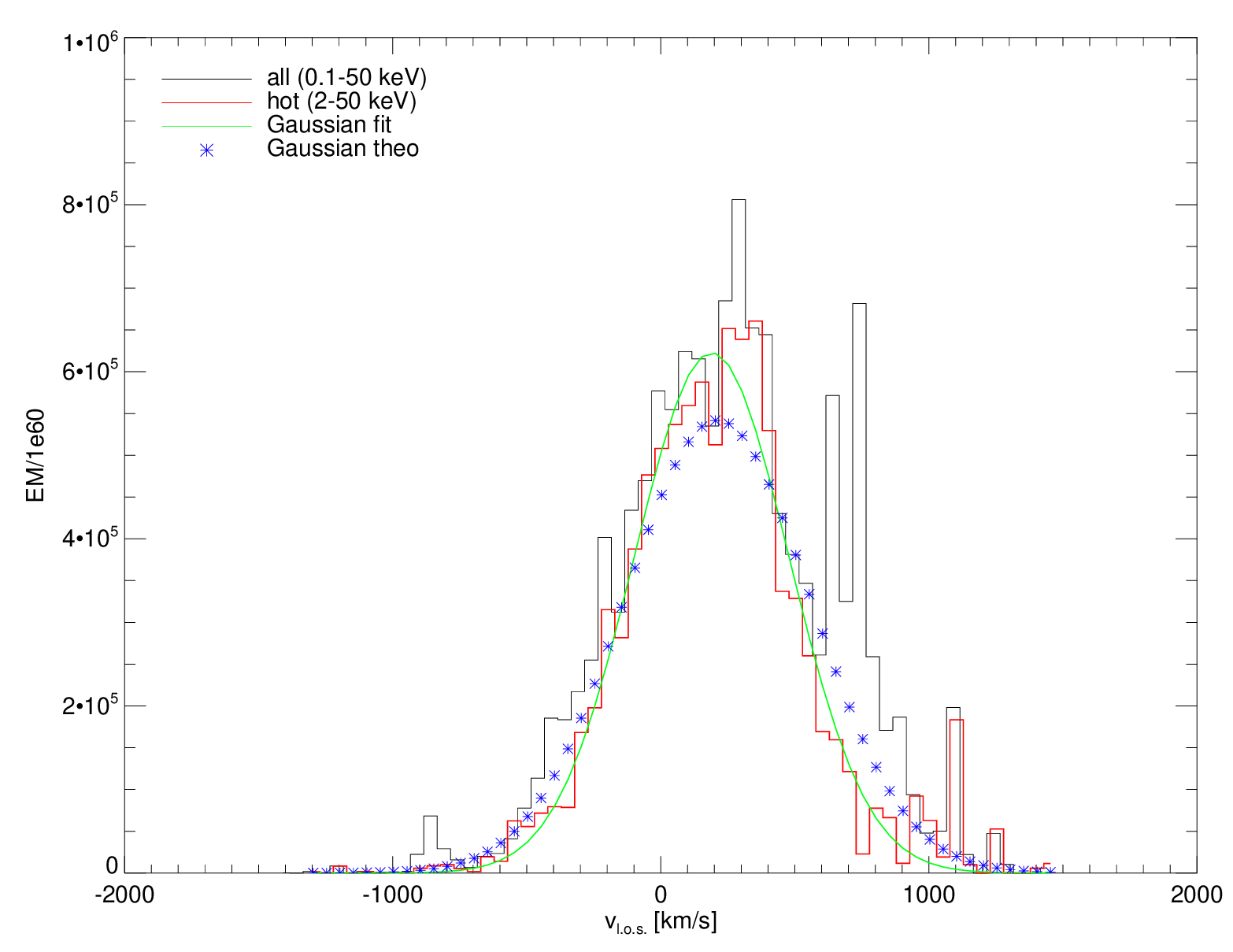}
          \includegraphics[width=0.22\textwidth]{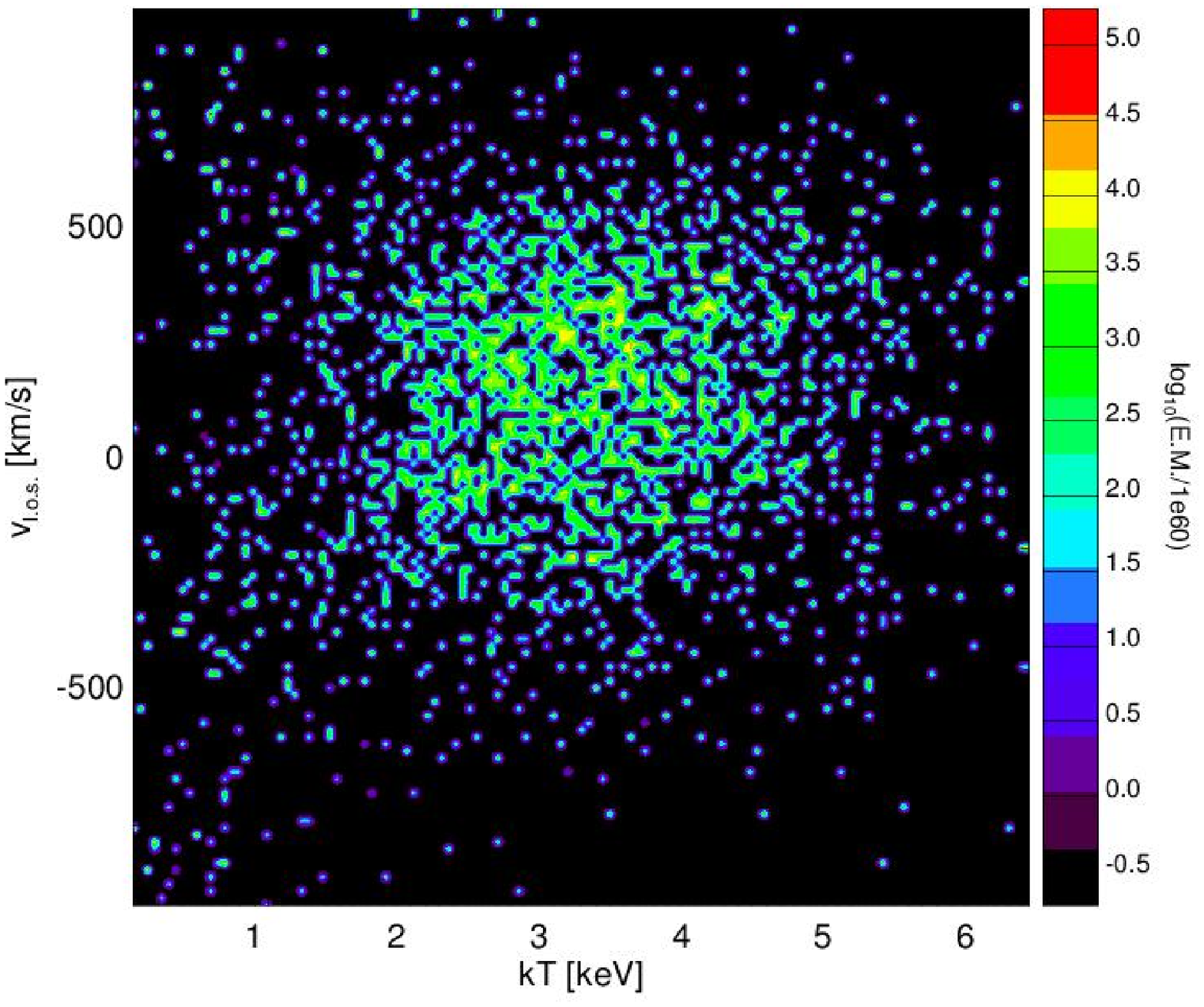}\\

          \includegraphics[width=0.22\textwidth]{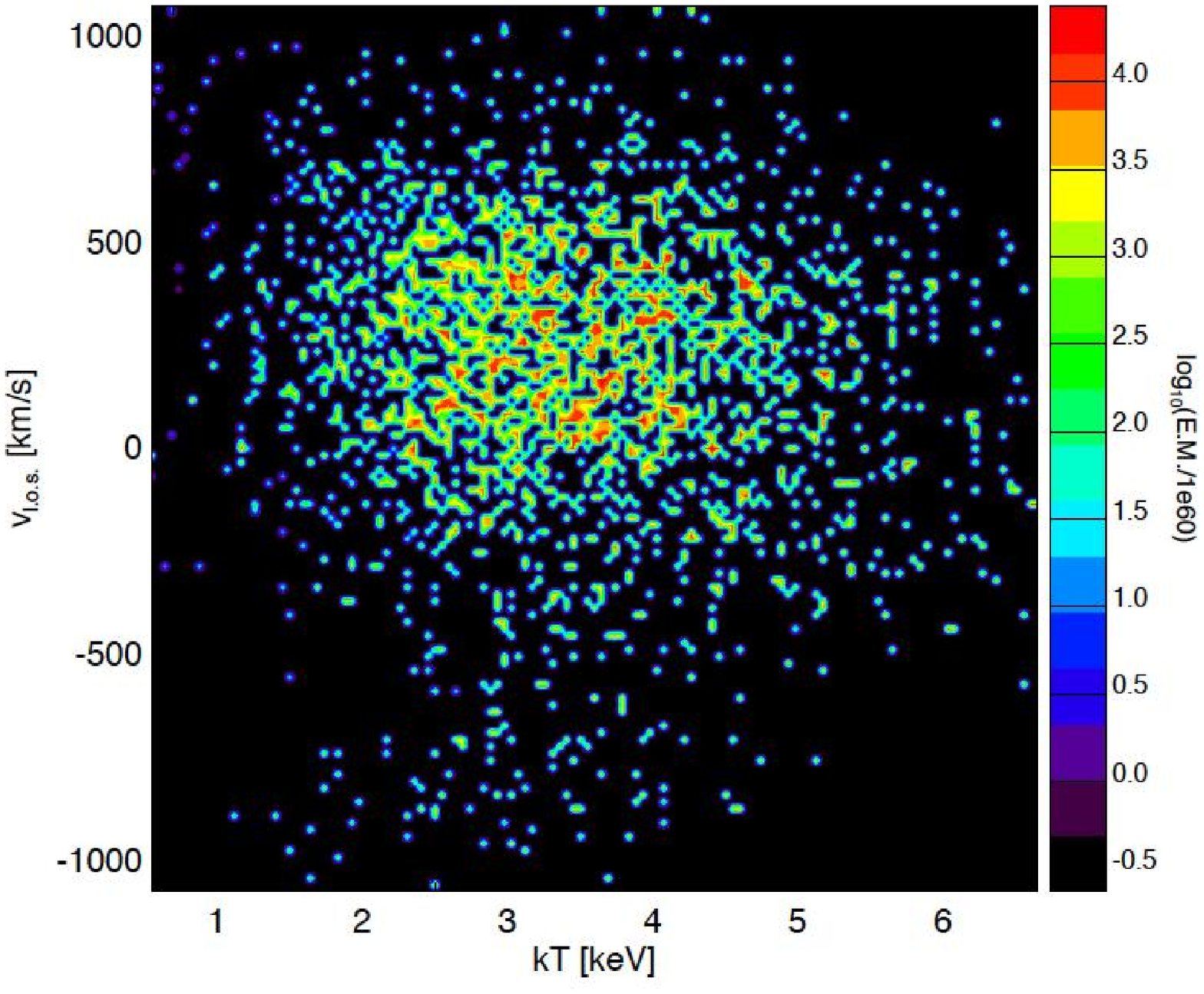}
          \includegraphics[width=0.24\textwidth]{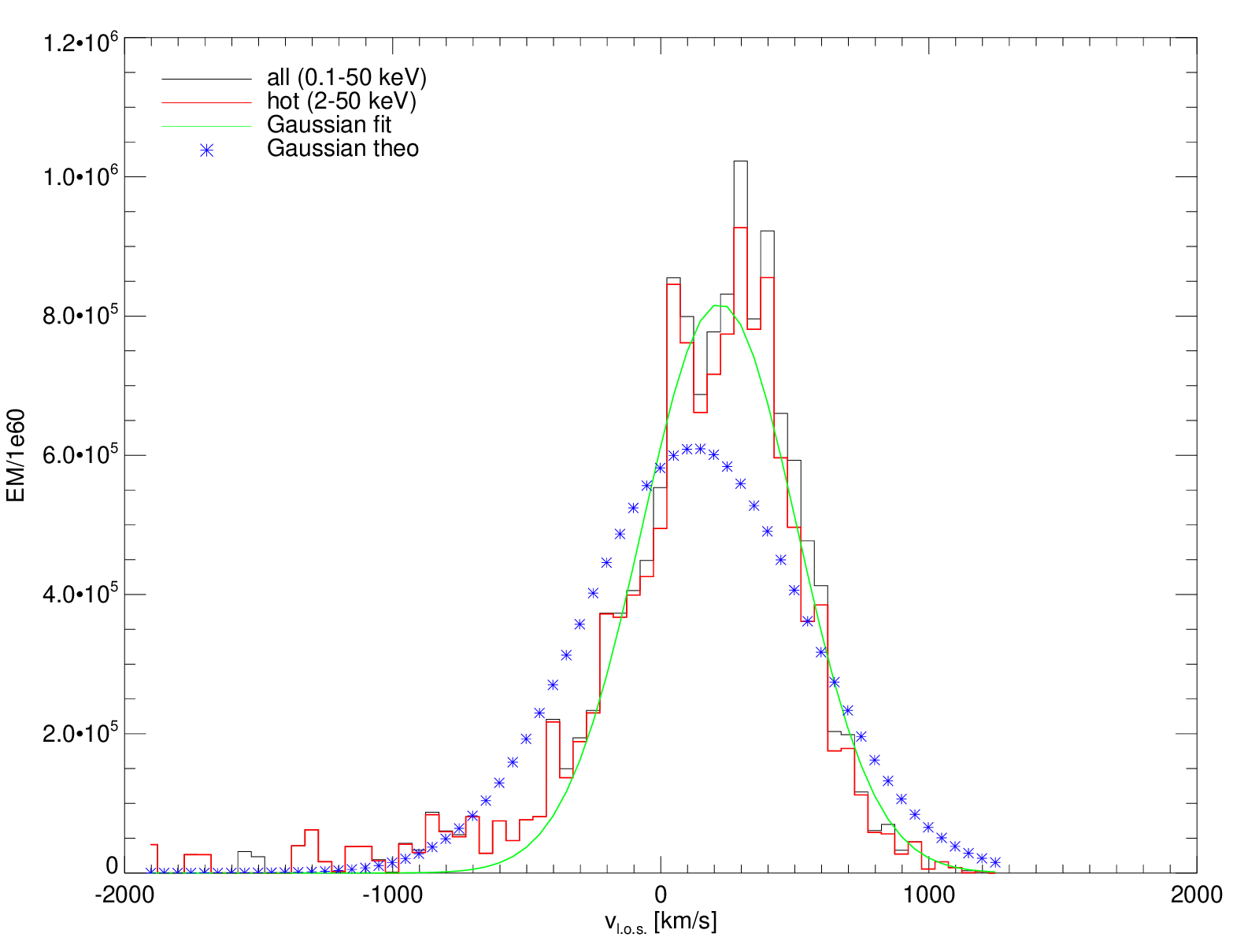}
          \includegraphics[width=0.24\textwidth]{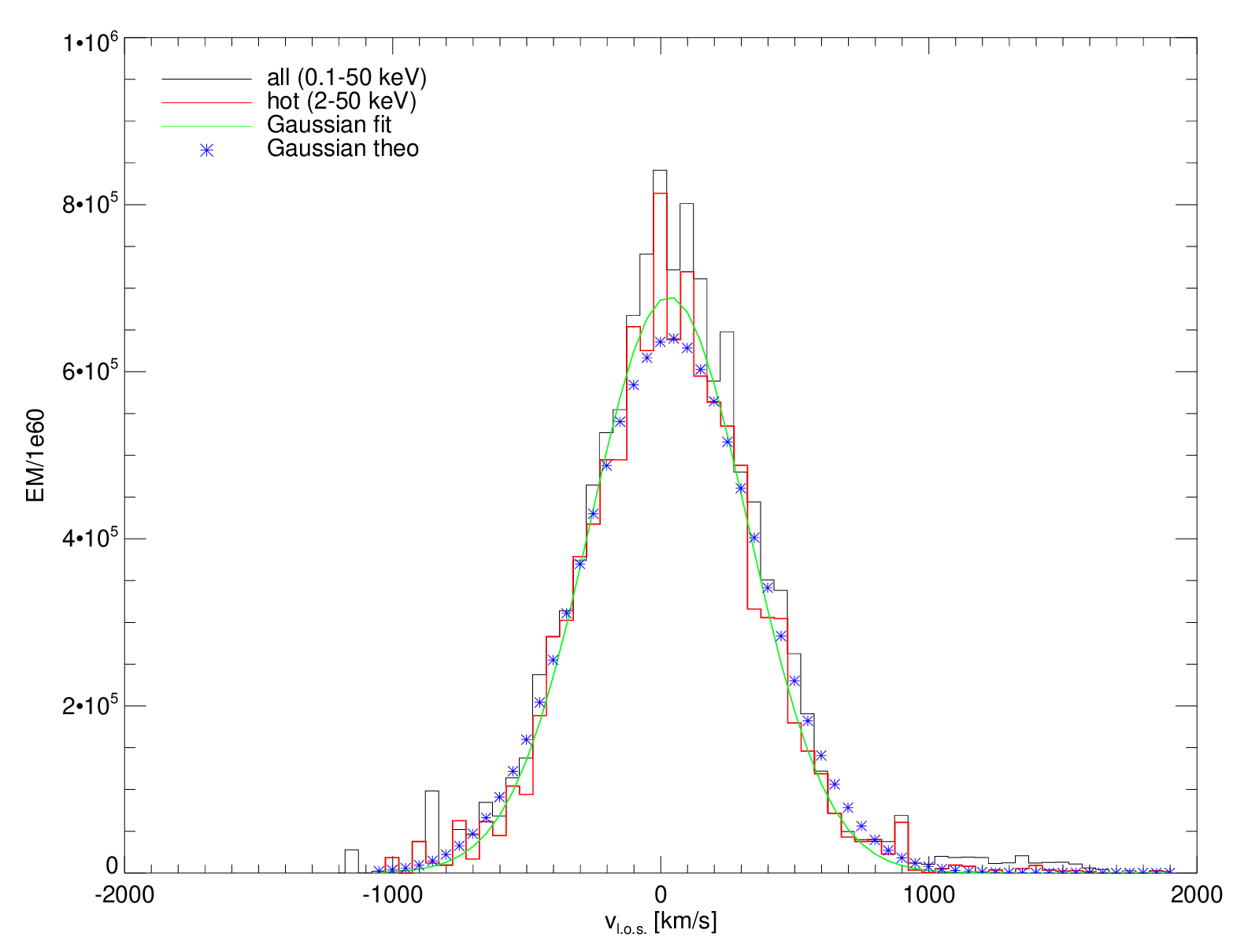}
          \includegraphics[width=0.22\textwidth]{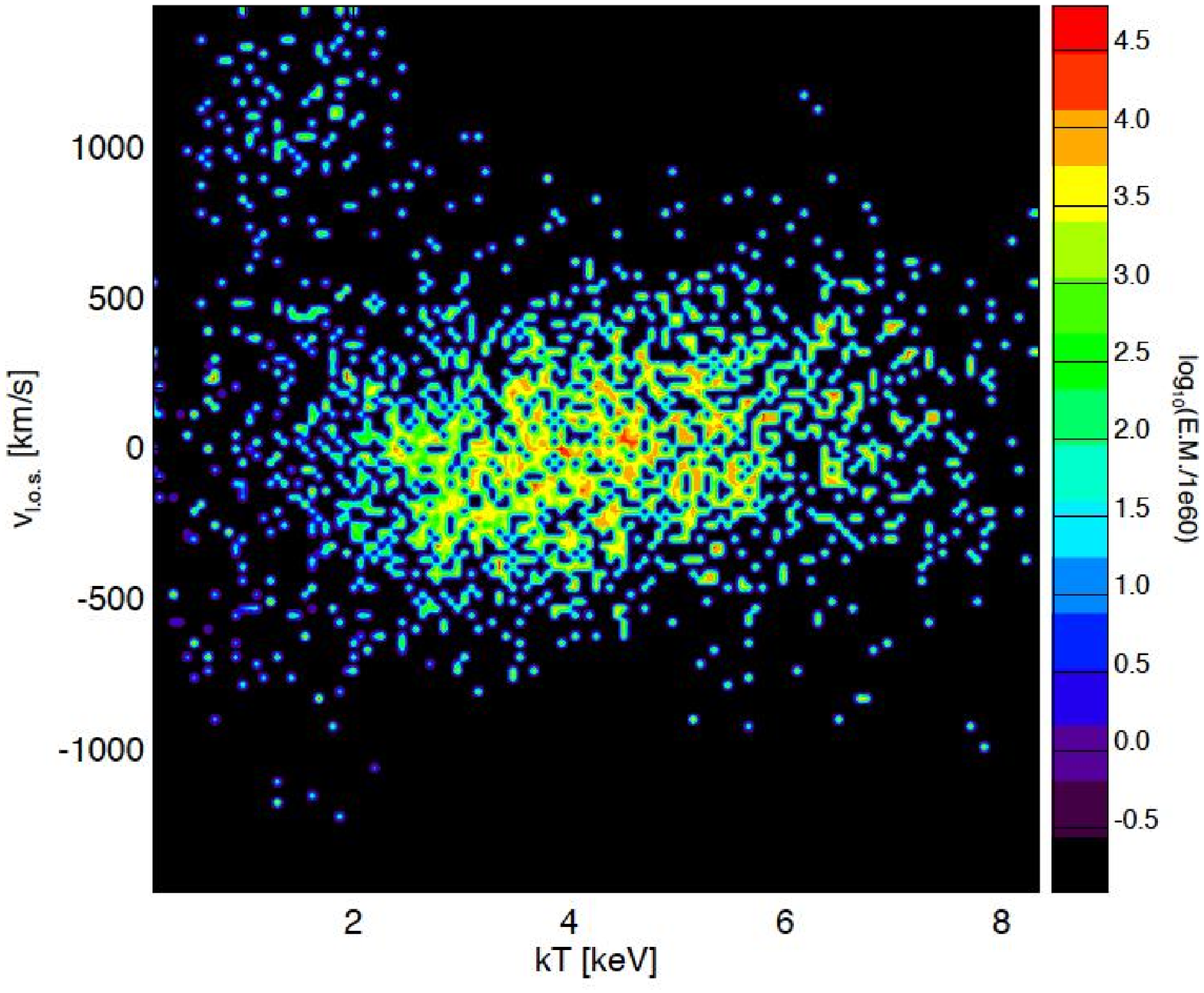}\\

          \includegraphics[width=0.22\textwidth]{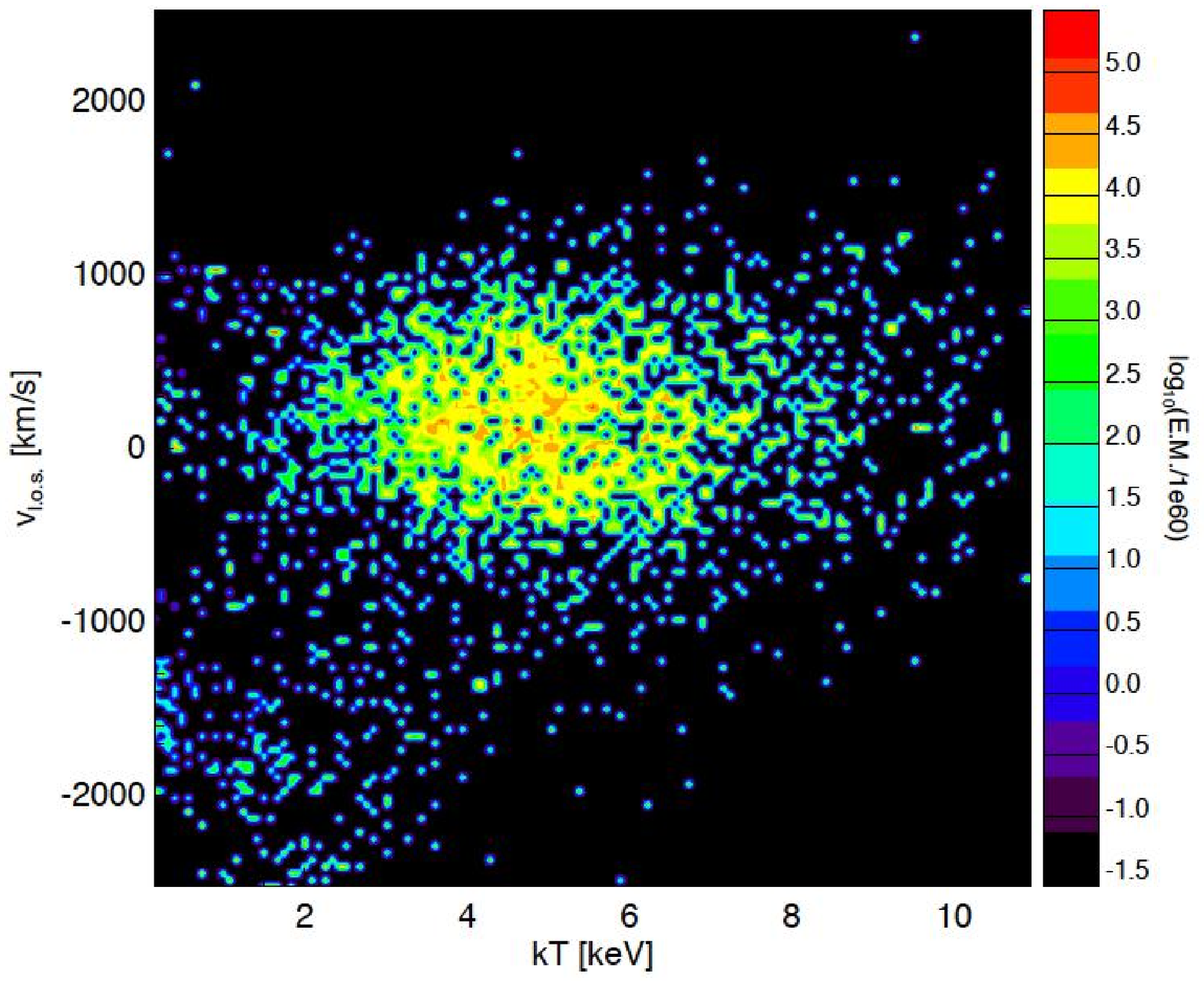}
          \includegraphics[width=0.24\textwidth]{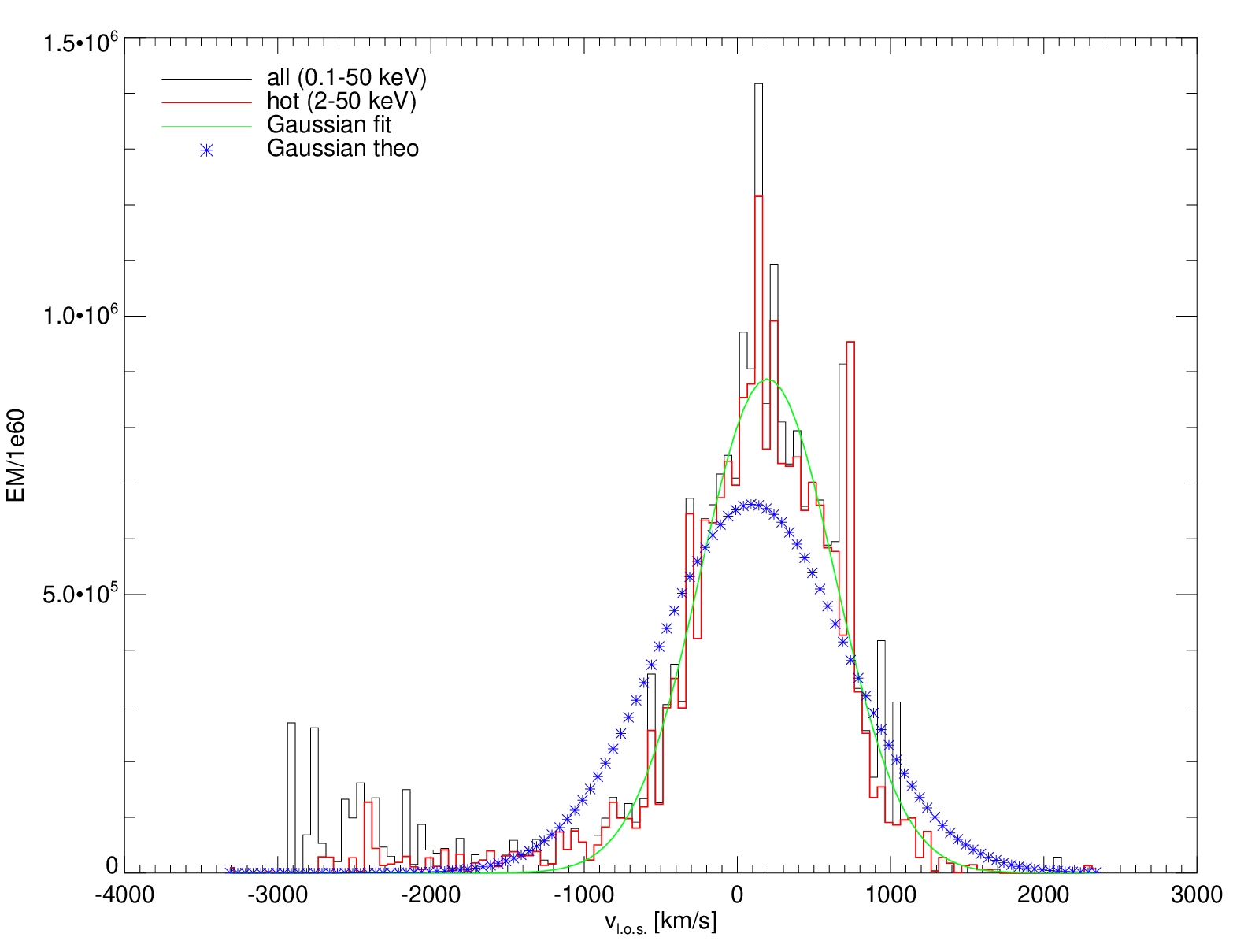}
          \includegraphics[width=0.24\textwidth]{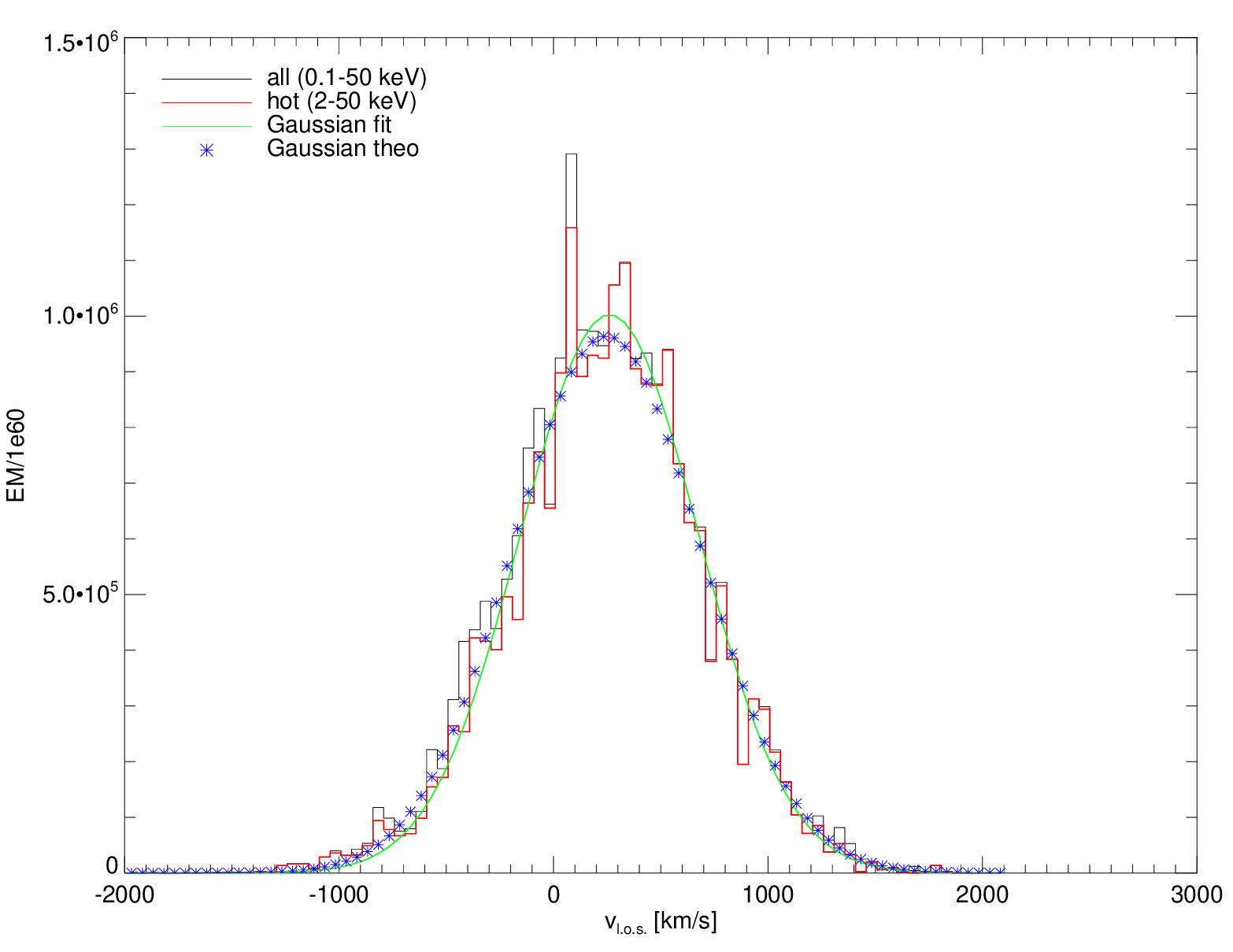}
          \includegraphics[width=0.22\textwidth]{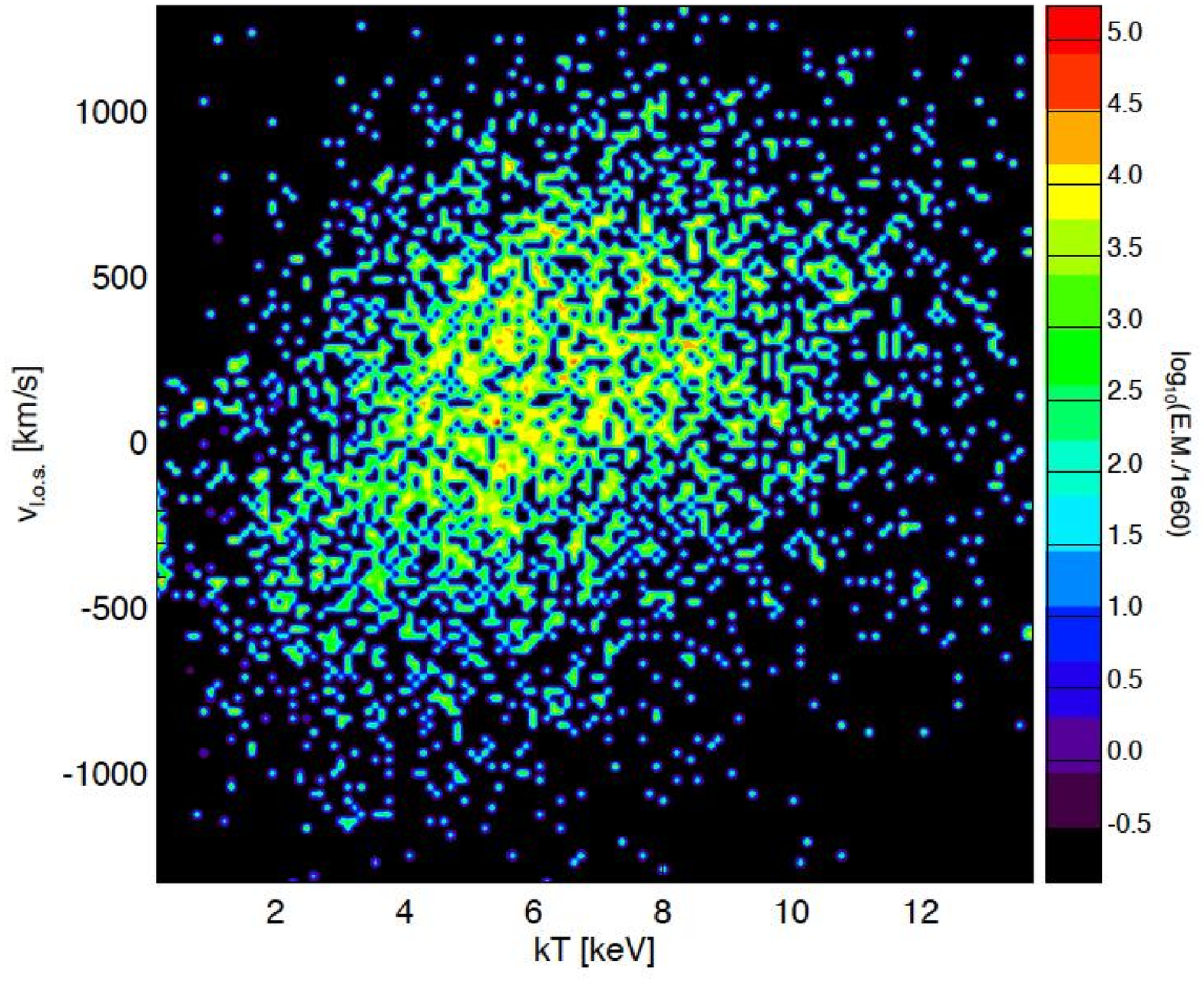}\\
 
          \caption{Comparison among the five most--deviating (left)
            and least--deviating (right) haloes,
            according to \fig\ref{fig:dev}, for the central region
            corresponding to the $2.4'\times 2.4'$ FoV. The level of
            deviation, or agreement, decreases from top to bottom.
            From left to right, the columns refer to: (1)
            $v_{l.o.s.}-kT$ distribution for the gas residing within
            the selected FoV, color--coded by EM, for the
            most--deviating haloes; (2) distribution of EM as function of gas
            l.o.s. velocity, for the five most--deviating haloes; (3)
            same as (2) and (4) same as (1), but considering the
            least--deviating haloes. In the histograms shown in
            columns (2) and (3), the curves refer to: all the gas
            particles (black); hot--phase gas ($kT>2\kev{}$, red); Gaussian
            best--fit to the hot--gas distribution (green); 
            theoretical Gaussian distribution reconstructed from the
            estimated dispersion of the hot--gas distribution
            (blue asterisks);}
          \label{fig:vhisto_maps}
	\end{figure*}

In \fig\ref{fig:vhisto_maps} we show the l.o.s. velocity structure as
calculated from the simulation directly, for both subsets.
In particular, the 5 haloes for which mock and simulated data most
disagree are shown in the first and second columns, while the
least--deviating haloes are presented on the right--hand--side of the
Figure (third and fourth columns).
The rows in the Figure correspond to a decreasing level of deviation,
or agreement, from top to bottom.

Histograms in the central columns of the Figure, report the EM
distribution as function of the 
l.o.s. velocity, for the gas particles in the ATHENA FoV. Black curves
refer to all the gas in the region, while the 
overplotted red histograms only account for the hot--phase gas,
i.e. particles for which $kT>2\kev{}$.
The reason for selecting the hot gas is that it mostly contributes to the 
iron line emission, from which the velocity broadening is measured.

It is clear that in the most--deviating cases there are
substantial substructures within the gas velocity field.
The red histograms for the haloes that show best agreement (right--hand--side
column), instead, reflect more regular distributions of the EM as
function of the $v_{l.o.s.}$, indicating more regular velocity fields.

Observationally, the value estimated from the broadening of the spectral lines
is assumed to be the dispersion of the Gaussian distribution that best
fits the line shape. 
Therefore, a more detailed comparison should
involve the dispersion of the Gaussian function matching the (red)
distribution shown, instead of the theoretical value calculated as in \eq\ref{eq:sigma}.
The green curves in \fig\ref{fig:vhisto_maps} define the Gaussian fits to
the red distributions, whose $\sigma_v^{gauss}$ is more directly
comparable to the mock spectral results.

As an additional comparison, we also overplot the Gaussian curve (blue
asterisks)
constructed from the theoretical estimation of the EM--weighted values
for gas velocity dispersion (\eq\ref{eq:sigma}) and mean l.o.s. velocity.
In the most--deviating clusters, the evident differences among the
green and blue Gaussian curves effectively
reflects the deviations discussed above (see, e.g., the halo with the
largest deviation, top row, left panels in \fig\ref{fig:vhisto_maps}). 
In particular, the green
best--fit Gaussian clearly fails to capture all the features of the
multi--component velocity distribution, as it most likely happens during the
spectral fit.
The low level of deviation found for the ``best'' halo set
(right--hand--side columns) is indeed shown by the good match between
best--fit (green curve) and theoretical (blue asterisks) Gaussian overplotted to the
EM--$v_{l.o.s.}$ distributions.

%---------------------------------------------
A visualization of the thermo--dynamical structure of these
clusters is also shown in the first and fourth columns
of \fig\ref{fig:vhisto_maps}, for the subsets of most--deviating and best 
haloes, respectively.
The maps show the $v_{l.o.s.}-kT$ distribution for the gas
residing within the selected FoV, color--coded by EM.

Especially for the most deviating haloes, the sub--structures in the velocity field
unveiled by the histograms are clearly visible, combined with
the gas thermal structure.

Despite the deviations discussed, a good overall agreement is found between the intrinsic
amplitude of the gas velocity dispersion and the velocity
broadening measured directly from mock spectra.
This confirms the promising potential of such well--resolved
observations, which would most likely allow us to derive reliable and
precise constraints on the ICM velocity field. 
 
Moreover, given that the simulation analysis suggests
the gas velocity structure in the innermost region (e.g. that
covered by the FoV of ATHENA) to be closely traced by that
within $\rfive{}$ (see \sec\ref{sec:results_sim}), we will use the latter
for our further investigation.
Eventually, limitations due to small fields of view could be observationally overcome
by multiple pointings covering a larger region.
%-------------------------------------------------------------------------------------------
\subsection{$L_X-T$ scaling relation}\label{sec:lt}
Here, we investigate the impact of the ICM velocity structure on X--ray
global properties by focusing on the $L_X-T$ relation.
The main motivation behind this choice is that, on one hand, the luminosity $L_X$, is very well
measured in X--ray surveys (e.g. with Chandra, XMM--Newton, or the
up--coming eRostita), and, on the other hand, the temperature $T$ 
provides a good mass proxy, since it is tightly related to
the total gravitating mass, which is the most fundamental quantity
to characterise a cluster.

$L_X$, also denoted as ``bolometric X--ray luminosity'',
is usually the X--ray luminosity extrapolated to the whole X--ray
band, $0.1-100\kev{},$ instead of being calculated in a narrow energy
band.
Nonetheless, in our case, the computation of the luminosity is limited to the largest energy band defined by the
ACIS--I3 response matrix (i.e. $0.26-12\kev{}$), since the difference
introduced with respect to the expected
bolometric X--ray luminosity is found to be minor.
	\begin{figure}
	\centering
	\includegraphics[width=0.46\textwidth]{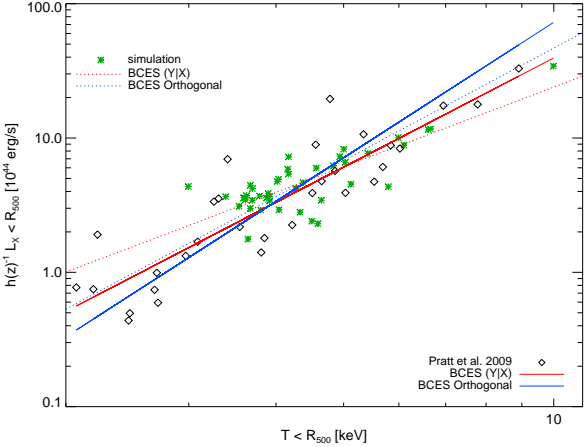}
	\caption{$L-T$ scaling relation from Chandra mock spectra
          (green asterisks), for 
          the region of the clusters encompassing
          $\rfive{}$. Overplotted are the results from P09
          (black diamonds). Best--fit relations are
          also overplotted (solid lines for P09; dotted lines for
          simulated haloes of the present work) and correspond to the
          two linear regression fitting methods adopted: 
          BCES~(L$\mid$T) (red) and BCES Orthogonal (blue).}
         \label{fig:lt_rel}
	\end{figure}

In \fig\ref{fig:lt_rel} we show the $L_X-T$ relation calculated for
the simulated haloes (asterisks), from the Chandra synthetic observations of the
$\rfive{}$ regions through the l.o.s., obtained with \phox{}. 
As a comparison, we also show the data presented in \cite[][also, P09]{pratt2009} for a
sample of $31$ nearby galaxy clusters (black diamonds), selected only in
X--ray luminosity from the Representative
XMM--Newton Cluster Structure Survey (REXCESS).

\subsubsection{Fitting procedure}
The $L_X-T$ relation was fitted with a power--law functional form, that is
\begin{equation}
  h^{-1}(z) L_X = C\,(T/T_0)^{\alpha},
\end{equation}
where $L_X$ is given in units of $10^{44}\ergs{}$, $T_0 = 5\kev{}$ and
the assumed cosmology is accounted for in $h^2(z)=\om (1+z)^3 + \Omega_{\Lambda}$.
The fit was performed using linear regression in the log--log plane.

As done in \cite{pratt2009}, as well as in several similar works
\cite[e.g.][]{reiprich2002,arnaud2005,mittal2011}, we adopted the BCES
(Bivariate Correlated Errors and intrinsic Scatter) regression method by
\cite{bces1996}, which accounts for both the errors in $L_X$ and in
$T$, as well as the intrinsic scatter in the data.
The BCES algorithm allowed us to calculate the best--fit values of
slope ($\alpha$) and normalization ($C$) for four different regression
methods, amongst which we restrict our attention to the
BCES~(L$\mid$T) and the BCES Orthogonal 
methods.
Our primary goal is to find the best fit which minimizes the residuals
of both variables at the same time, orthogonally to the linear
relation. This is given by the BCES Orthogonal method.
Additionally, we also explore the results given by the BCES~(L$\mid$T)
fitting method (analogously to \cite{pratt2009}), which minimises
the residuals in $L_X$. 
Reasons for this rely on the fact that $L_X$ can be
treated as the dependent variable, while the temperature can be assumed
to be the ``independent'' one, as closely related to the cluster mass, 
which is the intrinsic quantity characterizing the system.

Given the statistical uncertainties\footnote{In the log space, errors
  are transformed as $\Delta{\rm log}x={\rm log}e \times (\Delta x)/(2x),$
where $\Delta x$ is the difference between the upper and lower boundary of the
error range around the quantity.} on both variables,
$\sigma_{Y_i}$ and $\sigma_{X_i}$, the raw scatter was estimated
using the orthogonal distances to the regression line, weighted by the
errors.
Namely, for a linear relation of the form $Y_i=\alpha X_i + \beta$ in
the log--log space, the raw scatter is
\begin{equation}
\sigma^2_{raw} = \frac{N/(N-2)}{\Sigma_{i=1}^N
  1/\sigma_i^2}\sum_{i=1}^N(1/\sigma_i^2)(Y_i - (\alpha X_i + \beta))^2
\end{equation}
where $N$ is the number of data points in the sample,
$Y_i = {\rm log}(L_{X,i})$, $Y_i = {\rm log}(T_{i})$, $\beta = {\rm log}C$,
and
\begin{equation}
\sigma_i^2 = \sigma_{X_i}^2 + \alpha^2 \sigma_{Y_i}^2.
\end{equation}
The intrinsic scatter was estimated from the difference between the
raw and the statistical scatter, in quadrature ($\sigma^2_{raw} =
\sigma^2_{stat} + \sigma^2_{intrinsic}$).

Best--fit relations are overplotted in \fig\ref{fig:lt_rel} for both observational data
\cite[solid red and blue lines, for the BCES~(L$\mid$T) and the BCES
Orthogonal  method respectively, from][]{pratt2009} and for Chandra
synthetic observations of the simulated sample (dotted lines). 
The linear relations found for the simulated clusters are overall
shallower than the observed ones, and, among the two fitting methods
considered, the BCES~(L$\mid$T) method still provides a shallower slope than 
the BCES Orthogonal case.

In particular, we find a slope $\alpha_{(L\mid T)} = 1.97 \pm 0.23$ for
the BCES~(L$\mid$T) fit and $\alpha_{Ortho} = 2.78 \pm 0.3$ for the BCES
Orthogonal fit. As for the normalization of the best--fit relations, we
find $C_{(L\mid T)} = 6.81 \pm 0.39 (10^{44}\ergs{})$ and $C_{Ortho} =
6.11 \pm 0.34 (10^{44}\ergs{}),$ respectively.

We note, however, that our cluster sample probes a smaller dynamical
range with respect to observations and, in particular, 
lacks low--temperature haloes,
whose presence might contribute providing tighter constraints on
the slope of the relation.
%
%--------------------------------------------------
\section{Discussion} \label{sec:discussion}
\subsection{$L_X-T$ relation: effects of the velocity structure}
%---------------------------------------------
	\begin{figure*}
	\centering
        \includegraphics[width=0.82\textwidth]{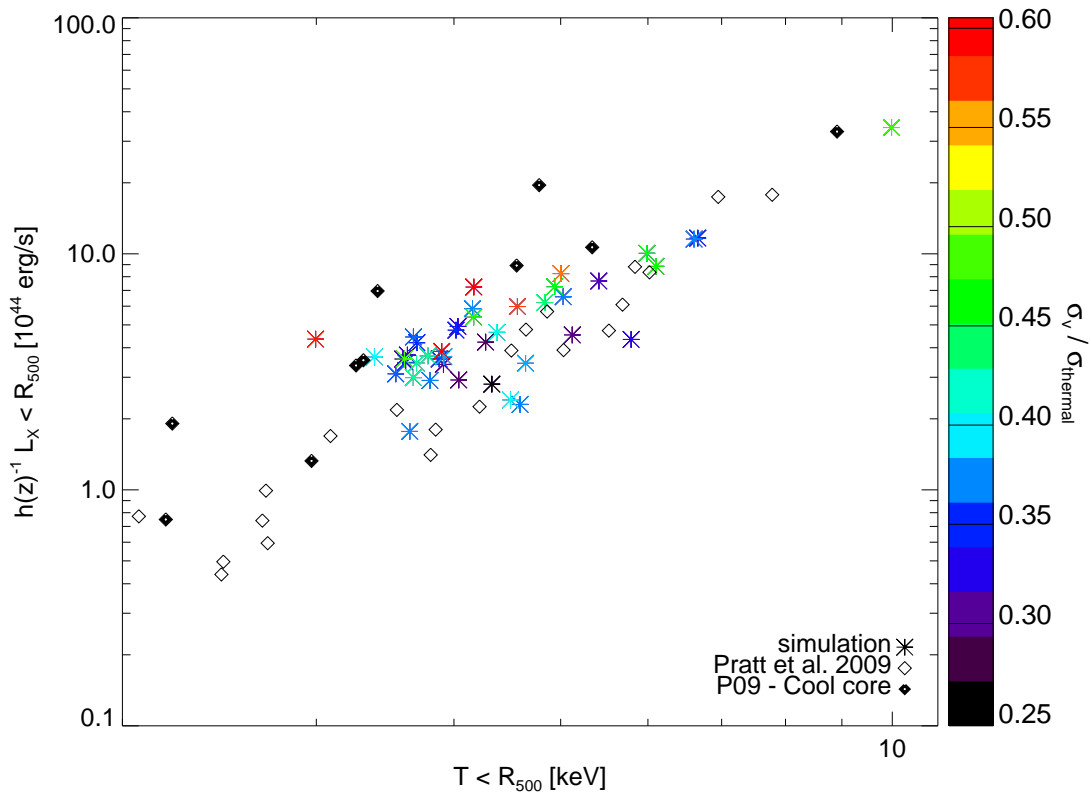}
	\caption{$L_X-T$ scaling relation from Chandra mock spectra, for 
          the region of the clusters encompassing
          $\rfive$. Color code: value of $\mu =
          \sigma_v(<\rfive)/\sigma_{thermal}$, calculated from gas
          velocities in the region encompassed by
          $\rfive$. Observational data points (black diamonds) are from 
          P09, with cool core clusters marked with thicker black diamonds.}
         \label{fig:lt_rel_sig}
	\end{figure*}
%-------------------
In \fig\ref{fig:lt_rel_sig} we include an additional parameter to
further characterize the cluster state. In particular we show the
$L_X-T$ relation for the simulated haloes, as in \fig\ref{fig:lt_rel},
color--coding the data points according to the increasing ratio of
\begin{equation}\label{eq:sig_ratio}
\mu = \sigma_v(<\rfive)/\sigma_{thermal},
\end{equation} 
calculated for the region encompassed by
$\rfive{}$.

The velocity dispersion $\sigma_v$ corresponds to the EM--weighted
value calculated directly from the gas particles in the simulation
(see \sec\ref{sec:results_sim}).
The thermal velocity dispersion, $\sigma_{thermal}$, is the expected
three--dimensional value for the ICM temperature $T$, reported in the
$x$--axis of the relation. 
To this purpose of normalizing the
non--thermal velocity dispersion to a value characteristic for the halo
potential well, the choice of the
three--dimensional thermal velocity is equivalent to the 1D value,
apart from the constant scale factor $\sqrt{3}$. 
Small values of $\mu$ indicate a low level of non--thermal velocity
with respect to the characteristic thermal velocity dispersion of the
gas\footnote{In fact, the thermal component of the velocity is not
  included within $\sigma_v$, which traces macroscopic motions of the
  gas elements in the simulation. Analogously, the value measured from
the velocity broadening of spectral lines, used for comparison in
\sec\ref{sec:results_mock}, does not include the thermal component either.}.
For the extreme case of $\mu \sim 1$, the non--thermal velocity
dispersion would equal the thermal value. 

%-------------------
\begin{figure}
  \centering
  \includegraphics[width=0.46\textwidth]{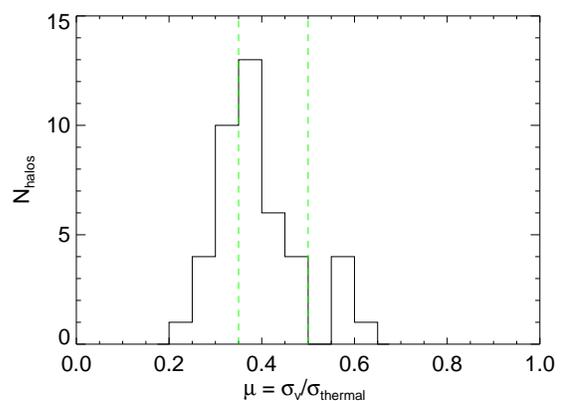}
  \caption{Distribution of
    $\mu = \sigma_v(<\rfive)/\sigma_{thermal}$, for the sample of
    $43$ simulated clusters. Green, dashed lines correspond to the
    values $\mu_{max}={0.5,0.35}$ used to identify the two
    sub--samples.}
  \label{fig:mu}
\end{figure}
The distribution of $\mu$ for the $43$ simulated clusters is shown in
\fig\ref{fig:mu}. A
significant number of haloes ($\sim35\%$) are characterised by a gas velocity
dispersion which is larger than $0.4\sigma_{thermal}$.

In particular, by examining the distribution of $\mu$ within
the sample,
we decide to extract two additional sub--samples from the $43$ haloes,
selected to have a maximum of $\mu_{max}={0.5,0.35}$, respectively.
The first sub--sample is intended to exclude the most prominent
outliers in the $\mu$ distribution, for which the velocity dispersion
$\sigma_v$ exceeds $50\%$ of $\sigma_{thermal}$.
The smaller sub--sample contains the haloes with the smallest fraction
of $\mu$, indicating that their expected thermal velocity is dominant with
respect to $\sigma_v$.

%--------------
\begin{figure}
  \centering
  \includegraphics[width=0.46\textwidth]{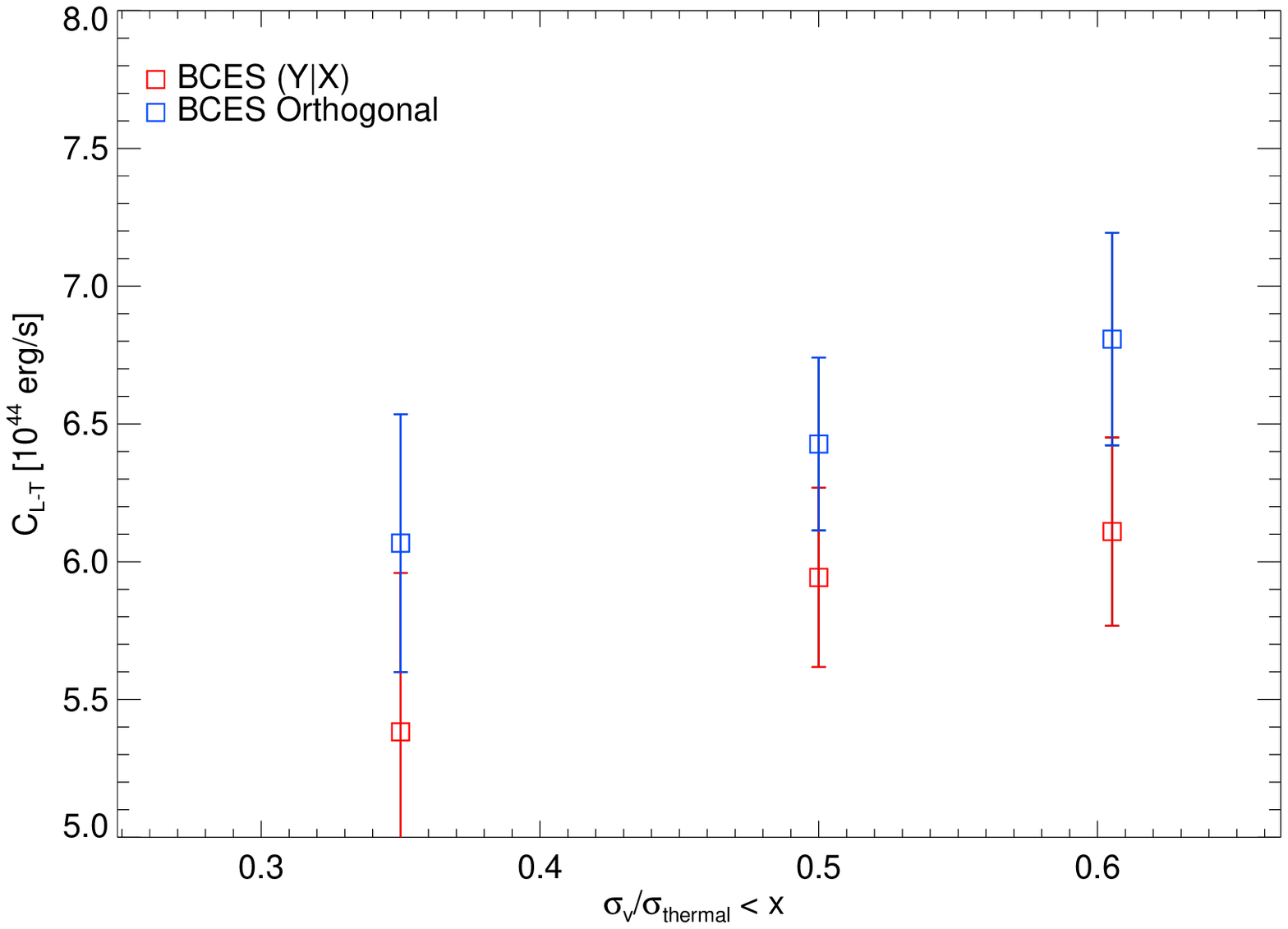}
  \includegraphics[width=0.46\textwidth]{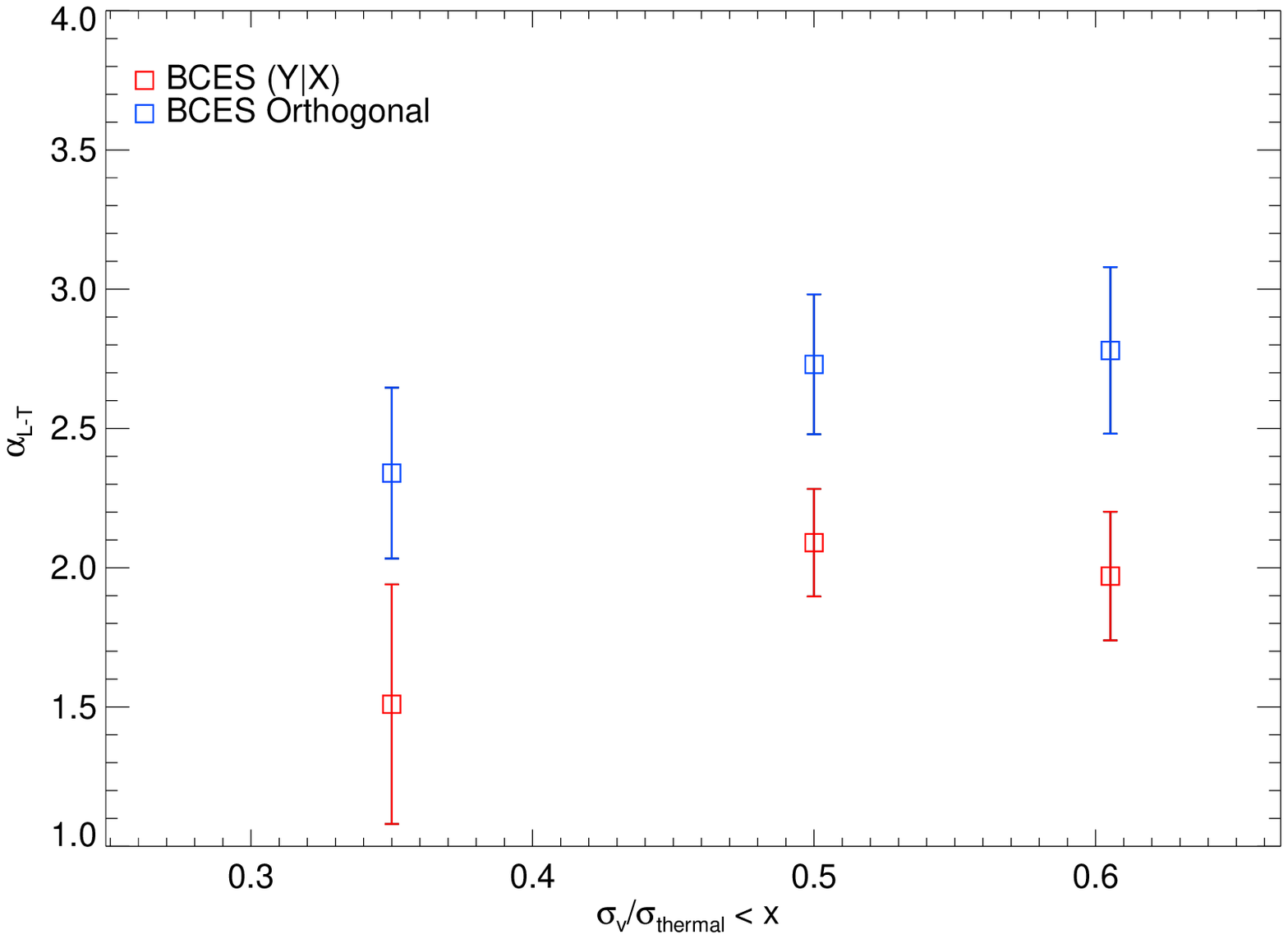}
  \includegraphics[width=0.46\textwidth]{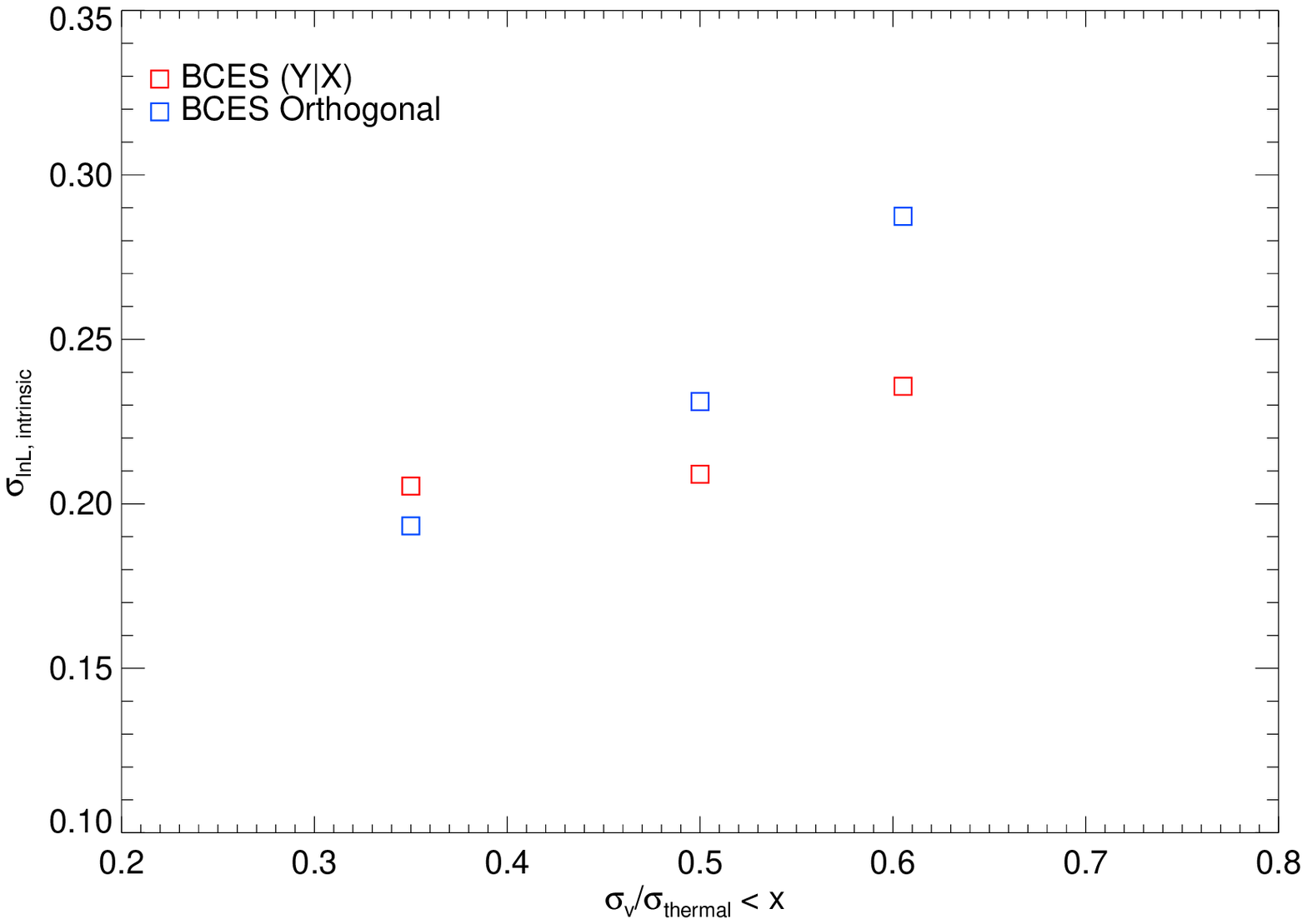}
  \caption{Dependence of normalization ($C$, top), slope ($\alpha$,
    middle) and 
    intrinsic scatter ($\sigma_{{\rm ln}L, intrinsic}$, bottom)
    of the $L-T$ relation on the maximum  $\mu$ used to select the
    corresponding sub--sample of clusters.}
  \label{fig:sigv_trends}
\end{figure}
%---------
In order to quantify the effects of l.o.s. velocity structure on the
resulting scaling relation, we investigated the dependences of
best--fit slope, normalization and scatter on the value of
$\mu$, by fitting the $L_X-T$ relation for the original sample and for the
two sub--samples, separately.
Results are shown in \fig\ref{fig:sigv_trends},
where we report both the BCES~(L$\mid$T) (red squares) and the BCES Orthogonal (blue squares) values.

For all the quantities studied, the best--fit values generally increase
with the increasing amplitude of velocity dispersion, relative to the
characteristic thermal value.
This has significant implications for the intrinsic
scatter (calculated here for the luminosity, $\sigma_{{\rm ln}L,
  intrinsic}$\footnote{We report the final value by considering the natural
  logarithm, ln, for comparisons to \cite{pratt2009}, although the
  equations and the logarithm space mentioned in the paper usually
  refer to ${\rm log}={\rm log}_{10}$}).
In fact, the trend found reflects how the scatter in luminosity of
the $L_X-T$ scaling relation is sensitive to the baryonic physics and
can be closely related to complex or disturbed configurations of the
ICM velocity field, which are quantified by large values of $\mu$.
Essentially, we find that {\it the introduction of clusters with significant
non--thermal velocity dispersion, with respect to their typical
thermal velocity, augments the scatter in the sample about the
best--fit $L_X-T$ relation.}\\

We note from the relation shown in \fig\ref{fig:lt_rel_sig}, that the additional
characterization of the cluster through the line--of--sight non--thermal
velocity dispersion defines a different picture with respect to observations.
The observed behaviour of cool--core and disturbed galaxy
clusters shows in fact a distinct separation of the two populations
in the $L_X-T$ relation (see data points reported in the Figure from
\cite{pratt2009}), where cool--core clusters are generally found to 
occupy the upper envelope of the best--fit relation (black, thick
diamonds in \fig\ref{fig:lt_rel_sig}), while morphologically disturbed clusters
mostly reside in the lower one. 
In \fig\ref{fig:lt_rel_sig}, however, we find that haloes with the
most significant degree of non--thermal motions populate
mostly the region above the best--fit curve. According to this
velocity diagnostics, these haloes might be classified as disturbed.
We interpret this apparent inconsistency with observations as a
different probe for deviation from the regularity of the haloes.
Indeed, a deeper analysis of the simulated sample indicates that all
the clusters have central electron densities which are sufficiently low to be
identified as non--cool core clusters, which remove the discrepancy
with observed clusters entirely, as further discussed in \sec\ref{app:cool_cores}.
Namely, within the population represented by our sample, the
quantification of non--thermal velocity dispersion of the gas along
the line of sight consitutes an additional, complementary method to
further discriminate the halo dynamical state.

\subsection{Velocity diagnostics: prominent outliers}\label{sec:vel_outliers}
	\begin{figure*}
	\centering
        {\bf \Large VELOCITY--DIAGNOSTICS OUTLIERS}\\
        \vspace{0.3cm}
        % halos with highest values of non-thermal gas velocity dispersion
        {\bf \Large halo 29}\\
	\includegraphics[width=0.24\textwidth]{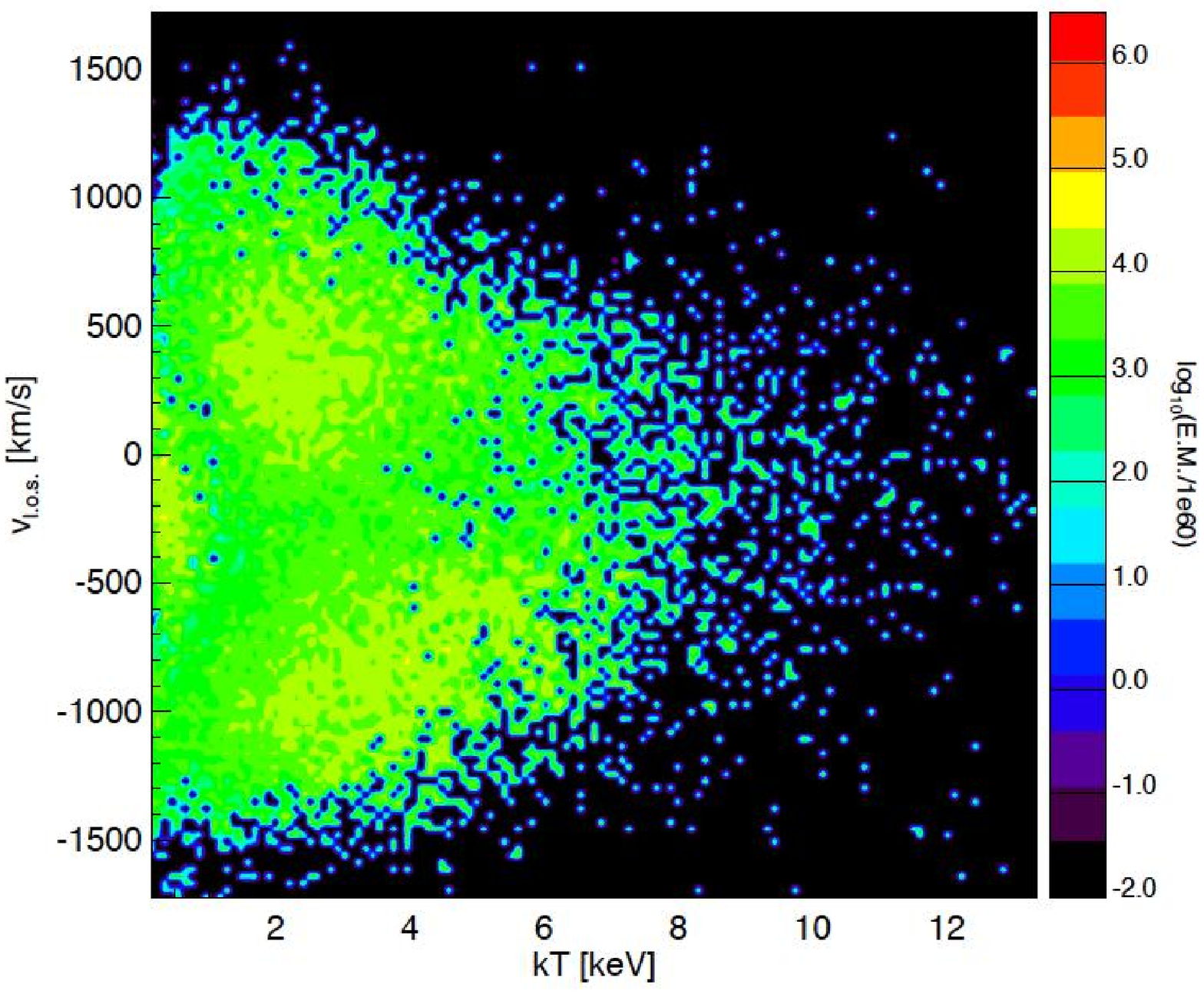}
	\includegraphics[width=0.24\textwidth]{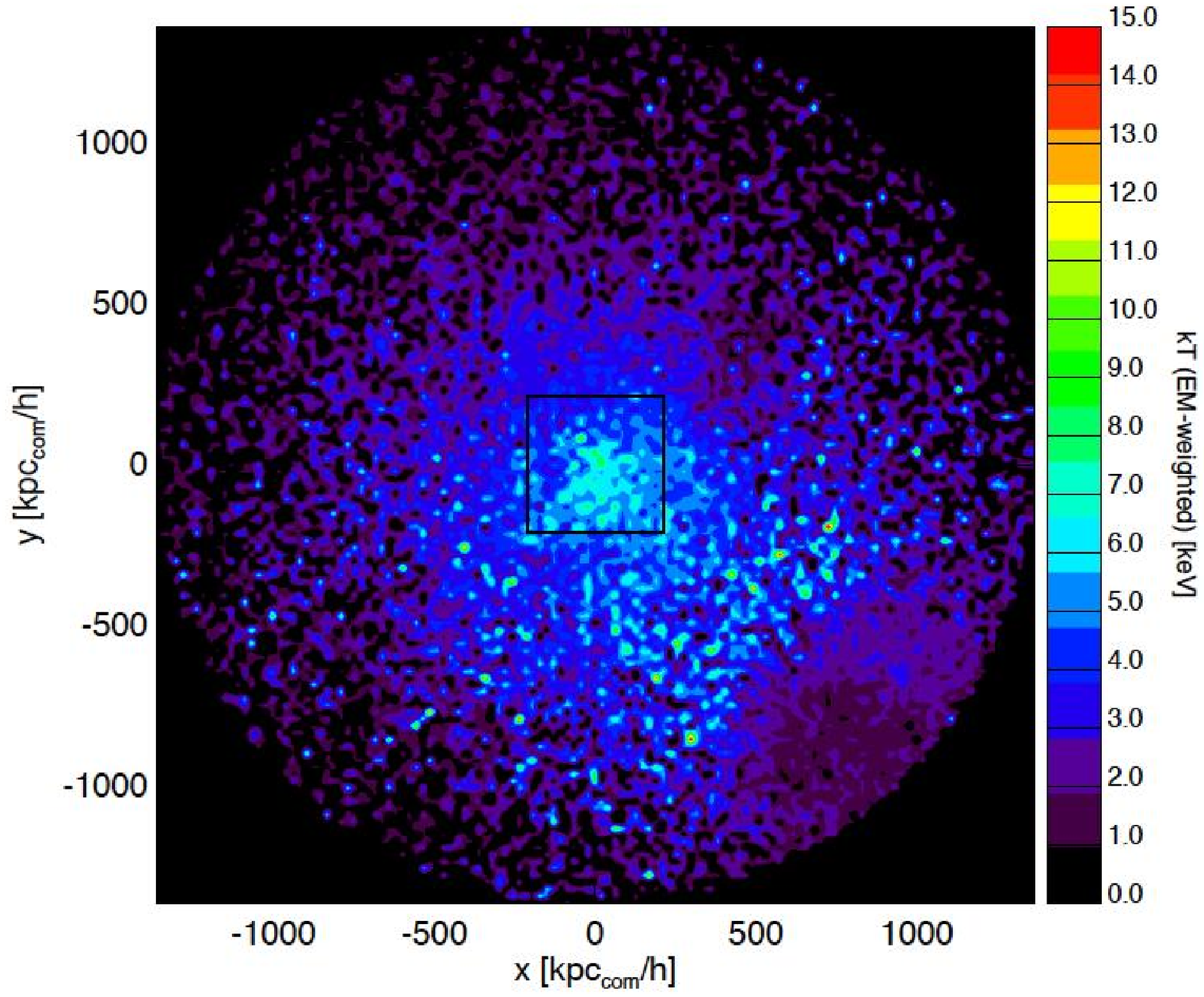}
	\includegraphics[width=0.24\textwidth]{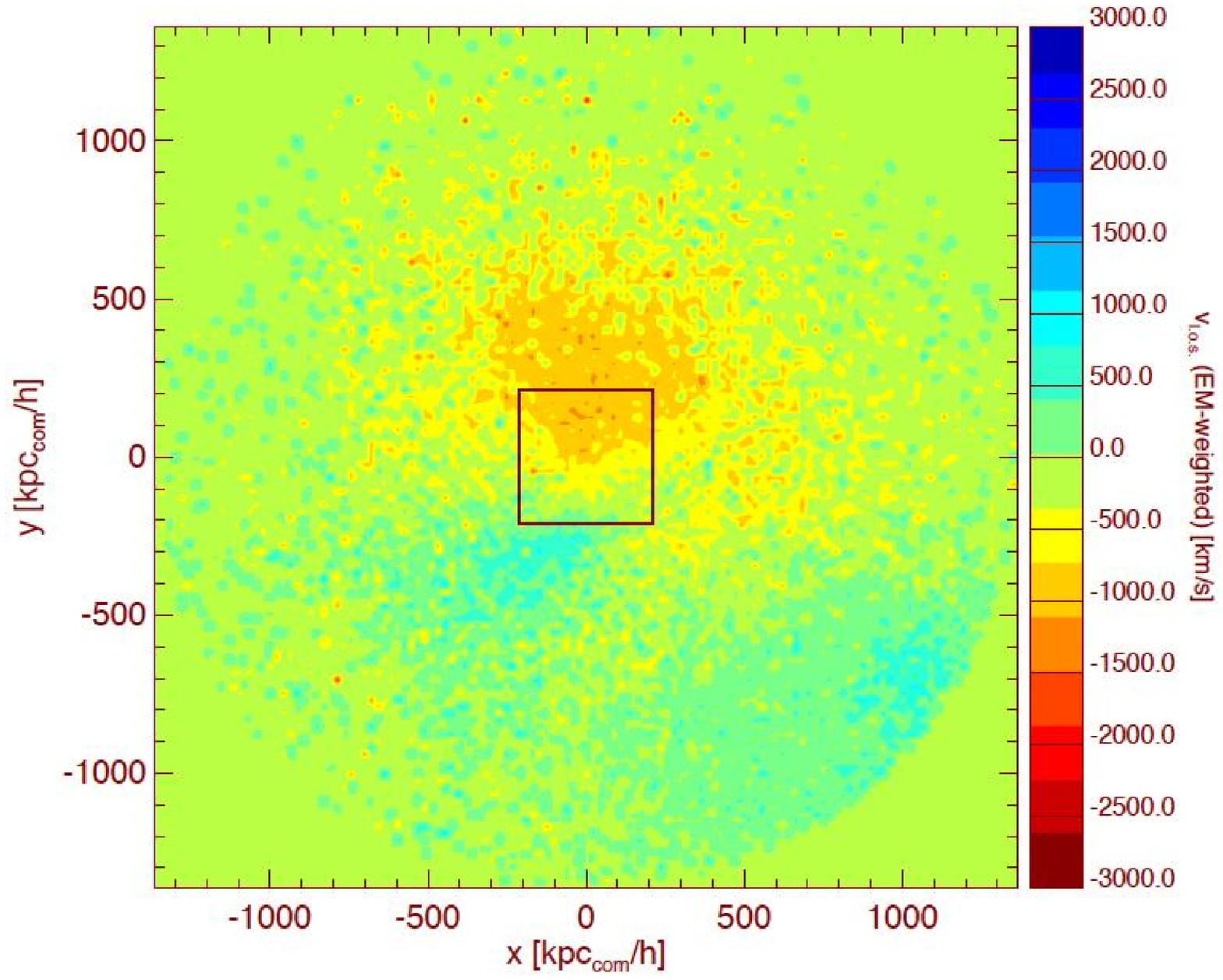}
	\includegraphics[width=0.24\textwidth]{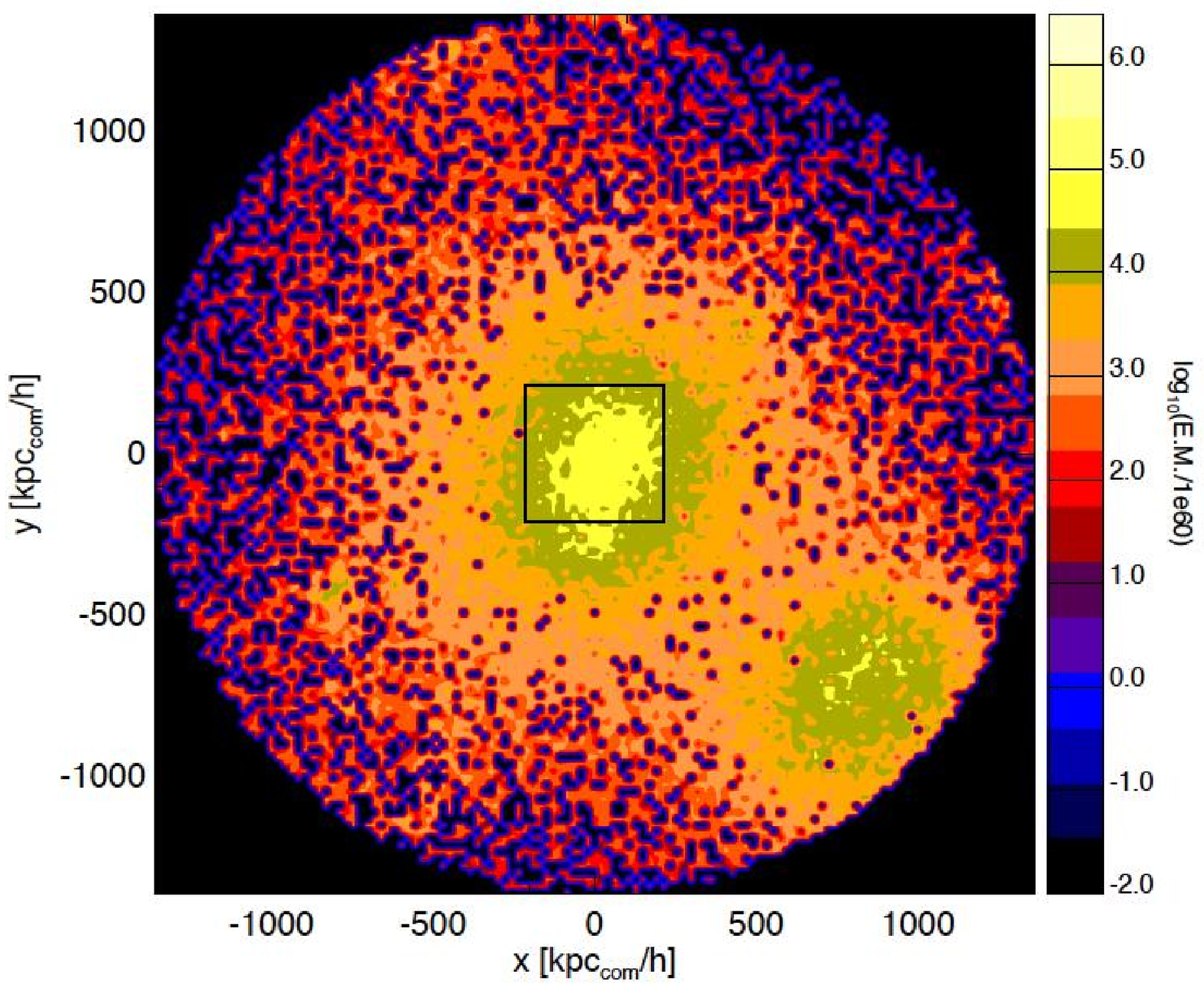}
        \vspace{0.2cm}
        {\bf \Large halo 28}\\
	\includegraphics[width=0.24\textwidth]{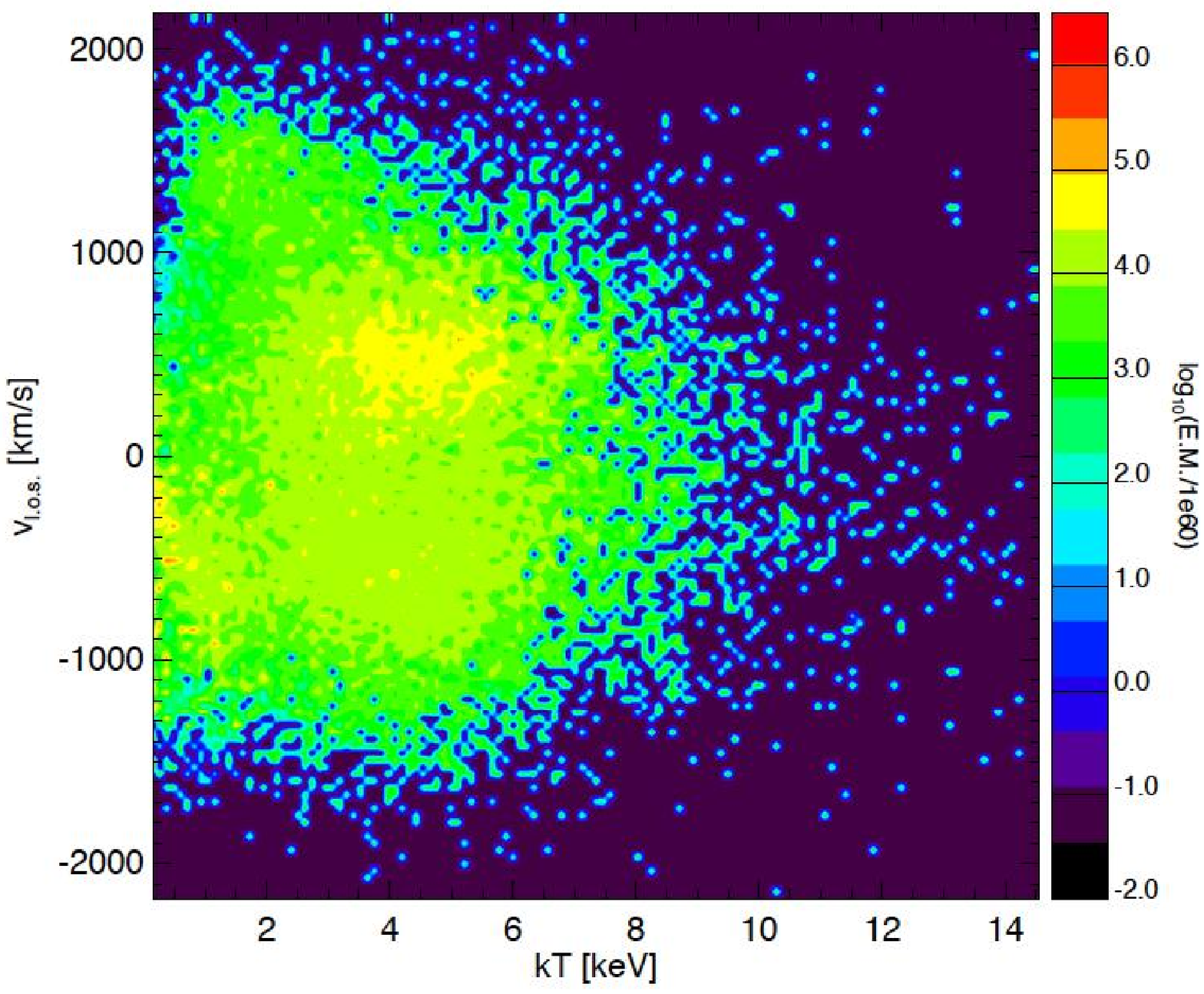}
	\includegraphics[width=0.24\textwidth]{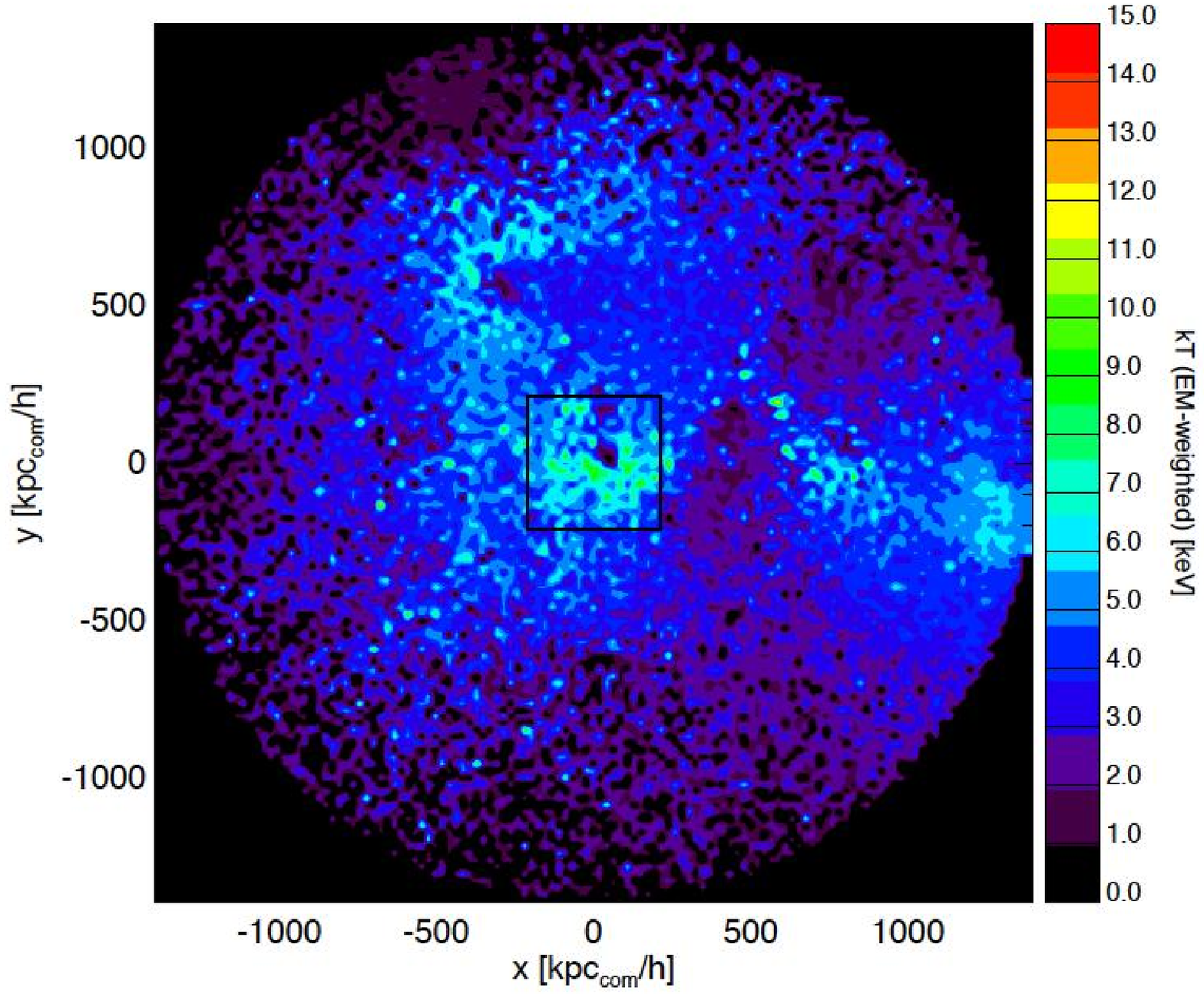}
	\includegraphics[width=0.24\textwidth]{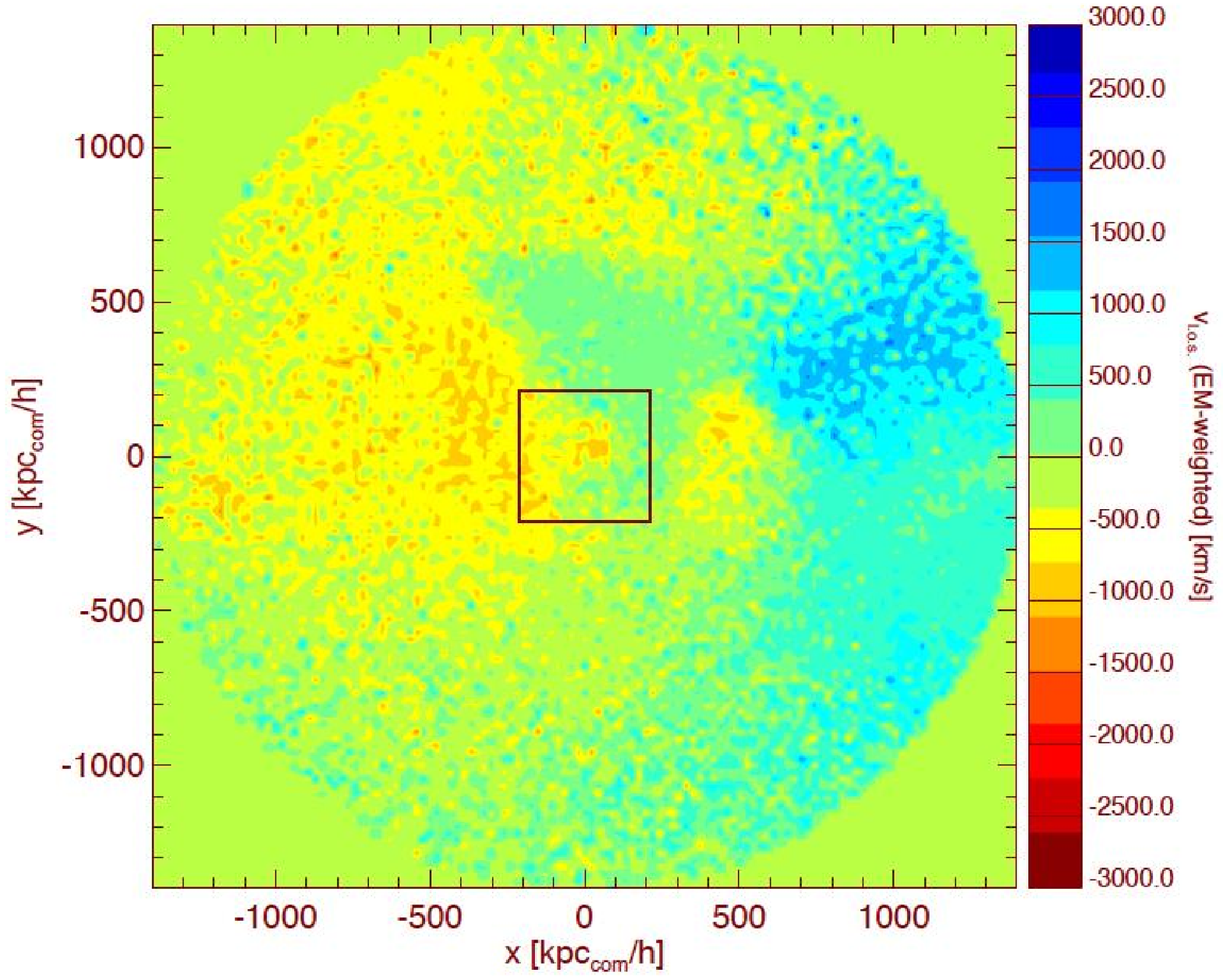}
	\includegraphics[width=0.24\textwidth]{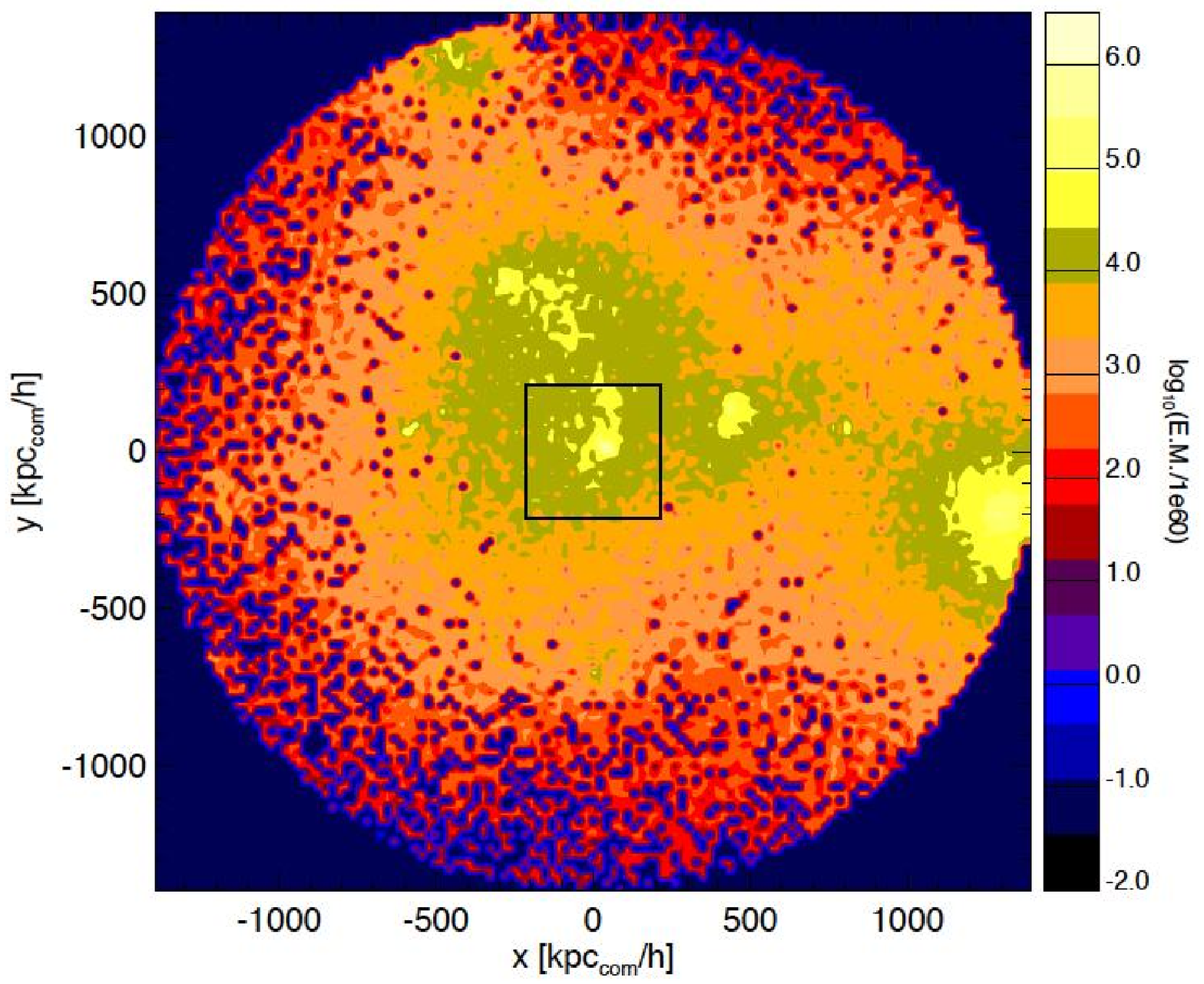}
        \vspace{0.2cm}
        {\bf \Large halo 37}\\
	\includegraphics[width=0.24\textwidth]{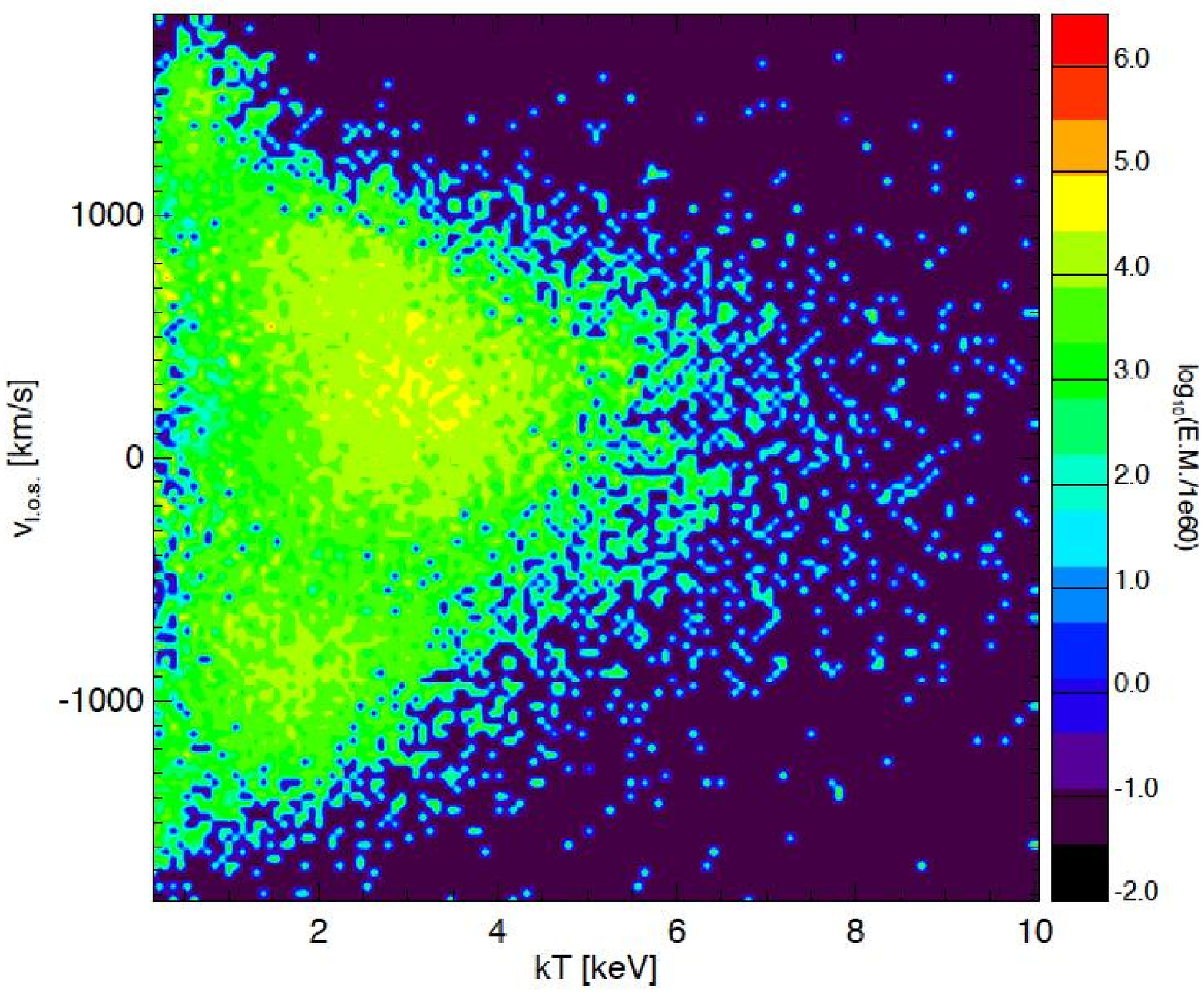}
	\includegraphics[width=0.24\textwidth]{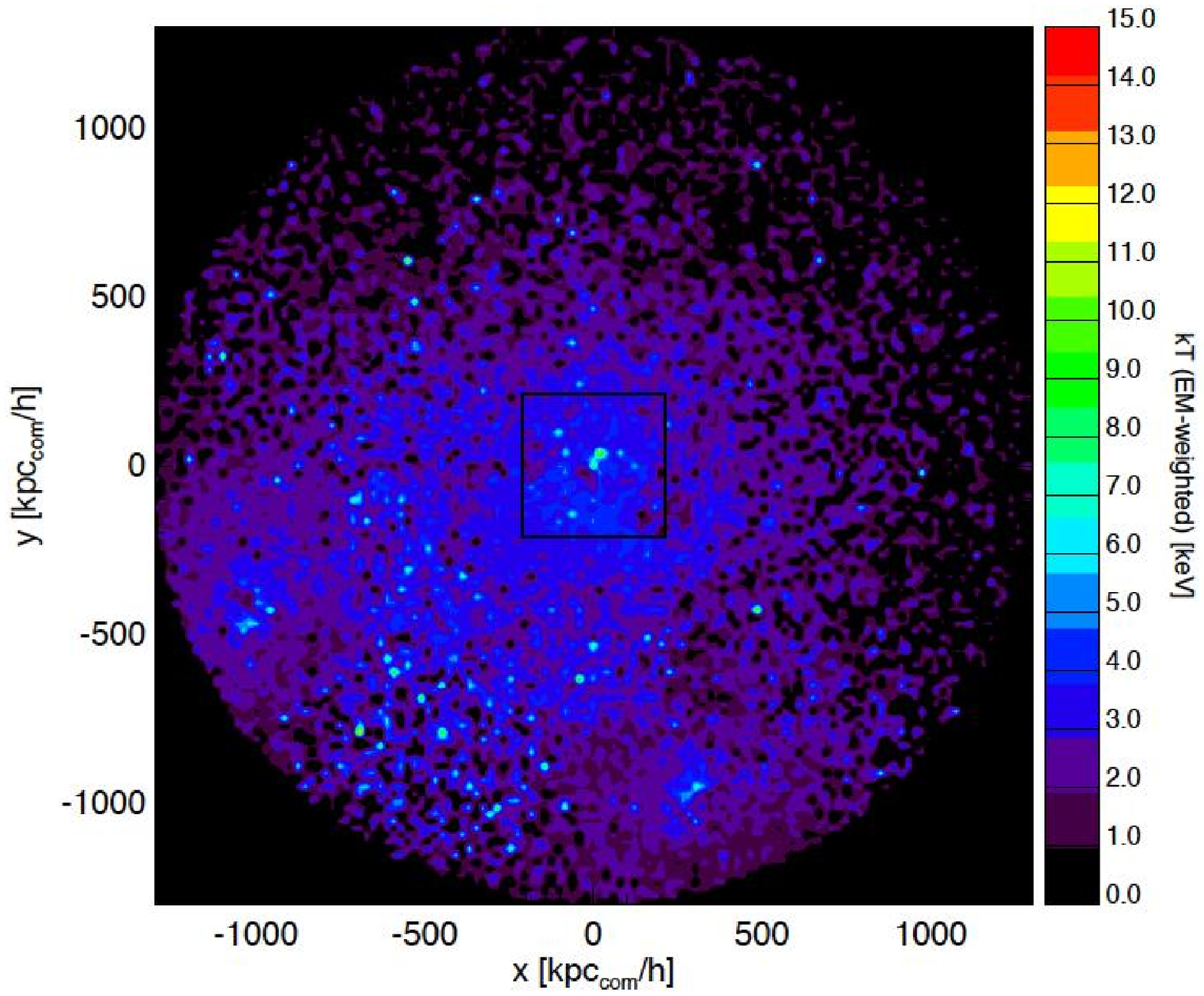}
	\includegraphics[width=0.24\textwidth]{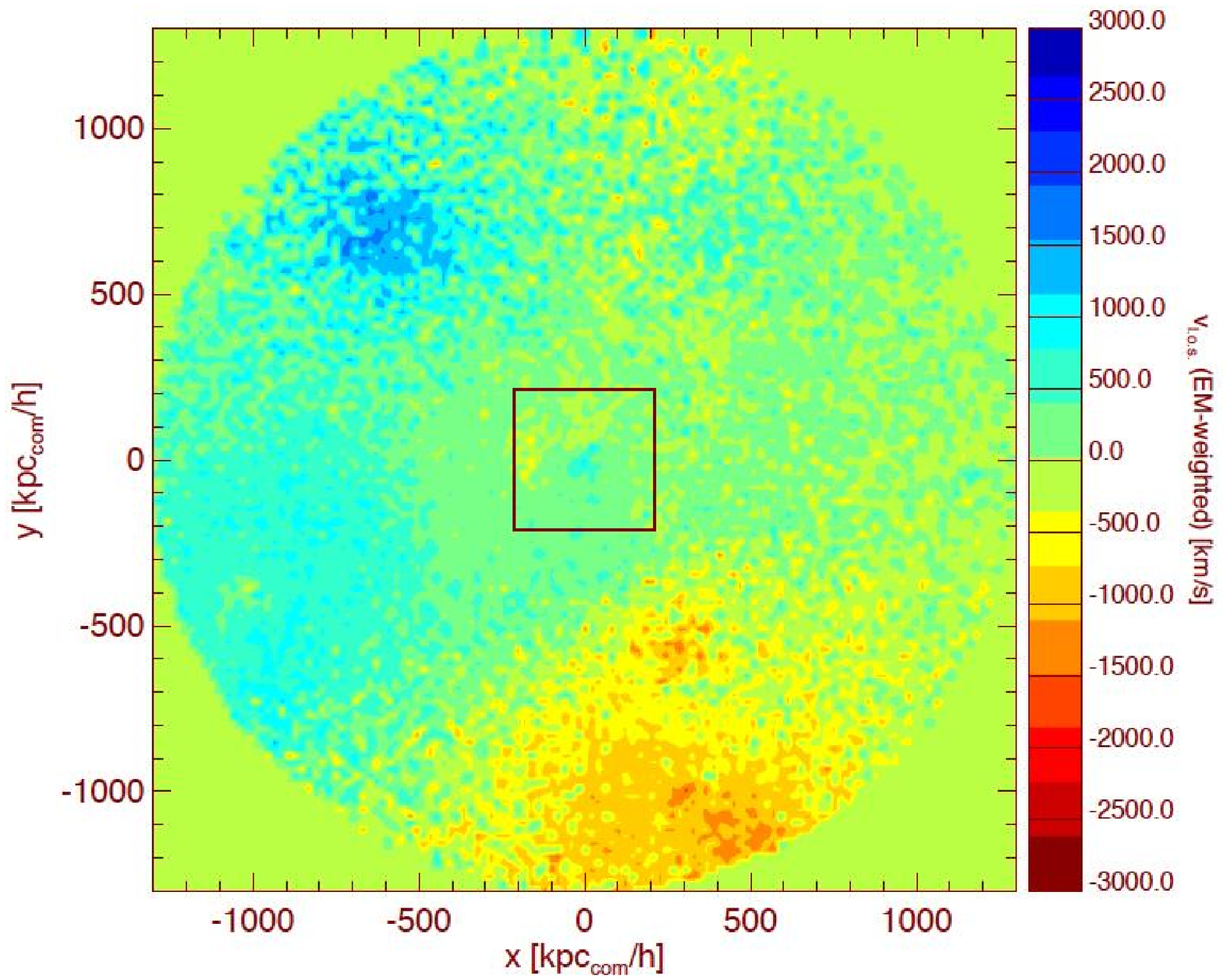}
	\includegraphics[width=0.24\textwidth]{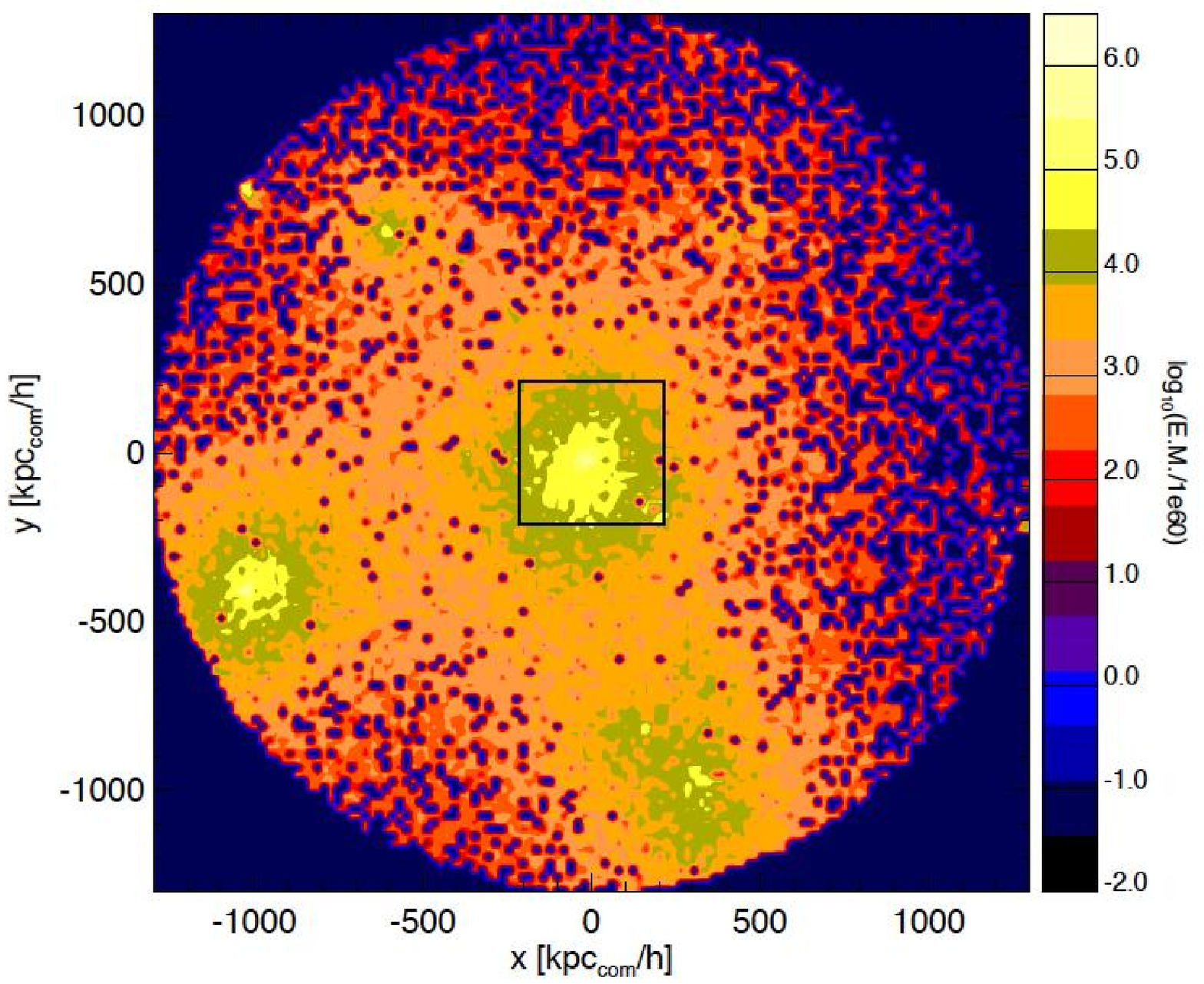}
        \vspace{0.2cm}
        {\bf \Large halo 27}\\
	\includegraphics[width=0.24\textwidth]{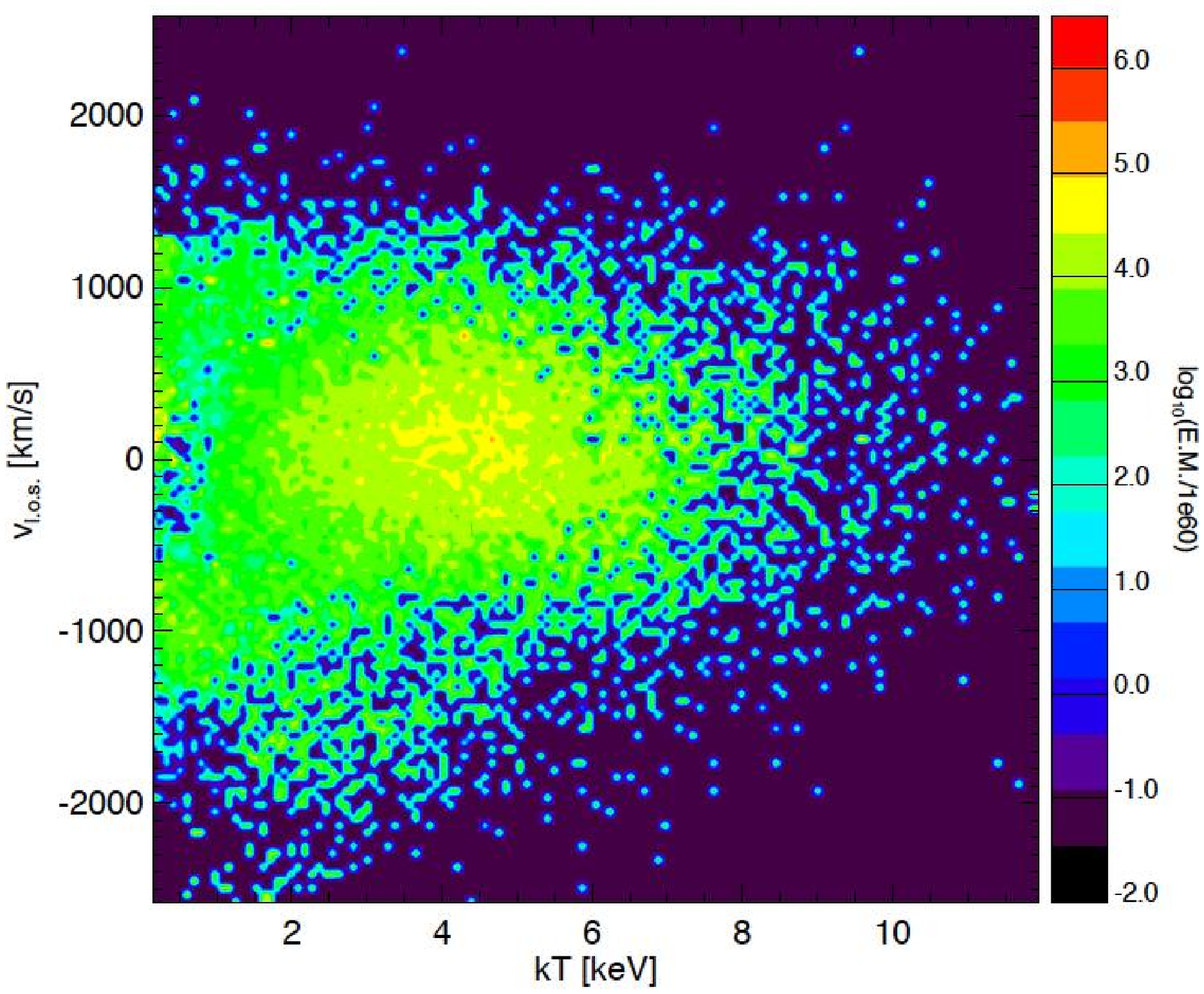}
	\includegraphics[width=0.24\textwidth]{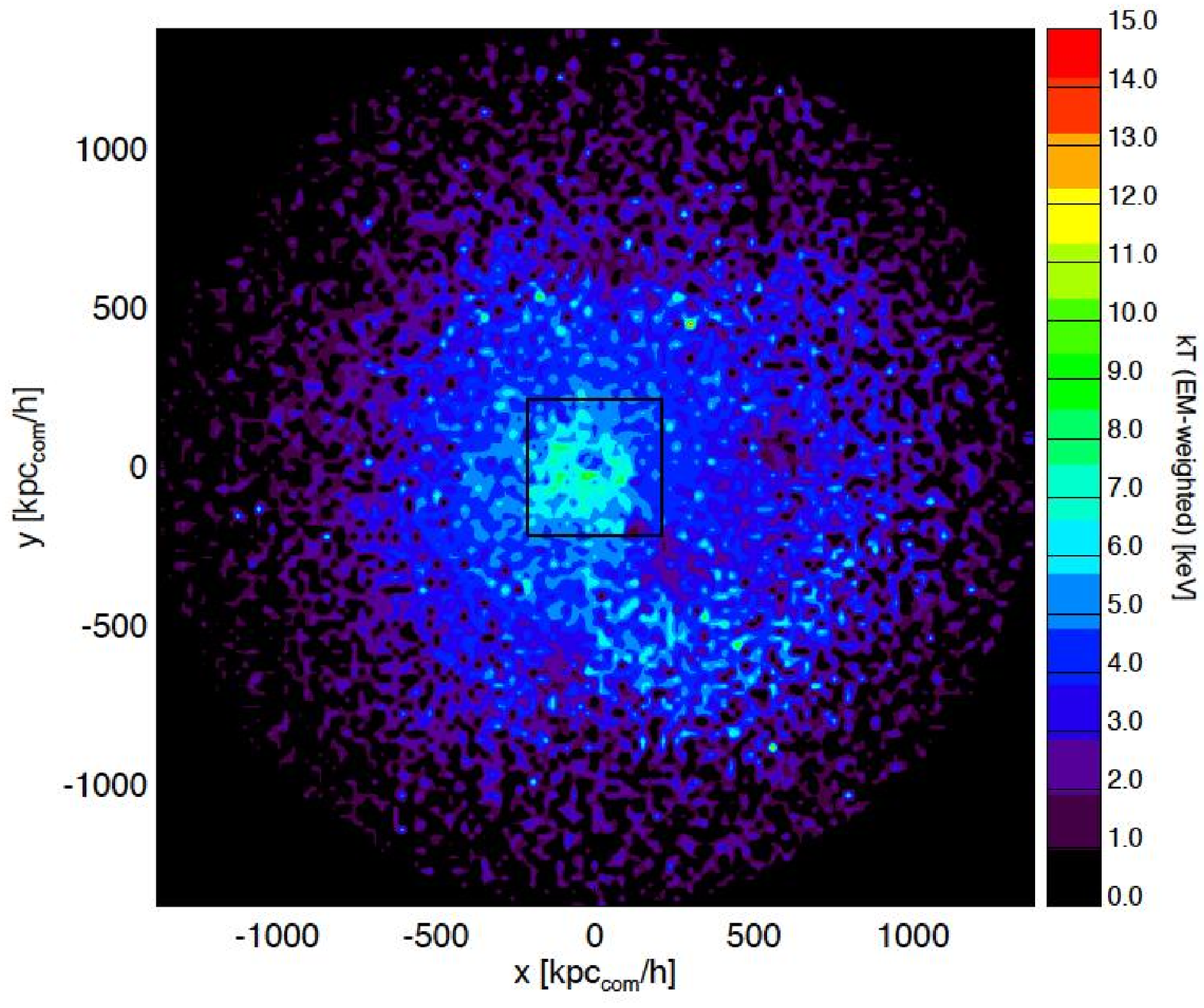}
	\includegraphics[width=0.24\textwidth]{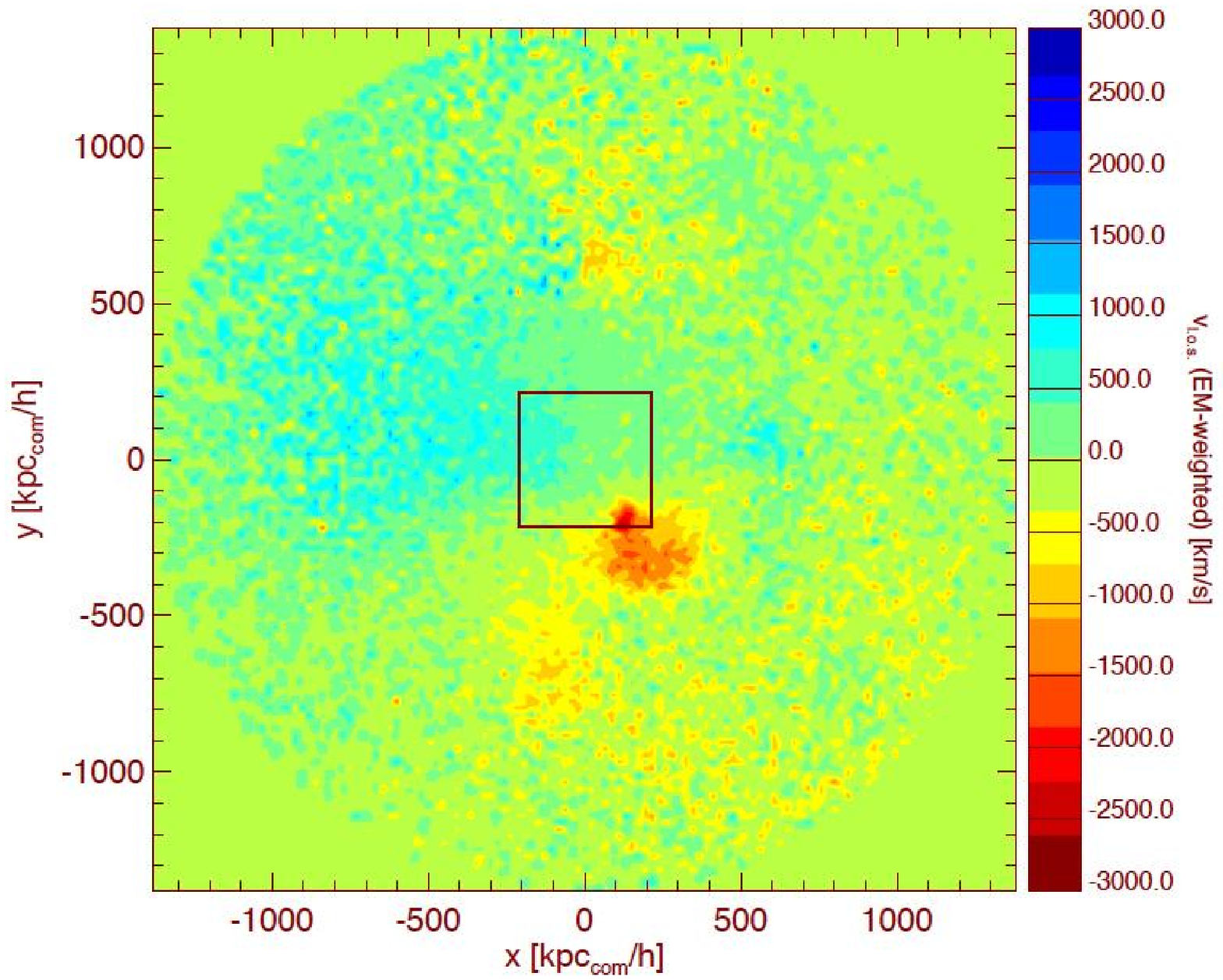}
	\includegraphics[width=0.24\textwidth]{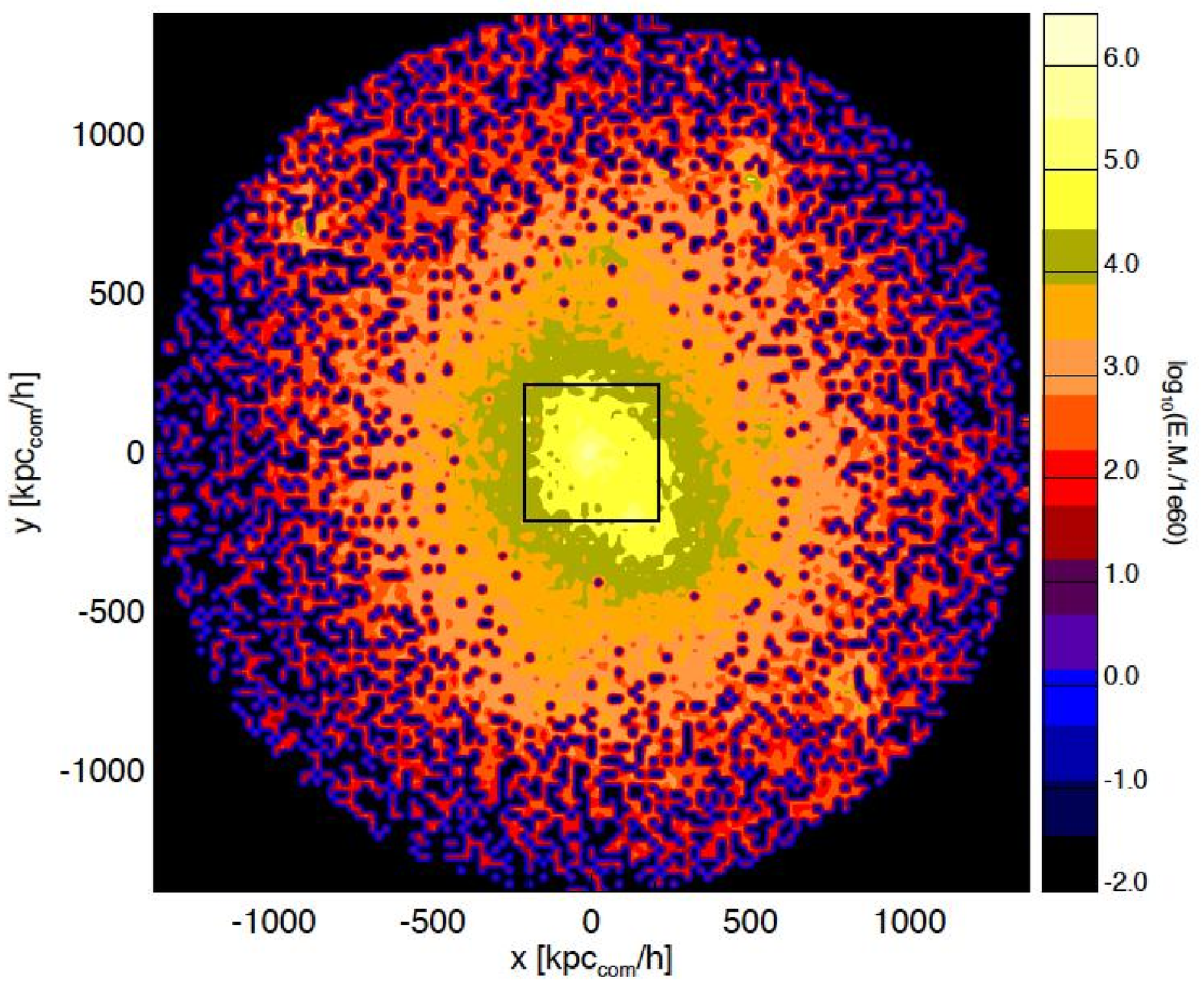}
        \vspace{0.2cm}
        {\bf \Large halo 36}\\
	\includegraphics[width=0.24\textwidth]{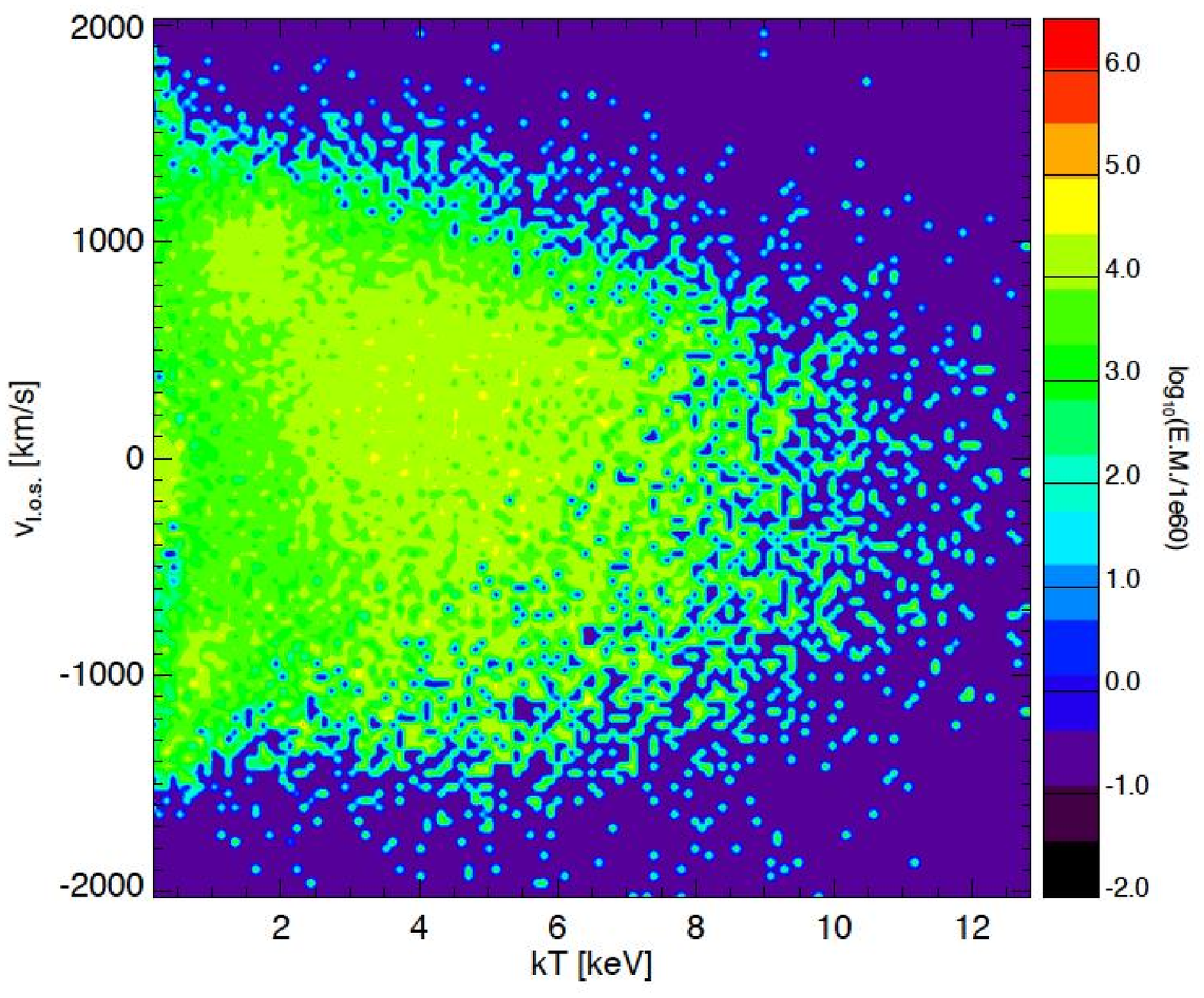}
	\includegraphics[width=0.24\textwidth]{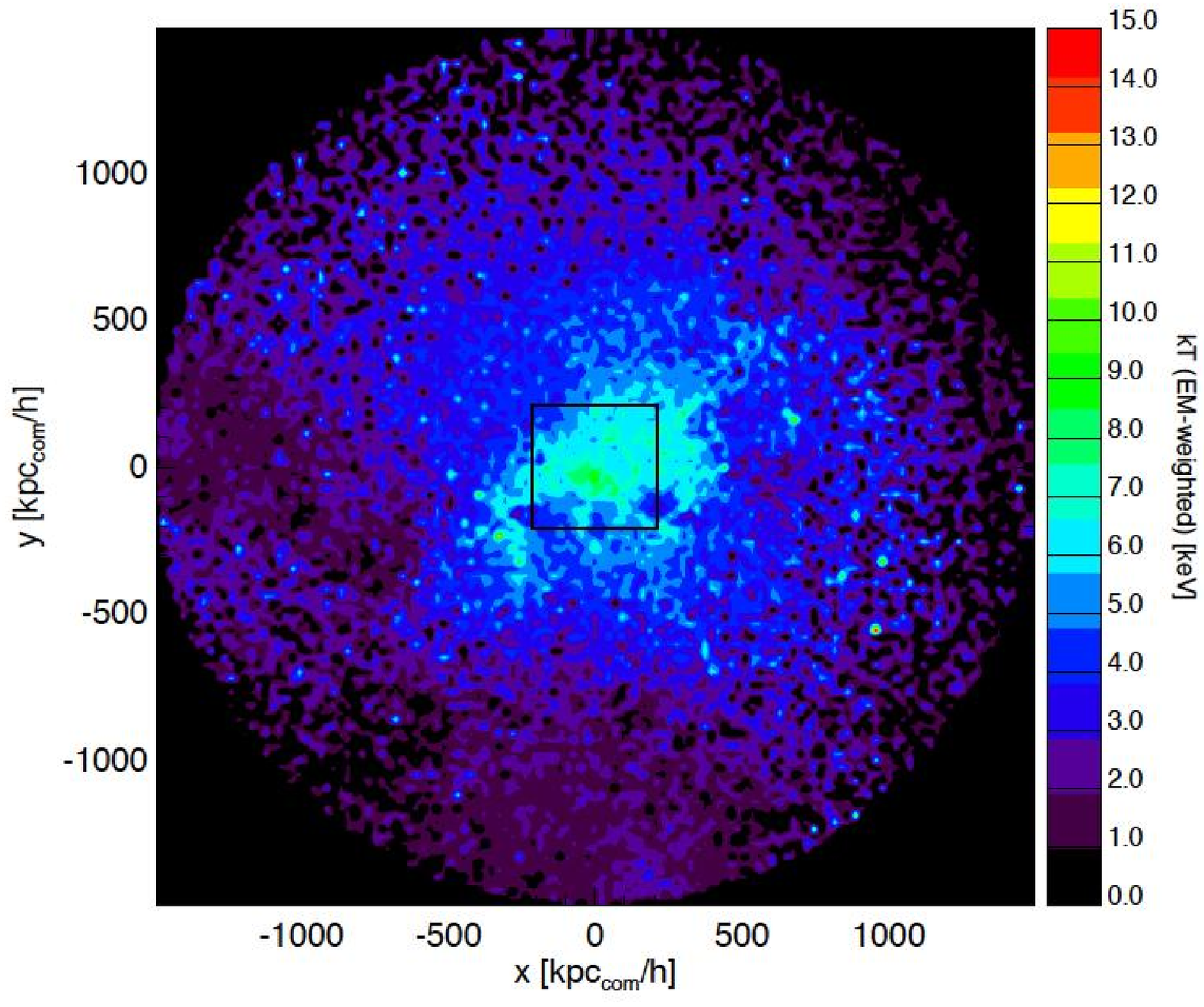}
	\includegraphics[width=0.24\textwidth]{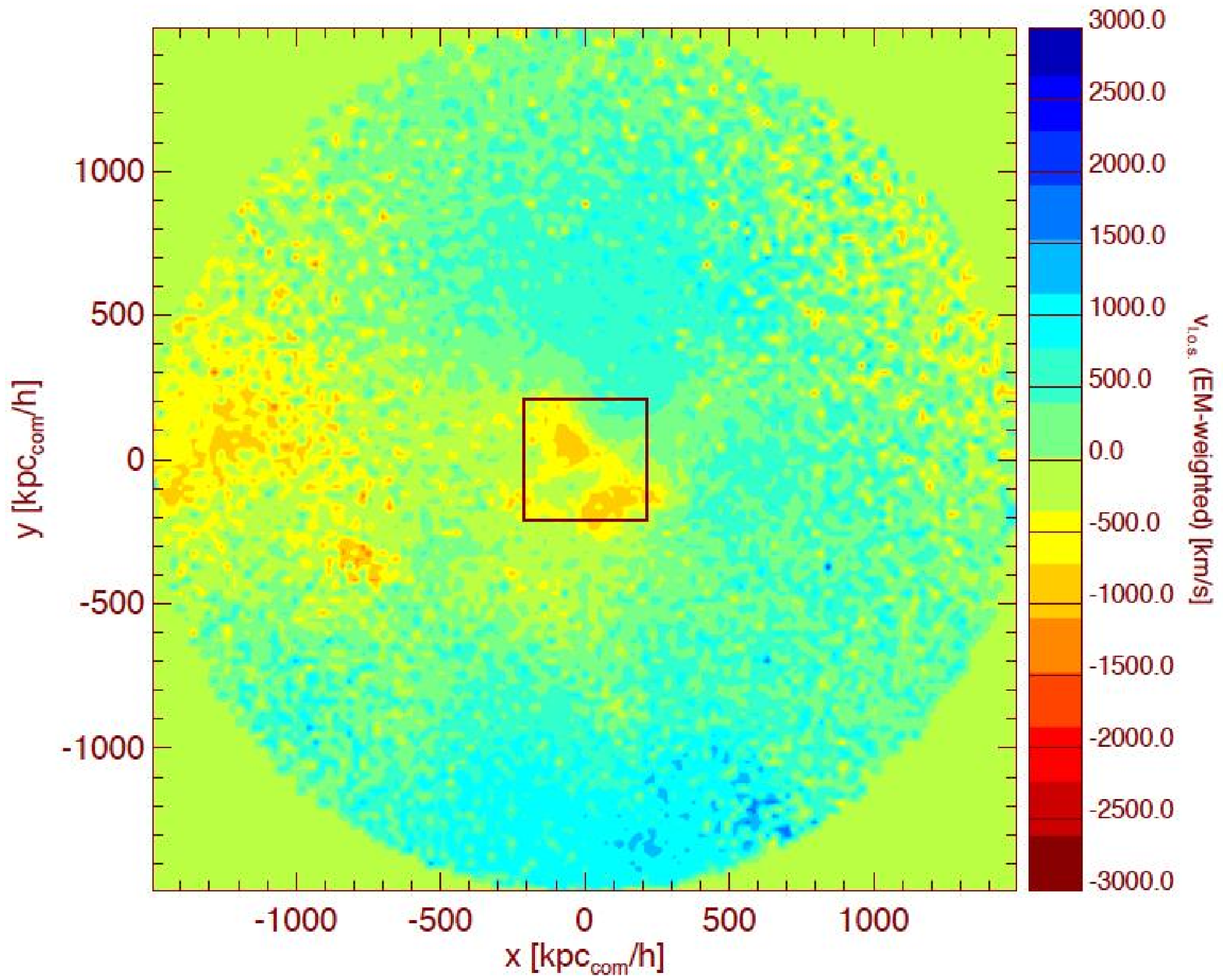}
	\includegraphics[width=0.24\textwidth]{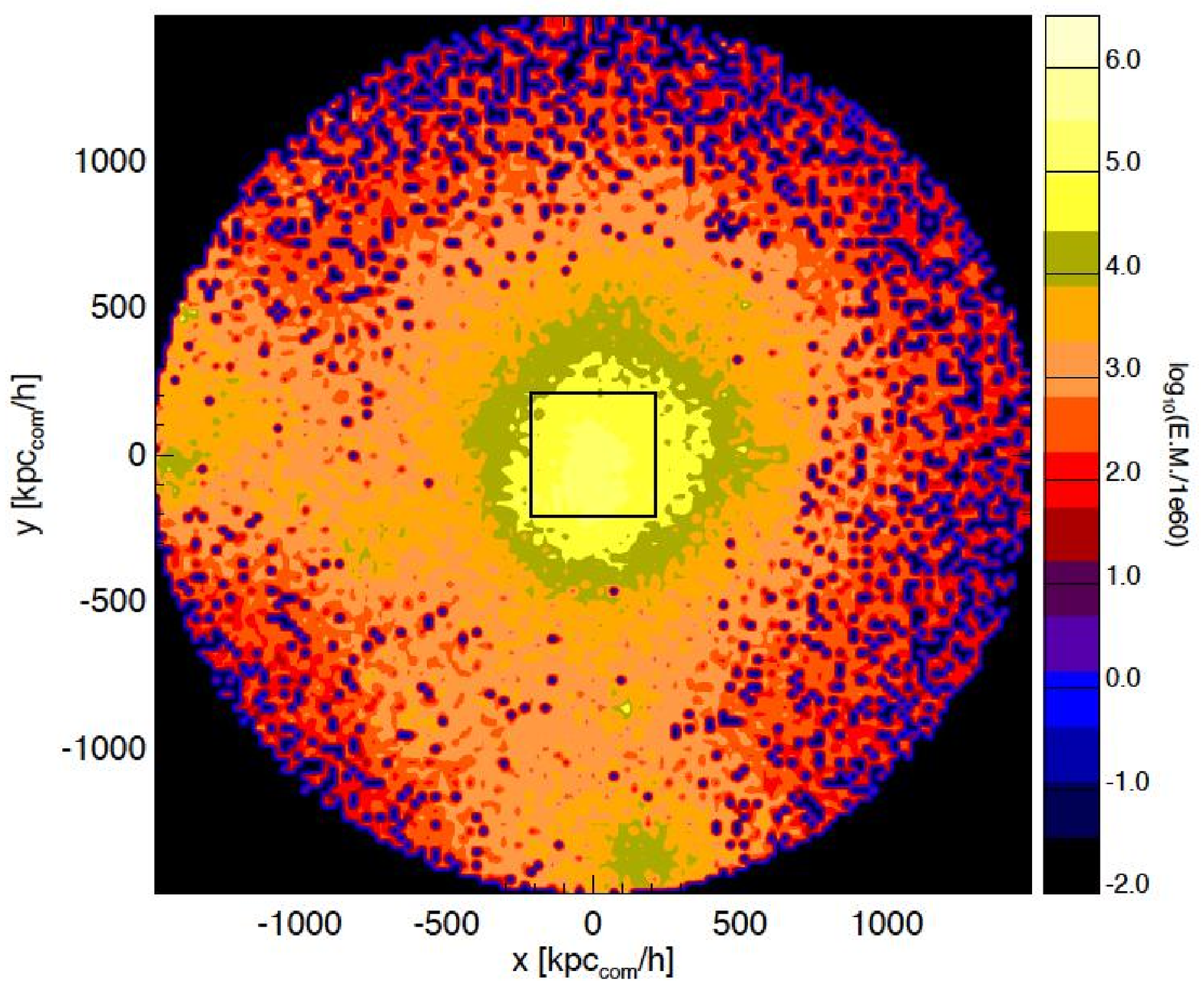}
        \vspace{0.2cm}
        {\bf \huge \hspace{4cm} T \hspace{3.5cm} $v_{l.o.s.}$ \hspace{3.5cm} EM}\\
	\caption{Maps of the ICM thermo--dynamical status for the five
          most-significant outliers singled out by the velocity
          diagnostics. Columns, from left to
          right: (1) $v_{l.o.s.}-kT$ map, color--coded by EM
          (considering gas particles within $\rfive{}$); (2)
          EM--weighted, temperature map, projected along the l.o.s.;
          (3) EM--weighted, l.o.s.--velocity map projected onto the
          $xy$ plan (i.e. along the l.o.s.); (4) EM map projected along
          the l.o.s., white circles mark the most significant
          substructures residing within $\rfive{}$, in the plane
          perpendicular to the l.o.s.. Spatial maps in columns (2)-(3)-(4) refer to the whole $\rfive{}$
          region and, overplotted, is the ATHENA--like FoV for
          comparison. From top to bottom the haloes are ordered such
          that the value of $\mu = \sigma_v(<\rfive)/\sigma_{thermal}$
          (\eq\ref{eq:sig_ratio}) is {\it decreasing}.}
	\label{fig:maps_sigv_outliers}
	\end{figure*}
The velocity diagnostics performed and the distribution of the $\mu$
value clearly single out some haloes in the high--velocity tail.
This subsample of $5$ clusters includes the most prominent outliers in
the $L_X-T$ relation, and the exclusion of these haloes
turns out to reduce the scatter in the best-fit relation.
Therefore, we further 
discuss here their
thermo--dynamical structure from the simulation directly, 
in order to
provide an interpretation of the high gas velocities found. 

In \fig\ref{fig:maps_sigv_outliers} we show the spatial distribution
of temperature (kT), l.o.s. component of the gas velocity ($v_{l.o.s.}$) and emission
measure (EM) of the $5$ outliers, as well as the $v_{l.o.s.}-kT$ phase
diagram, color--coded by EM. The region of the clusters considered is
always that enclosed within the projected
$\rfive$, along the line of sight through
the simulation box.

Despite the fact that all of these clusters show significant
non--thermal motions compared to the characteristic thermal velocity
expected, 
they show different evolutionary states of such velocity patterns.

Clearly, the maps show very different internal structures and the
presence among these velocity outliers of both substructured haloes
and more regular roundly--shaped ones, like halo $36$ especially.
The latter, in partiuclar, shows very regular features, no significant
substructures and relatively low cooling time (as will be shown in
\sec\ref{app:cool_cores}), suggesting that the non--thermal velocities
measured are not due to merging or infalling subhaloes but rather
characterize intrinsically the ICM. 
Similar is the case of halo $27$, which shows a very regular
morphology and structure from all the maps in
\fig\ref{fig:maps_sigv_outliers} and, however, is 
charaterized by both large cooling times and significant non--thermal
velocities. 

Such examples indeed represent the problematic case where no
self--bound substructures are visible in the emission maps, being
probably dissolved but not yet fully thermalized, which causes
high--velocity streams to persist in the ICM. Here the usual
classification of the halo as a regular, not disturbed cluster and the
assumption of thermal motions to dominate the gas pressure support
might be misleading.

Nonetheless, we also find more trivial cases, like halos $37$ and $28$, that
show significant substructures in the emission maps, reflected in their
velocity and temperature structure. 
This suggests a non--relaxed dynamical configuration, where the
subhaloes are still in the form of self--bound objects with 
an own dark matter potential and are most likely
responsible for driving high--velocity motions.
In fact, these clusters deviate
significantly from the best--fit $L_X-T$ relation, meaning that the kinetic
energy of the infalling substructures has not yet thermalized and the
boost in luminosity is not yet followed by the increase of
temperature.
Similarly, the high velocity measured for halo $29$ might be related
to a disturbed state of the cluster, for which a subhalo is clearly
present from the emission map.

In an observational case of this kind, though, clearly disturbed
haloes can be identified and prominent substructures can be
explicitely excluded from the analysis.
%--------------------------------------------------
\subsection{Cool-core clusters in the simulated
  sample}\label{app:cool_cores}
\begin{figure}
  \centering
  \includegraphics[width=0.46\textwidth]{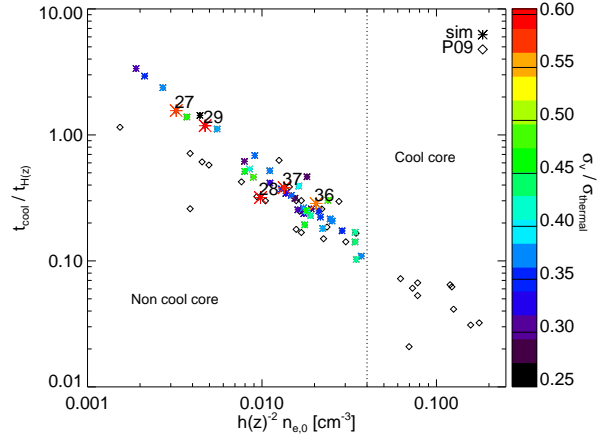}
  \caption{Cooling time versus central electron density, with color
    code as in \fig\ref{fig:lt_rel_sig}. Asterisks refer to simulated
    data, while black diamonds represent the observational sample
    investigated in Pratt et al. (2009) (P09). The dotted line marks the
    threshold adopted by P09 to define cool core systems,
    i.e. $h(z)^{-2}\,n_{e,0}>4\times 10^{-2}\cm^{-3}$.}
  \label{fig:cool_core}
\end{figure}

In \fig\ref{fig:cool_core} we plot the central cooling time,
normalized to Hubble time, as
function of central electron density, $n_{e,0}$, for all the cluster in the
sample.
As in \cite{pratt2009}, we define the central values as those
calculated at $0.03\rfive$\footnote{While for global properties
  as $L_X$ or $T$ the difference in the definition of $\rfive{}$ is
  not crucial, here we also
  calculate the $\Delta=500$ overdensity with respect to the critical
  density of the Universe, in order to compare as consistently as
  possible against the observational data. In fact, the definition is
  particularly relevant in order to evaluate the values of cooling
  time and density at exactly the same radius, being not mean values within a
  region but rather local values.}.
The data reported have been derived from the three--dimensional profiles of the
haloes, calculated directly from the simulation.

Opposite to the overcooling problem of standard numerical simulations,
we show with this Figure that no real cool cores are present in our
sample, as a consequence of the inclusion in the simulations of AGN
feedback.
In fact, this has the effect of preventing the gas in the hot phase to
remain in the central region long enough time to cool and form stars.
Comparing to observational data, we do not find haloes populating the
region of very short cooling times and high central electron density.

Here, we adopt the same color code as in
\fig\ref{fig:lt_rel_sig} in order to identify the most prominent
velocity outliers previously discussed. 
We note that the five outliers singled out by the velocity
diagnostics (\sec\ref{sec:vel_outliers}) comprise both haloes with very long
cooling times and halos for which the cooling time is only some
tenths of the Hubble time. 
For the latter, although not classified as strong cool cores (CCs), cooling
might be nevertheless important in the central region. 
With respect to our analysis, this can also be the case of
substructured haloes, whenever the approaching subhaloes have not merged yet,
namely the first passage through the cluster central region has still to
occur. In such circumstances, the cooling time calculated for the cluster core can also be
relatively small, since the innermost region has still to be affected by the
interaction, while the luminosity and velocity fields already are.

Among these haloes with relatively short cooling time, 
we find two haloes ($37$ and $28$) out of the five clusters with significant
non--thermal motions, which reside in the region above the
best--fit $L_X-T$ relation.
This suggests that there is no real tension between their
position in the $L_X-T$ plane and the observational evidence of CC
clusters occupying that region of the relation.
The velocity diagnostics proposed can actually probe different aspects 
of the cluster state in addition to the usual dinstinction between CCs
and morphologically disturbed clusters. 
%
%-----------------
\section{Conclusions} \label{sec:conclusion}
In this paper we have presented the study of the ICM velocity
structure and its impact on the $L_X-T$ scaling relation and cluster status, from
hydrodynamical simulations and synthetic X--ray observations of a set
of galaxy clusters.
Numerical simulations have been performed by means of the treePM/SPH,
parallel code P--GADGET3, including a number of physical processes
describing the baryonic component with a level of
detail never reached so far, namely cooling, star formation and supernova--driven winds \cite[][]{springel2003},
chemical enrichment from stellar population, AGB stars and SNe \cite[][]{tornatore2004,tornatore2007},
low--viscosity SPH \cite[][]{dolag2005}, 
black--hole growth and feedback from AGN. The set of $43$ simulated cluster--like haloes has been
selected from a medium--resolution cosmological box by requiring
$\mfive{} > 3\times10^{14}h^{-1}\msun$. 
The mock X--ray observations have been obtained with the virtual
photon simulator \phox{} \cite[][]{biffi2012} for both Chandra
ACIS--I3 and the XMS spectrometer originally designed for ATHENA,
and used here as prototype for any next generation X--ray spectrometer
capable of reaching very high energy resolution. 

From the direct analysis of the gas component in the central part of
the simulated haloes, in the projection plane perpendicular to the
chosen line of sight, we find tight relations among different
definitions and selection criteria used to calculate the
line--of--sight component of the gas velocity dispersion.
In order to consistently compare and use results from simulations and
mock observations, we always refer to cylinder--like regions which
extend throughout the $z$ coordinate of the simulated box, i.e. along
the l.o.s. direction.
In particular, we find that mass-- and emission--measure-- (EM--) weighted
velocity dispersion calculated for the gas particles with $\rfive{}$
both correlate closely to the halo $\mfive{}$ and, therefore, can be equivalently
employed.
Between the two definitions, we concentrate on the EM-weighted velocity
dispersion since it is more sensitively related to the gas X--ray
emission.

We also find that $\sigma_{500,EM}$ probes fairly well the velocity
dispersion in inner, smaller regions of the clusters, such as the
region covered by a small FoV like that of ATHENA. 
As a fair approximation, we can safely assume the results to be representative
of the whole region out to $\rfive{}$, even though single--pointing
observations with a telescope of similar FoV of clusters
at the considered redshift would only cover their
innermost part.
On the other hand, small--scale spatial features of the ICM velocity
field could be possibly unveiled by the detailed comparison between
the large--scale amplitude of the l.o.s. velocity dispersion and the
value detectable with ATHENA--like observations.

By applying our virtual telescope, \phox{}, to the simulated haloes,
the l.o.s. velocity dispersion is derived from the velocity broadening
of the K$\alpha$ iron complex around $6.7\kev{}$ in the synthetic X--ray
spectra, provided the high energy resolution expected to be reached by
such a spectrometer.

Moreover, predictions on $\sigma_v$
provided by the simulation and detections mimicked with \phox{}
show a level of deviation of less than $20\%$, for $74\%$ of the clusters in the sample.
The largest deviations of observed velocities from expected values 
are significantly dependent on the complicated distribution of the gas
velocities and thermal structure (see \fig\ref{fig:vhisto_maps}).
In principle, a more detailed modelling of the shape of the spectral
emission lines, such as the iron complex used here, rather than the
standard Gaussian fit, should permit to derive more accurate
measures of the l.o.s. velocity dispersion of the gas, accounting for
multiple (thermal) components in the velocity field.

Given the ideal case of a similar X--ray spectrometer at high
resolution, we base our further investigation on the simulation results.
Indeed, this is further suggested by the good agreement between mock
data and simulation and therefore we utilise the $\sigma_{500,EM}$
obtained directly from the simulation to constrain the impact of
velocity structure onto X--ray observables, such as luminosity and
temperature.

In principle, as soon as high--precision X--ray spectroscopy will
become available for real observations, we will be able to safely employ the detected
values of the ICM non--thermal velocity dispersion for studies of real
galaxy clusters as well. Possibly, this will be already explored, for
instance, in the case of the
brightest clusters by means of the upcoming ASTRO--H satellite, due to
be launched in 2014, for which a energy resolution of $7{\rm eV}$ at $0.3-12\kev$
is expected.

From the synthetic Chandra spectra obtained with \phox{}, we were able to estimate
observed bolometric luminosity, extrapolated to the entire X--ray
band, and temperature, for the ICM residing within $\rfive{}$ in the
projection plane.
The $L_X-T$ scaling relation constructed from these mock data
generally agrees with real observations, e.g. from \cite{pratt2009}.
By using the BCES linear regression fitting method
\cite[][]{bces1996}, we assume a linear relation between $L_X$ and $T$
in the log--log space and determine the best--fit slope and normalization of the $L_X-T$
for our $43$ simulated clusters, finding a shallower slope with
respect to the results reported in \cite{pratt2009}.
In the fitting procedure, we explore both the BCES~(L$\mid$T) and BCES
Orthogonal methods, where the former treats the luminosity as
dependent variable and minimizes its residuals, while the
latter implies both variables, $L$ and $T$, as independent variables
and minimizes the orthogonal distances to the linear relation.

As a step forwards, we include the information obtained about the ICM
velocity dispersion along the l.o.s. in order to investigate the
effects on slope, normalization and intrinsic scatter of the best--fit
$L_X-T$ relation.

Interestingly, the exclusion from the original sample of the haloes
with largest velocity dispersion (normalized to the characteristic
thermal value associated to their temperature) has the effect of reducing
the scatter in the $L_X-T$ relation, as shown by
the trend in \fig\ref{fig:sigv_trends} (bottom panel).
The increasing trend of slope and scatter, in particular, evokes a
dependence on the contamination of the sample by haloes which show
significant fraction of non--thermal velocity, although the results
would need a larger statistics to be more strongly confirmed,
especially at the low--temperature end of the relation, which is not
probed by our sample.

Nevertheless, an interesting indication of our velocity diagnostics is
that even haloes with fairly regular appearance or relatively low
central cooling time can be characterised by high non--thermal
velocities of the gas along the line of sight, deviating therefore
from the expected scaling relations among global properties (e.g. the $L_X-T$ scaling law).
Additional information on the velocity structure of
the clusters might therefore be helpful to constrain better their
state.
Observationally, for the clusters where the velocity measurements are
indeed achievable, this can be used as complementary characterisation
to the usual morphological or cooling--time--based approach. 

As a promising, future perspective, we expect high--precision X--ray
spectroscopy to provide valuable information on the non--thermal
velocity structure of the ICM in galaxy clusters along the line of
sight, which can be safely assumed to trace the intrinsic gas motions
of the cluster despite the effects due to projection and instrumental
response.
This will be certainly fundamental to correctly determine the total
mass in clusters and characterise in more detail the deviation of real
clusters from the hydrostatic equilibrium at the base of X--ray mass
estimates as well as the divergence from the self--similarity of the haloes expected from
theoretical models.
%--------------------------------------------------
\section*{Acknowledgments}
K.D. acknowledges the support by the DFG Priority Programme 1177 and additional
support by the DFG Cluster of Excellence ``Origin and Structure of the 
Universe''. 
The underlying cosmological simulations where performed within the
DEISA/DECI-6 initiative under the project ``MagPath''.
%--------------------------------------------------
%----------------- BIBLIOGRAPHY -------------------
\bibliographystyle{mn2e}
\bibliography{bibl.bib}

\appendix
\section[]{Treatment of chemical abundances in \phox{}}\label{app:metallicity}
Dealing with galaxy clusters, the main emission components in the
X--ray regime are constituted by bremsstrahlung continuum and emission
lines from heavy elements. 

Usually, simulations of the X--ray emission
from hydrodynamically simulated clusters assume the average value of
one third the solar metallicity, consistently to the average metallicity measured for real clusters.
With respect to previous hydro--simulations, however, the run analysed here
provides full information on the chemical enrichment of the gas,
tracing its composition with $10$ elements, namely: He, C, Ca, O, N,
Ne, Mg, S, Si, Fe. An additional field for each gas particle accounts also
for the total mass in all the remaining metals.

Therefore, the available abundances of the individual elements, He included, are
read and used to calculate the contribution of each element to the final
spectrum.

This is done, in \phox{}, via the external package \xspec{}, assuming a
single--temperature \vapec{} \cite[or, if desired, 
\vmekal{}\footnote{\scriptsize{See http://heasarc.nasa.gov/xanadu/xspec/manual/XSmodelMekal.html.}}][]{mewe1985,ka_me1993,liedahl1995}
model for each
element, where the corresponding abundance is fixed at $1\zsun$.
The total spectrum ($S$) of a gas particle, at its temperature $T$, will then be
obtained as the sum of the continuum, due to H, plus the contribution
of each element, weighted by its abundance in solar units (set, here,
according to the values given by \cite{angr1989}).
Explicitly:
\begin{equation}
S(T,Z_{tot}) = S(T,{\rm H)} + \Sigma_i Z_i\,[ S(T,{\rm
    Z}_i) - S(T,{\rm H}) ]
\end{equation}
The abundances of the \vapec{} (\vmekal{}) elements which are not
explicitly treated in the hydro--simulation are estimated from the
remaining mass in metals associated to the particle, redistributing
them according to the solar abundance pattern.

\section[]{The simulated sample}\label{app:maps_em}
In order to have a visual presentation of the sample discussed in this
work, we show here emission measure (EM) maps for the $43$ clusters, for
the cylinder--like region along the line of sight, encompassed
by $\rfive{}$ around each cluster.
	\begin{figure*}
	\centering
        \includegraphics[width=0.49\textwidth]{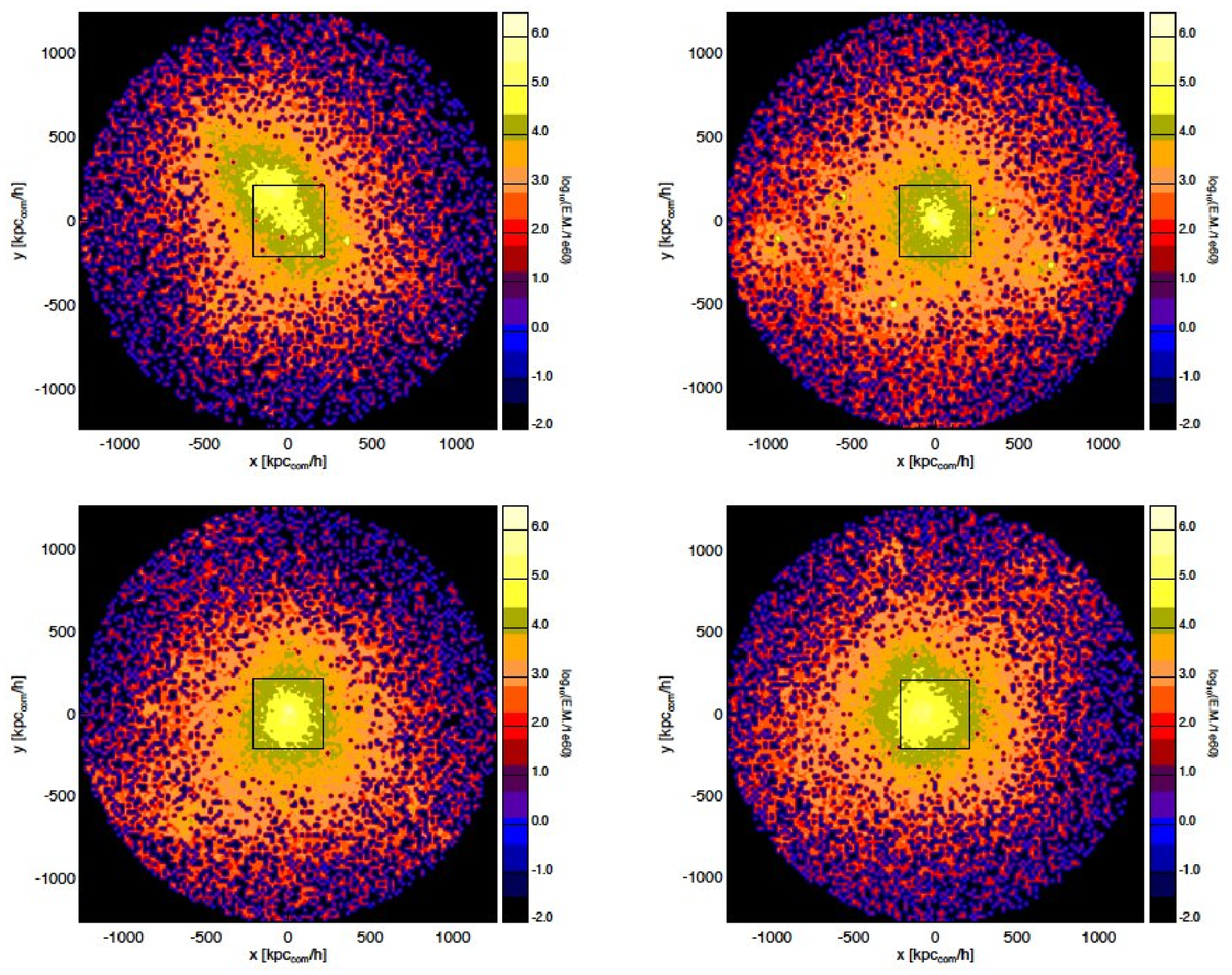}
        \includegraphics[width=0.49\textwidth]{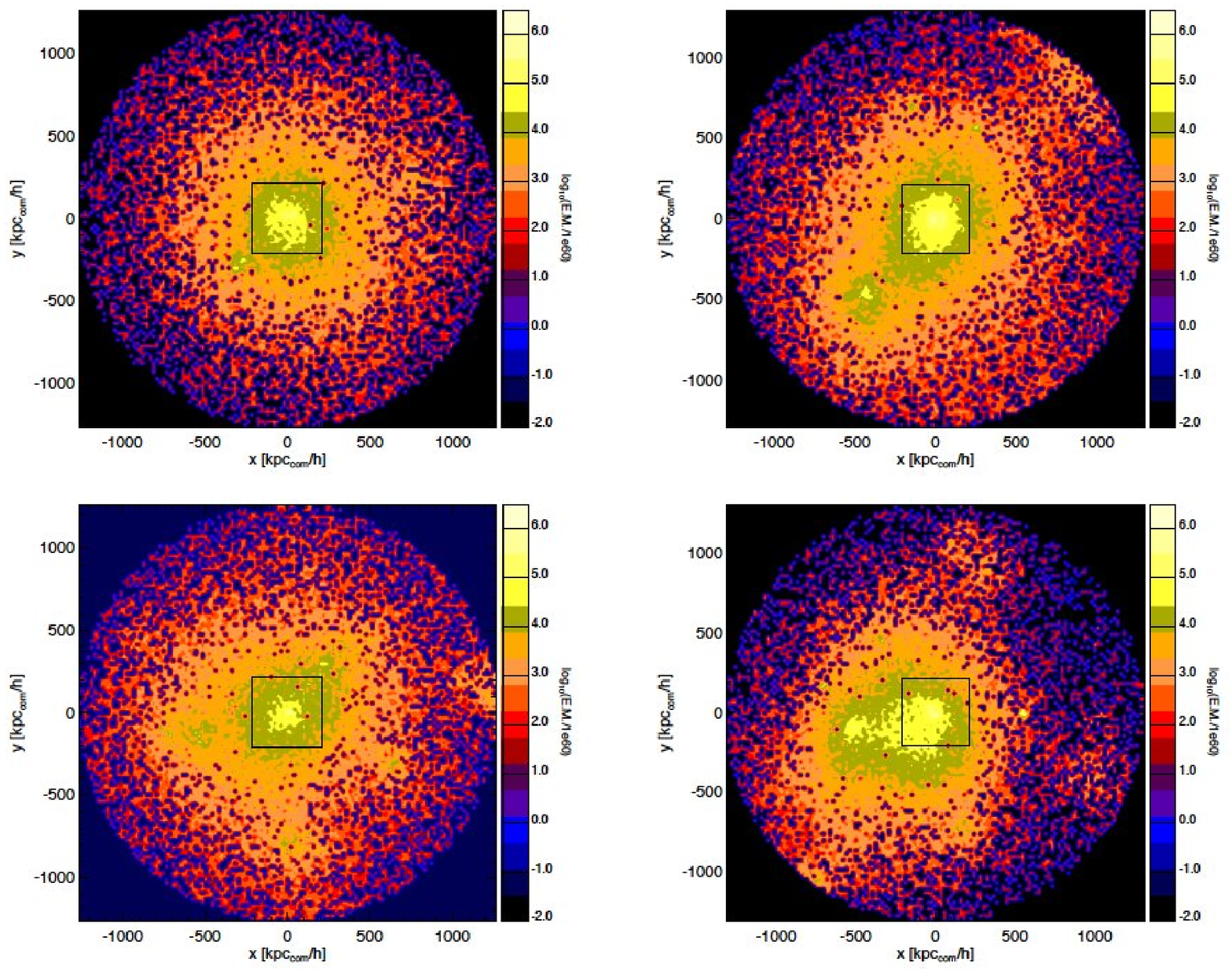}\\
        \includegraphics[width=0.49\textwidth]{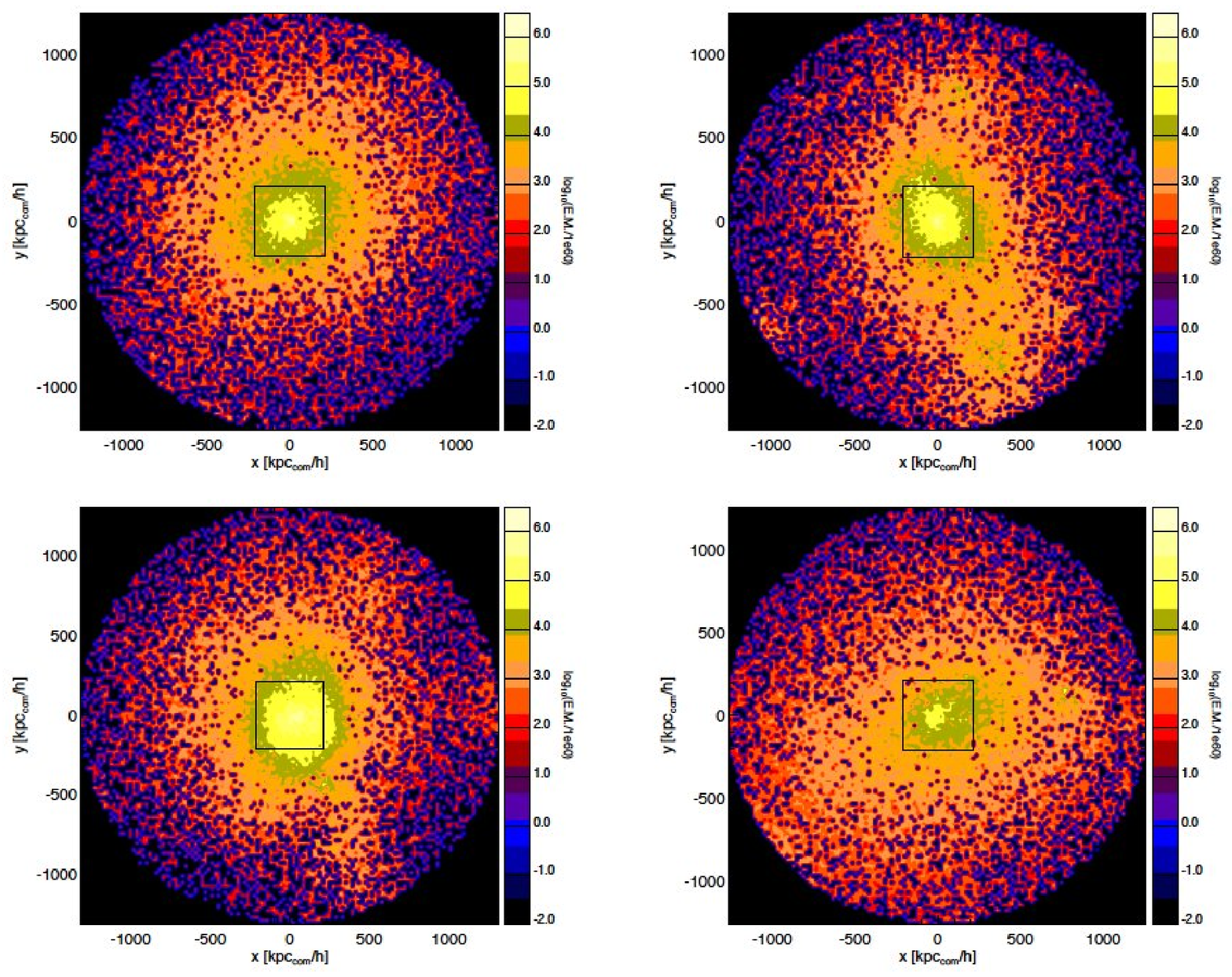}
        \includegraphics[width=0.49\textwidth]{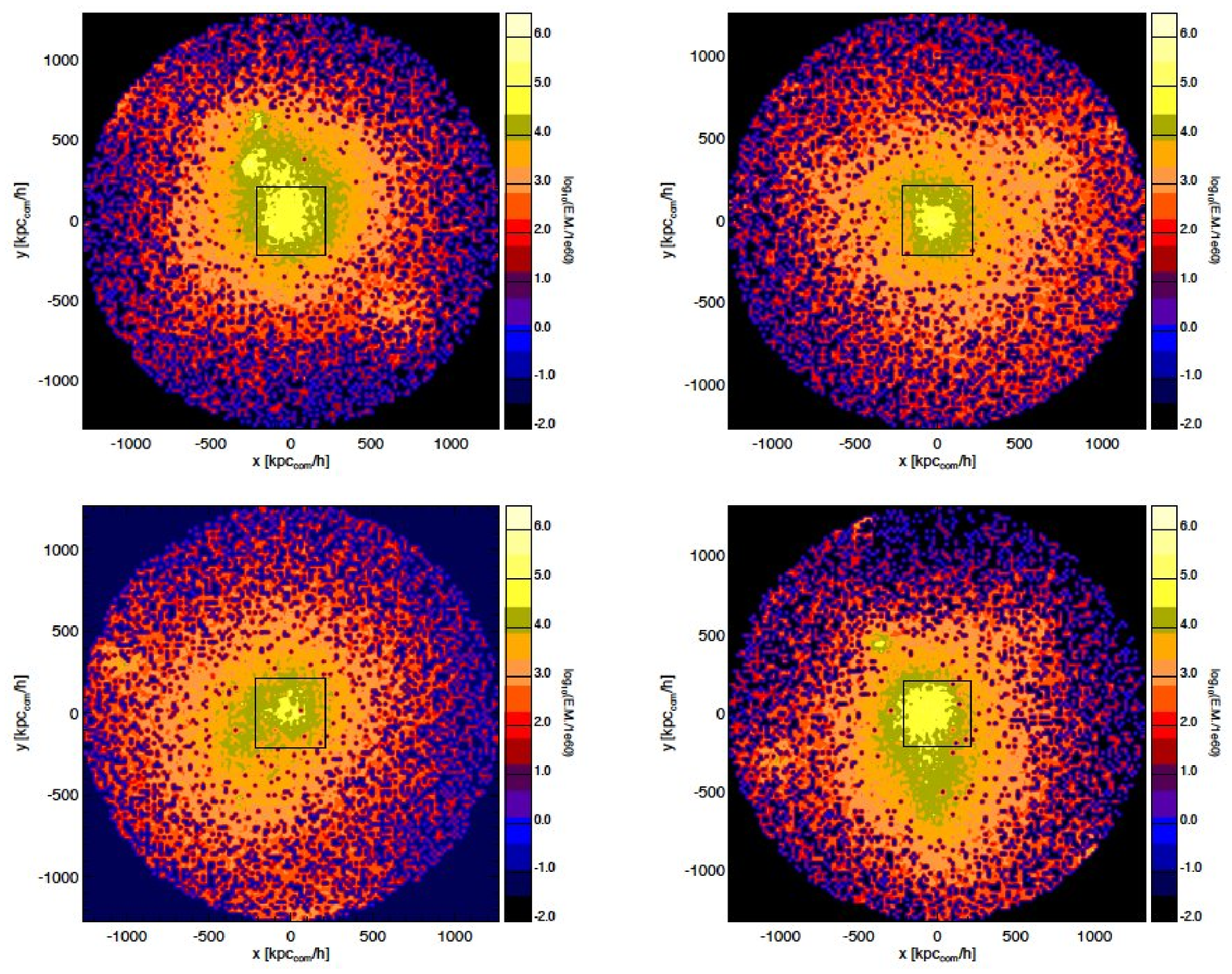}\\
        \includegraphics[width=0.49\textwidth]{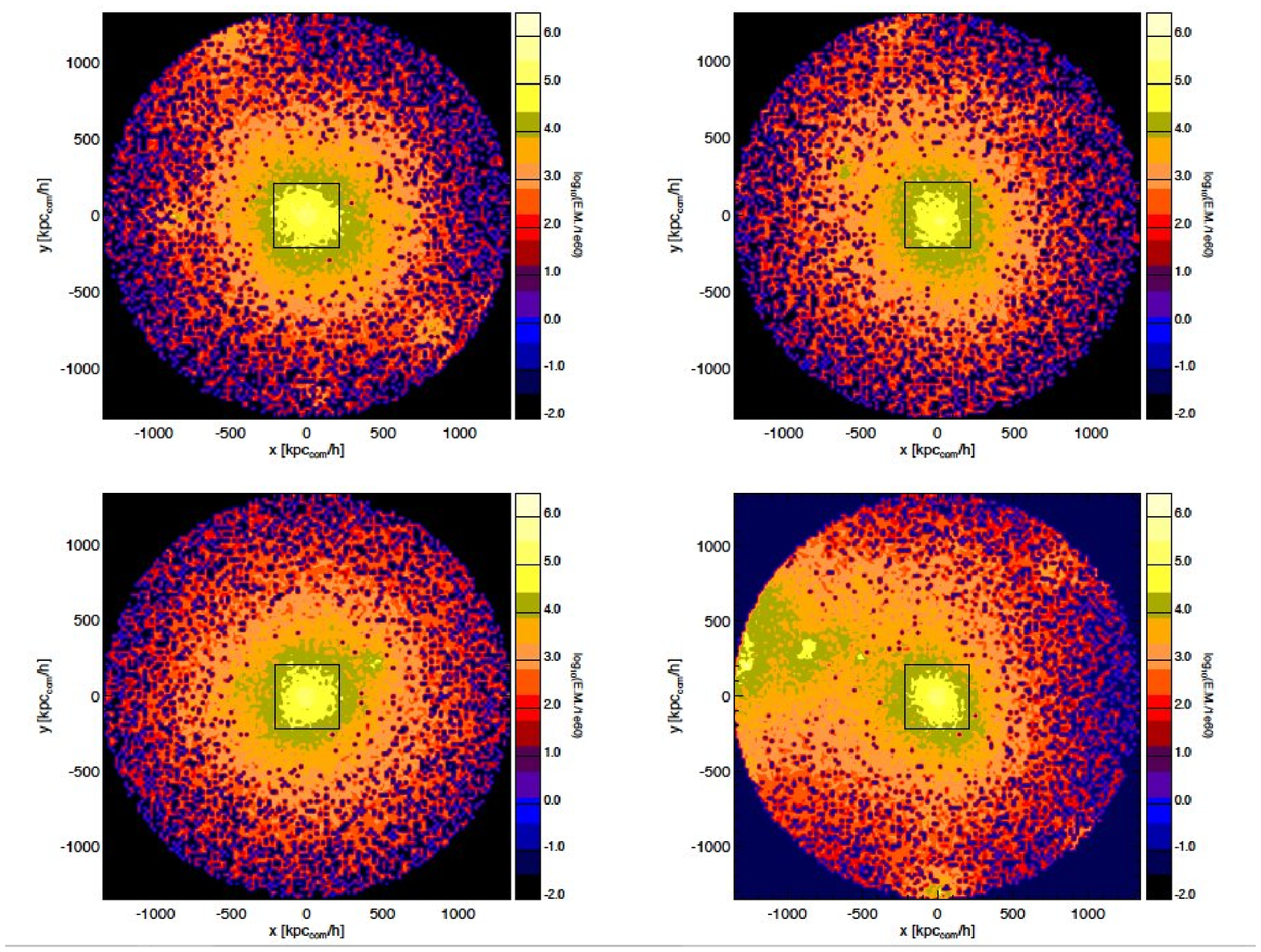}
        \includegraphics[width=0.49\textwidth]{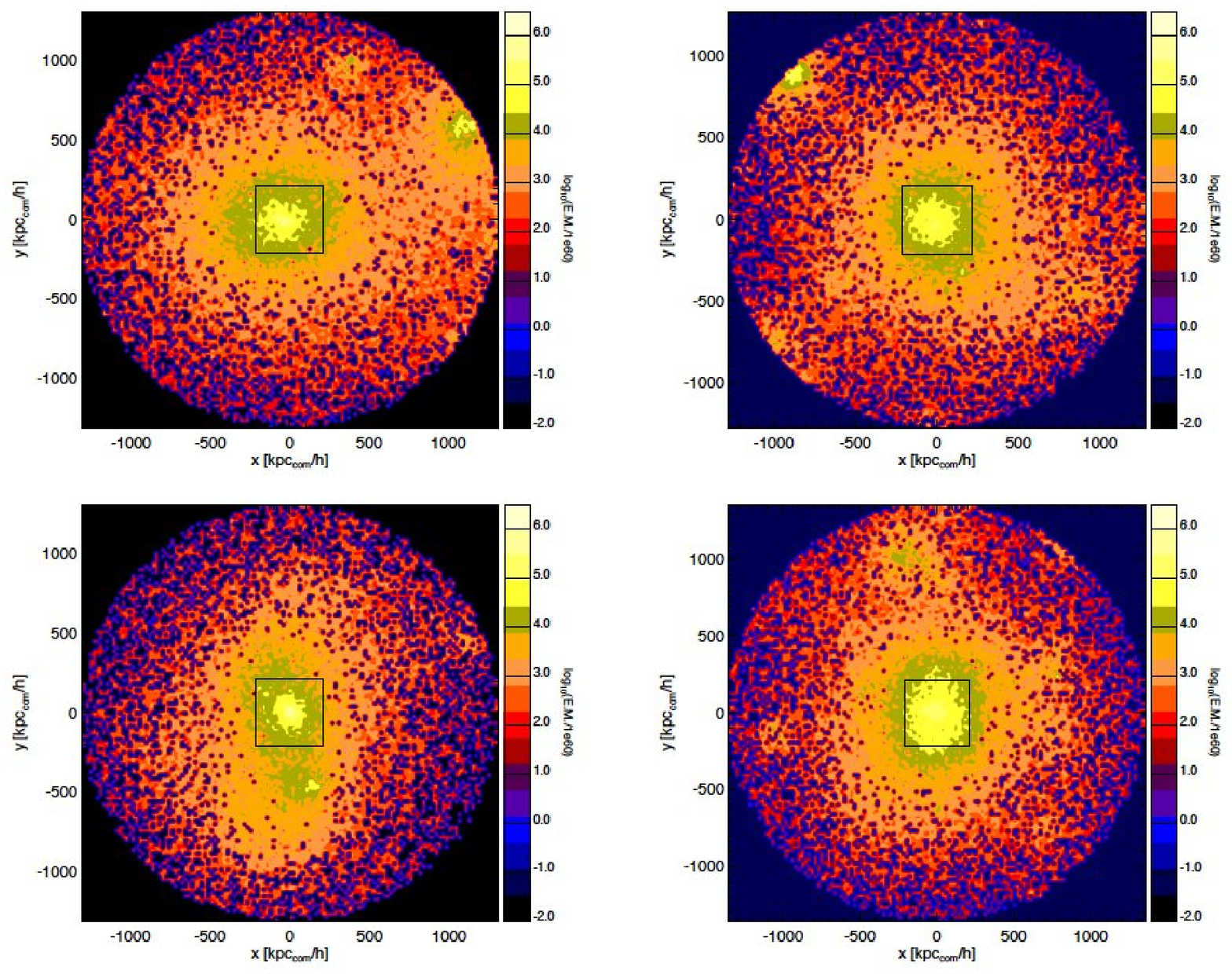}\\
        \includegraphics[width=0.49\textwidth]{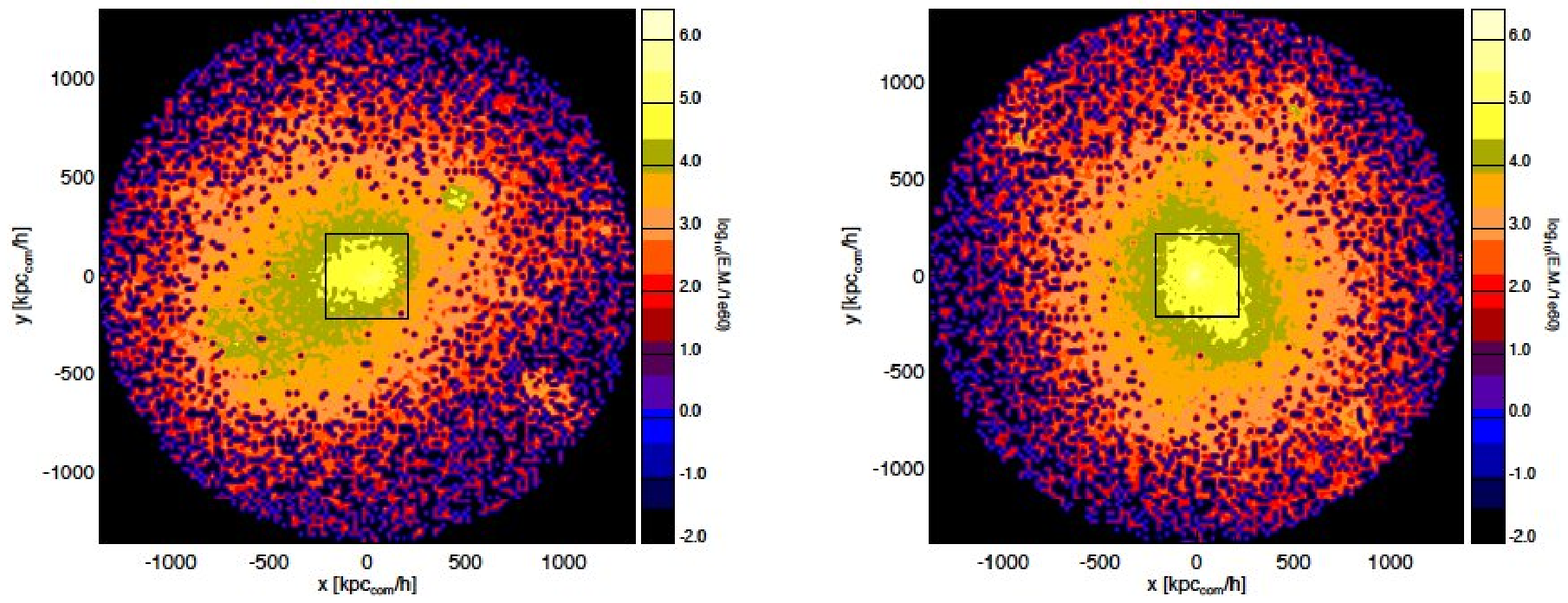}
        \includegraphics[width=0.49\textwidth]{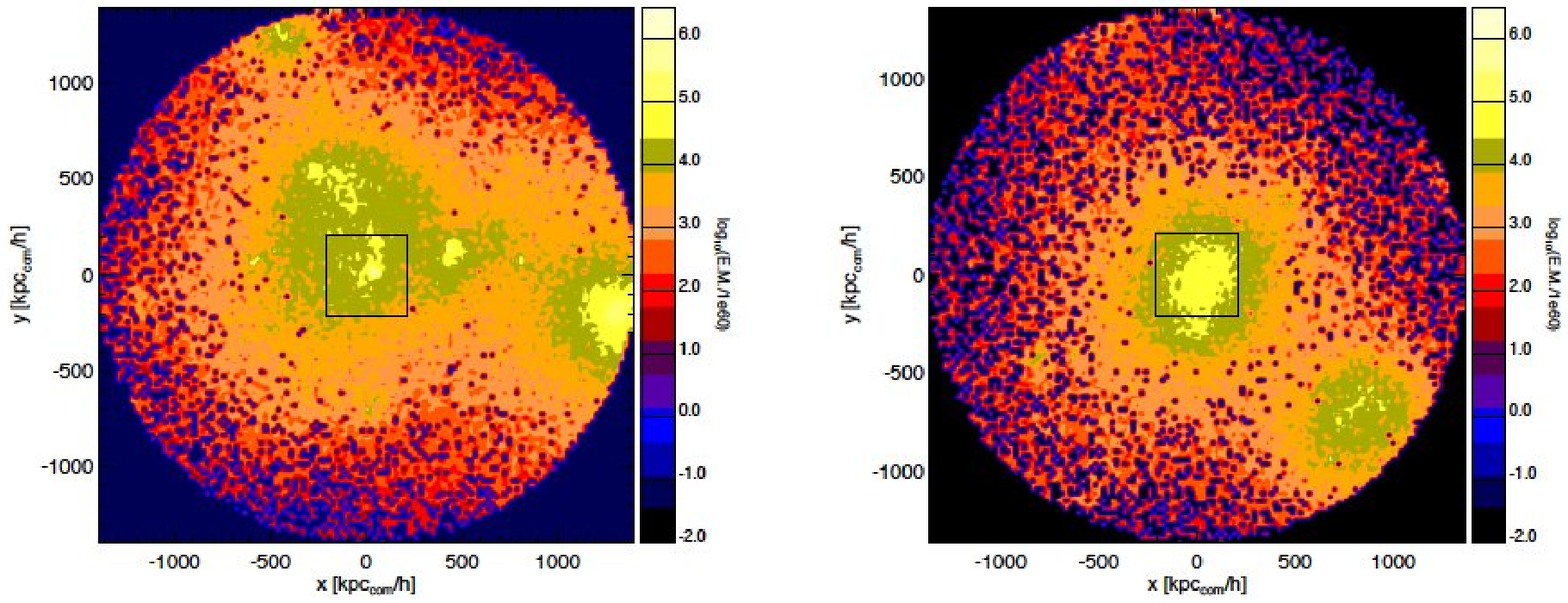}\\
        \caption{Visualization of the 43 most massive galaxy clusters
          in the simulated sample. The panels show the EM maps in the
          plane perpendicular to the line of sight, within the
          $\rfive$ region of each cluster.}
        \label{fig:maps1}
      \end{figure*}

	\begin{figure*}
	\centering 
        \includegraphics[width=0.49\textwidth]{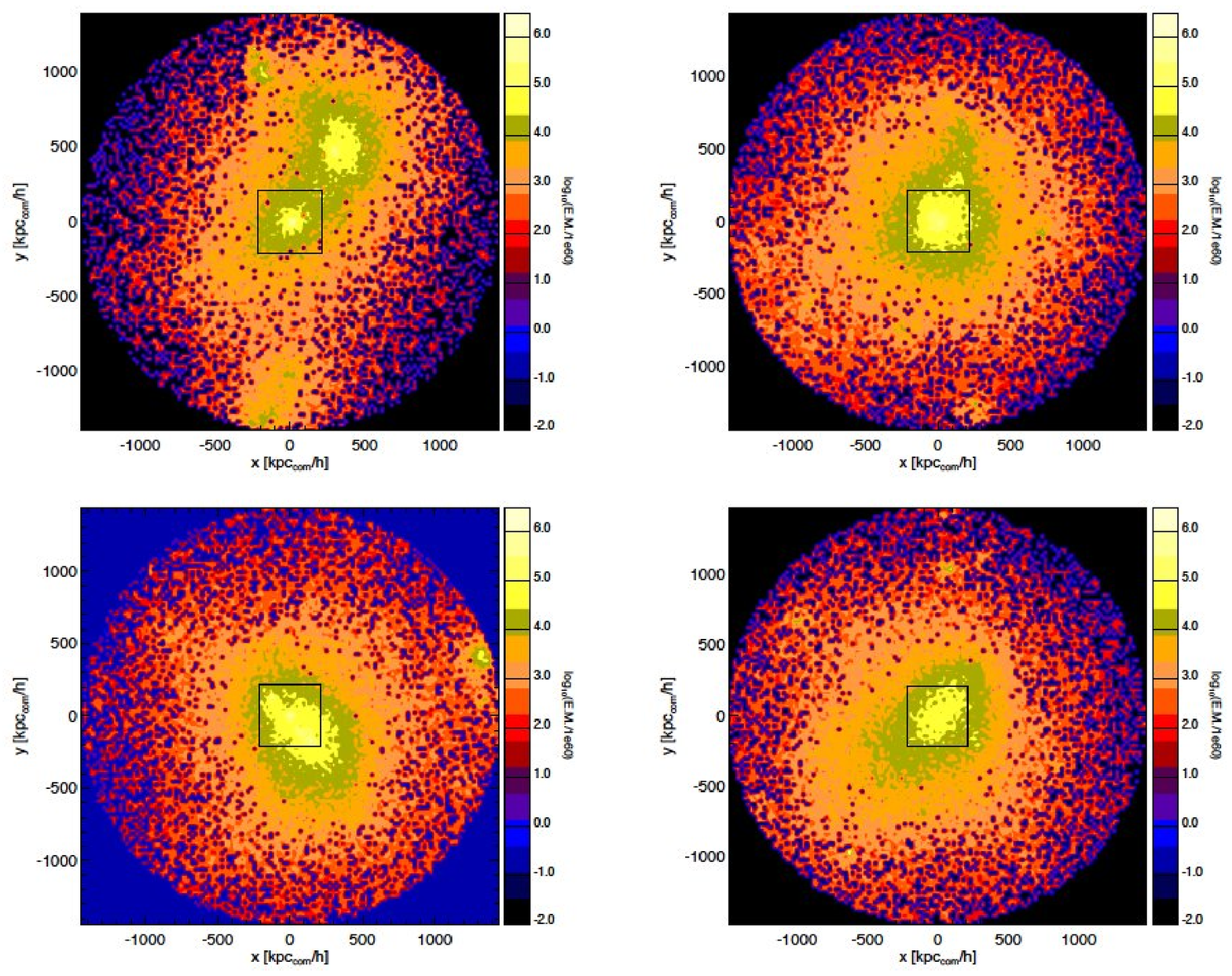}
        \includegraphics[width=0.49\textwidth]{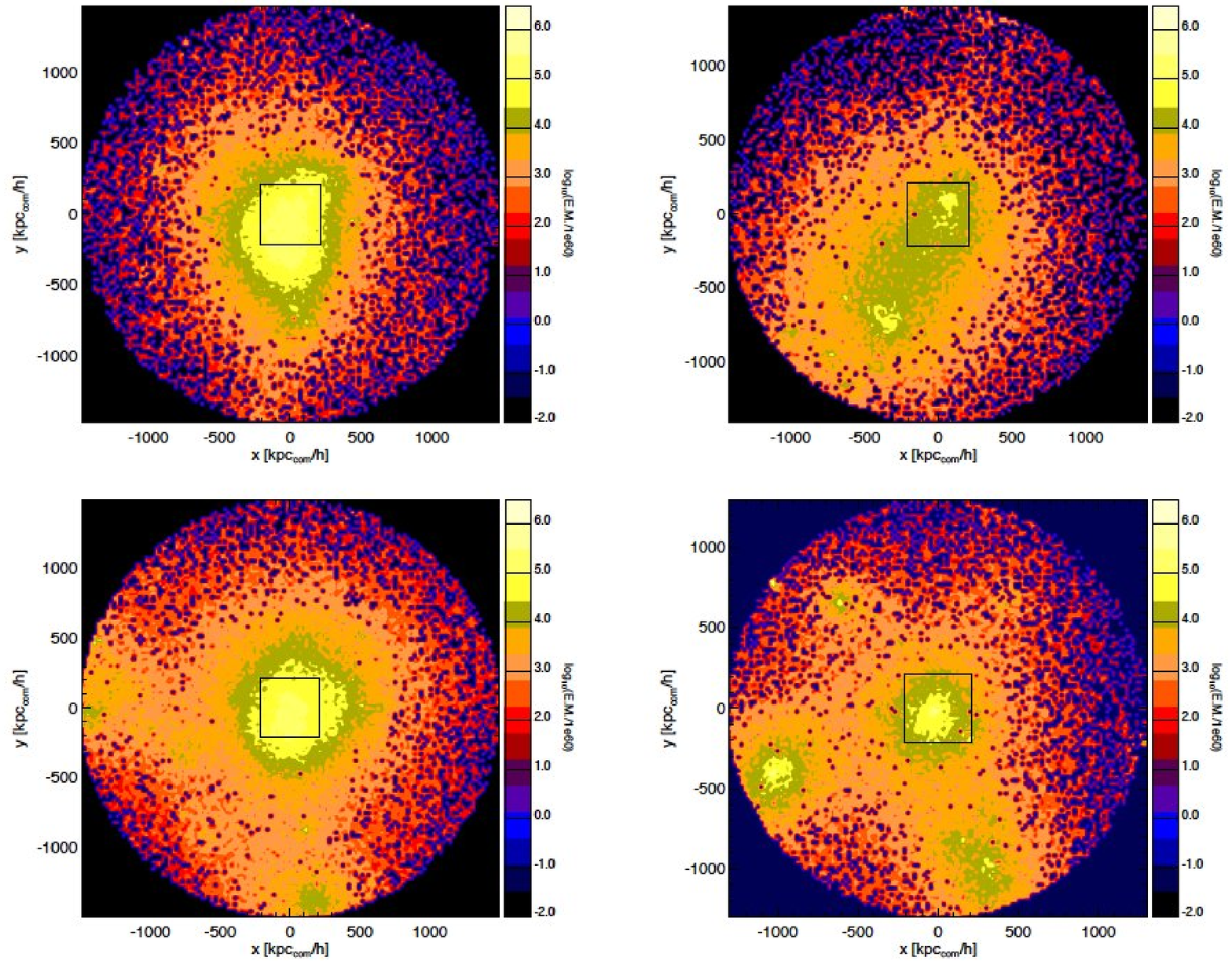}\\
        \includegraphics[width=0.49\textwidth]{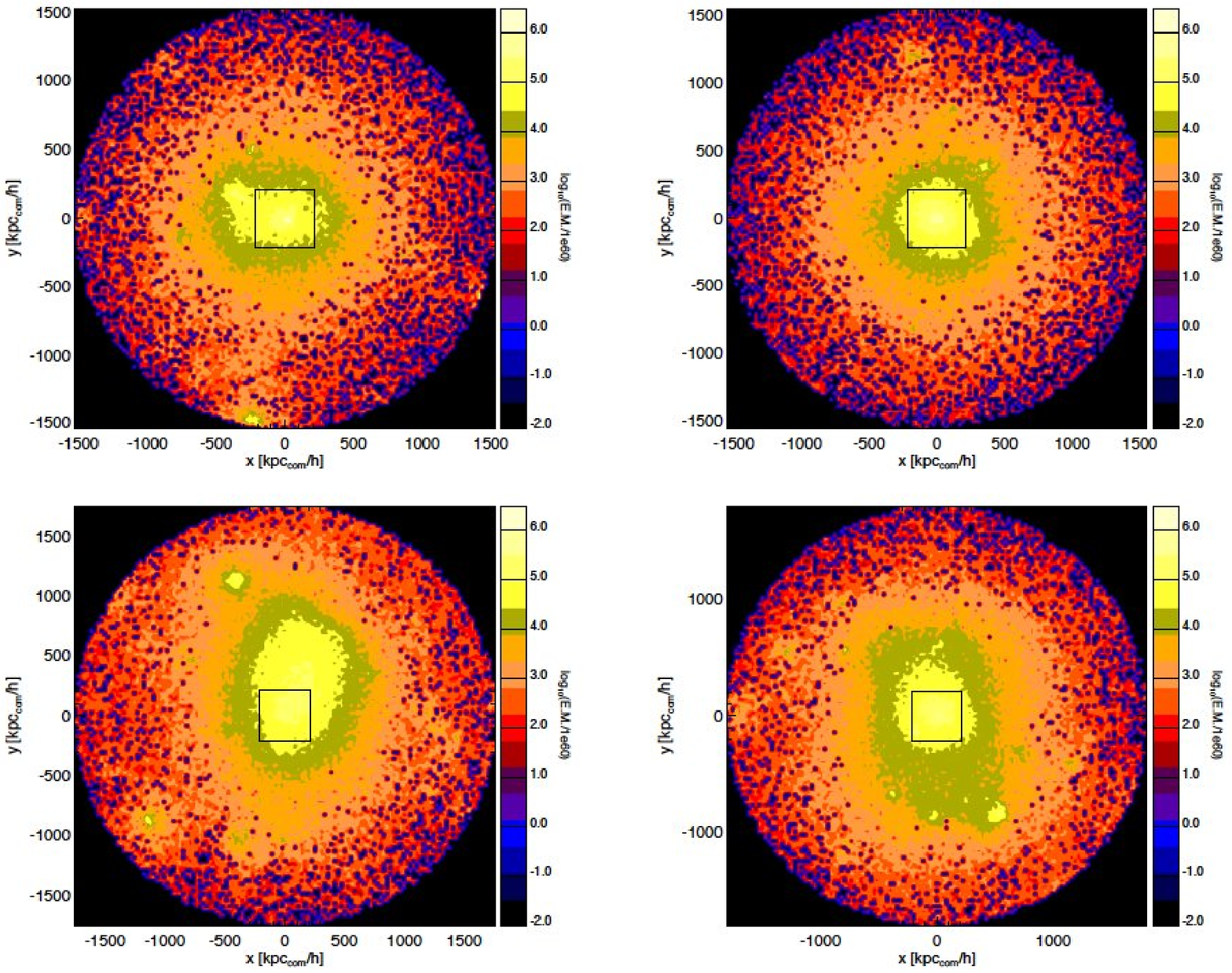}
        \includegraphics[width=0.49\textwidth]{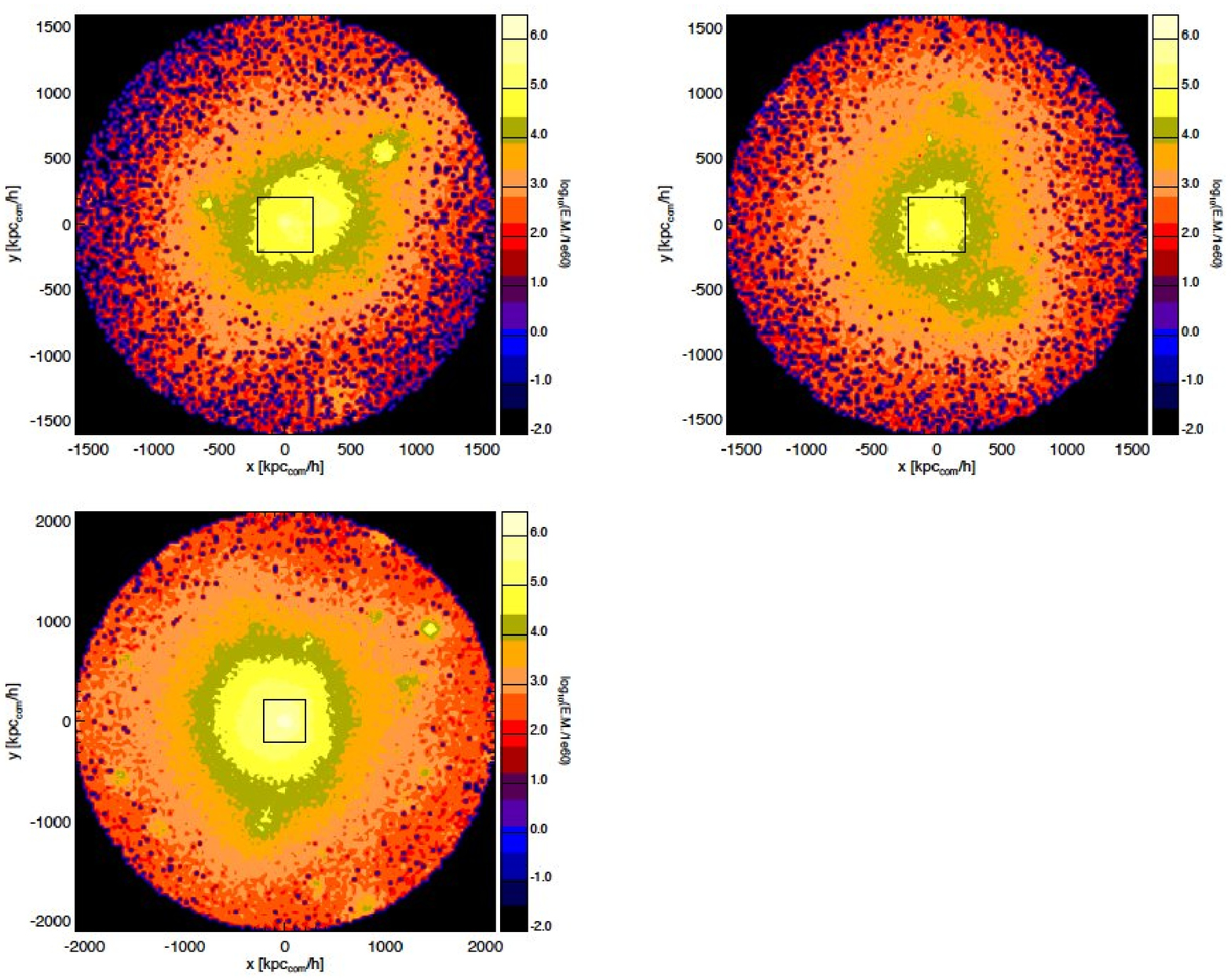}\\
% 	\includegraphics[width=0.24\textwidth]{figure/map_x_y_EM.30.ps}
% 	\includegraphics[width=0.24\textwidth]{figure/map_x_y_EM.31.ps}
% 	\includegraphics[width=0.24\textwidth]{figure/map_x_y_EM.32.ps}
% 	\includegraphics[width=0.24\textwidth]{figure/map_x_y_EM.33.ps}
% 	\includegraphics[width=0.24\textwidth]{figure/map_x_y_EM.34.ps}
% 	\includegraphics[width=0.24\textwidth]{figure/map_x_y_EM.35.ps}
% 	\includegraphics[width=0.24\textwidth]{figure/map_x_y_EM.36.ps}
% 	\includegraphics[width=0.24\textwidth]{figure/map_x_y_EM.37.ps}
% 	\includegraphics[width=0.24\textwidth]{figure/map_x_y_EM.38.ps}
% 	\includegraphics[width=0.24\textwidth]{figure/map_x_y_EM.39.ps}
% 	\includegraphics[width=0.24\textwidth]{figure/map_x_y_EM.40.ps}
% 	\includegraphics[width=0.24\textwidth]{figure/map_x_y_EM.41.ps}
% 	\includegraphics[width=0.24\textwidth]{figure/map_x_y_EM.42.ps}
% 	\includegraphics[width=0.24\textwidth]{figure/map_x_y_EM.43.ps}
% 	\includegraphics[width=0.24\textwidth]{figure/map_x_y_EM.44.ps}
% 	\phantom{\includegraphics[width=0.24\textwidth]{figure/map_x_y_EM.44.ps}}
        \caption{Continues from \fig\ref{fig:maps1}.}
        \label{fig:maps2}
      \end{figure*}
%--------------------------- END ------------------------------------

\bsp

\label{lastpage}

\end{document}